\documentclass[showpacs,aps,prd,nofootinbib,showkeys,superscriptaddress]{revtex4}
\usepackage{graphicx}
\usepackage{bm}
\usepackage{bbold}
\usepackage{amssymb}
\usepackage{float}
\usepackage{amsmath,latexsym}
\usepackage{amsfonts}
\usepackage[usenames]{color}
\usepackage{subfigure}
\usepackage{subfigure}
\usepackage{physymb}
\usepackage{slashed}
\usepackage{multirow,array}
\usepackage{mathtools}
\usepackage{mathrsfs}
\usepackage{dsfont}
\usepackage[colorlinks=false,linktocpage=true]{hyperref}
\usepackage{hyperref}
\usepackage[utf8]{inputenc}
\usepackage{aurical}
\usepackage[T1]{fontenc}
\usepackage{mathptmx}
\usepackage{soul}

\newcommand{\be}{\begin{eqnarray}}
\newcommand{\ee}{\end{eqnarray}}
\newcommand{\bea}{\begin{eqnarray}}
\newcommand{\eea}{\end{eqnarray}}
\newcommand{\bs}{\begin{subequations}}
\newcommand{\es}{\end{subequations}}
\newcommand{\beal}{\begin{align}}
\newcommand{\eal}{\end{align}}

\newcommand{\mc}[1]{\mathcal{#1}}
\newcommand{\cnl}{c_{nl}}
\newcommand{\cl}{c_{l}}

\newcommand{\aol}{\alpha_{0l}}

\newcommand{\gol}{\gamma_{0l}}

\newcommand{\nl}{\nonumber \\}
\newcommand{\pde}{\partial}

\DeclareUnicodeCharacter{2212}{-}

\begin{document}
\vskip 1.5in

\title{Dynamical systems and nonlinear transient rheology of the far-from-equilibrium Bjorken flow}
\author{Alireza Behtash}
\email[Email: ]{abehtas@ncsu.edu}
\affiliation{Department of Physics, North Carolina State University, Raleigh, NC 27695, USA}
\author{Syo Kamata}
\email[Email: ]{skamata@ncsu.edu}
\affiliation{Department of Physics, North Carolina State University, Raleigh, NC 27695, USA}
\author{Mauricio Martinez}
\email[Email: ]{mmarti11@ncsu.edu}
\affiliation{Department of Physics, North Carolina State University, Raleigh, NC 27695, USA}
\author{Haosheng~Shi}
\email[Email: ]{hshi3@ncsu.edu}
\affiliation{Department of Physics, North Carolina State University, Raleigh, NC 27695, USA}

\begin{abstract}
In relativistic kinetic theory, the one-particle distribution function is approximated by an asymptotic perturbative power series in Knudsen number which is divergent.  For the Bjorken flow, we expand the distribution function in terms of its moments and study their nonlinear evolution equations. The resulting coupled dynamical system can be solved for each moment consistently using a multi-parameter transseries which makes the constitutive relations inherit the same structure. A new non-perturbative dynamical renormalization scheme is born out of this formalism that goes beyond the linear response theory. We show that there is a Lyapunov function, aka dynamical potential, which is, in general, a function of the moments and time satisfying Lyapunov stability conditions along RG flows connected to the asymptotic hydrodynamic fixed point. As a result, the transport coefficients get dynamically renormalized at every order in the time-dependent perturbative expansion by receiving non-perturbative corrections present in the transseries.  The connection between the integration constants and the UV data is discussed using the language of dynamical systems. Furthermore, we show that the first dissipative correction in the Knudsen number to the distribution function is not only determined by the known effective shear viscous term but also a new high energy non-hydrodynamic mode. It is demonstrated that the survival of this new mode is intrinsically related to the nonlinear mode-to-mode coupling with the shear viscous term. Finally, we comment on some possible phenomenological applications of the proposed non-hydrodynamic transport theory.
\end{abstract}
\date{\today}
\maketitle

\section{Introduction}
\label{sec:intro}
Recent measurements of multiparticle cumulants in $d+Au$ , $p+Pb$, $p+ Au$, and $^3 He+Au$ systems have provided a compelling evidence for collective flow in heavy-light ions collisions~\cite{Chatrchyan:2013nka,Adare:2013piz,Aad:2012gla,Abelev:2012ola,PHENIX:2018lia}. Similar observations had been made previously in nucleus-nucleus collisions (cf. Refs.~\cite{Gale:2013da,Heinz:2013th} and references therein). Altogether, these measurements seem to imply that the collective evolution of a nuclear fireball created in these experiments can be understood in terms of hydrodynamics. 

The phenomenological success of hydrodynamic models in producing an accurate description of these extreme experimental situations have raised some questions as per the validity of the assumptions that hydrodynamics relies on. For instance, the local thermal equilibrium hypothesis has been usually thought of as a necessary and sufficient criterion for the applicability of the fluid-dynamical equations of motion. Different theoretical studies~\cite{Kurkela:2015qoa,Critelli:2017euk,Denicol:2014xca,Florkowski:2013lza,Florkowski:2013lya,Denicol:2014tha,Chesler:2009cy,Heller:2011ju,Wu:2011yd,vanderSchee:2012qj,Chesler:2016ceu,Casalderrey-Solana:2013aba,Martinez:2010sc} have shown that the equilibrium condition seems to be too restrictive, and that surprisingly no inconsistency arises when hydrodynamics is used for interpreting a system sitting far from equilibrium. These findings have repeatedly pointed out to the possibility of generalizing relativistic hydrodynamic theories for systems with the local thermal equilibrium removed~\cite{Romatschke:2017vte,Florkowski:2017olj,Romatschke:2017ejr}. Although the idea of employing fluid dynamics in non-thermal equilibrium physics is not entirely new~\cite{rheology1,rheology2}, little progress has been made to formulate it from first principles. 

In recent years we have learned more about certain generic properties of far-from-equilibrium hydrodynamics that follows suitably from the nonlinear nature of fluid dynamics. The breakdown of the naive perturbative asymptotic expansion is, therefore, imminent, meaning that a series ansatz fails to converge and accordingly it cannot be a full-blown solution to the underlying nonlinear differential equations. This has been long known in mathematics~\cite{costin1998}. An example of this fact in hydrodynamics was given in~\cite{Heller:2015dha} where the gradient expansion of the energy-momentum tensor of a conformal fluid was shown to be divergent. The divergences of the energy-momentum tensor ought to beg for the existence of additional degrees of freedom called \textit{non-hydrodynamic modes}.  Since then more examples have appeared in generic setups in both strong and weakly coupled regimes~
\cite{Casalderrey-Solana:2017zyh,Noronha:2015jia,Bazow:2016oky,Strickland:2017kux,Denicol:2017lxn,Denicol:2018pak,Romatschke:2017vte,Buchel:2016cbj,Spalinski:2018mqg,Kurkela:2018xxd,Kurkela:2018oqw,Heller:2016rtz,Heller:2018qvh,Heller:2014wfa,Heller:2015dha}.

A more relevant and important question to ask in this context has to do with how non-hydrodynamic modes can affect the transport properties of the system and what their possible phenomenological significance is. We addressed this question in our previous work~\cite{Behtash:2018moe} by investigating the nonlinear dynamics of a far-from-equilibrium weakly coupled plasma undergoing Bjorken expansion. Our work first treated the equations governing the time evolution of momentum moments of the one-particle distribution function as a coupled nonlinear dynamical system, the solutions of which are found to be multi-parameter transseries with {\it real} integration constants $\sigma$. These transseries carry exponentially small (non-perturbative) information in terms of the Knudsen $Kn$ and inverse Reynolds $Re^{-1}$ numbers which both play the role of expansion parameters in the transseries, whereas the relaxation time $\tau_r$ characterizes the size of exponential corrections. A linear combination of these moments defines the constitutive relations at every order in the perturbative expansion whose coefficients are indeed related to the transport coefficients. Then summing over all exponentially small factors and other monomials of the expansion parameters in the transseris renders an effective dynamical renormalization of the transport coefficients away from the values obtained at equilibrium using linear response theory. It should be noted that in this approach, `dynamical renormalization' means that there exists a  positive-definite monotonically decreasing function(al) also known as {\it Lyapunov function} in the stability theory, along every flow line in the space of moments, that can be easily derived from our dynamical system. 

One advantage that comes with studying dynamical systems is the ability to have full control over the late-time (or IR) as well as early-time (or UV) behavior of the flow lines. We should keep in mind that the alternative techniques such as Borel resummation heavily rely on the asymptotics of the divergent series. However, in a dynamical system of evolving moments, the resurgent multi-parameter transseries is directly found as a formal solution without the need for going to Borel plane where UV information is completely lost. The Borel singularities in this sense only encode IR data such as the size of exponential corrections, which in our approach corresponds to the eigenvalues of the linearization matrix commonly known as {\it Lyapunov exponents} around the asymptotic hydrodynamic fixed point. Therefore, the theory after Borel resummation is only asymptotically equivalent to the original theory, and drawing any conclusions on the fate of solutions of the original theory in the UV from a resumed series is, in general, not possible.  

The organization of this paper is as follows. We extend and generalize our previous findings~\cite{Behtash:2018moe} to include higher non-hydrodynamic modes in Sects.~\ref{sec:kinmod}-\ref{sec:coustin} while providing technical details of our derivations in the Appendices. In Sect.~\ref{sec:global}, we explain further the deep relationship between dynamical systems and the resurgence theory in the time-dependent (non-autonomous) system at hand, which was first discussed in the context of relativistic hydrodynamics in Ref.~\cite{Behtash:2017wqg}. Finally, in Sect.~\ref{sec:hydrodyn} it will be reported that a new high energy non-hydrodynamic mode exists whose first-order perturbative asymptotic term goes like $\sim (\tau T)^{-1}$, which obviously displays the same decay as the Navier-Stokes (NS) shear viscous tensor component. The survival of this new non-hydrodynamic mode is due to the nonlinear mode-to-mode coupling with the effective shear viscous tensor. Some of the quantitative and qualitative properties of this new mode will also be discussed. As a last couple of remarks, we should remind the readers that for the sake of keeping the paper self-contained, we have included a small number of mathematical definitions from dynamical systems in App.~\ref{app:dynsys} that will make the read easier. Also, as mentioned above, we will adopt the field theory language when talking about the early-time and late-time regimes, which then can be interchangeable with UV and IR, respectively.

\section{Kinetic model}
\label{sec:kinmod}

The statistical and transport properties of the weakly coupled systems are usually described by kinetic theory. Within this approach, it becomes important to understand the evolution of one-particle distribution function from the Boltzmann equation~\cite{Groot:1980,Cercignani,Cercignani2}. Fluid-dynamical equations of motion arise as a coarse-graining process of the distribution function where the fastest degrees of freedom are integrated out. The approximations made in the coarse-graining procedure do not necessarily lead to the same macroscopic evolution equations~\cite{Denicol:2012cn,Denicol:2012es}, and nor do they in general draw an accurate picture of the physics when compared against exact numerical solutions~\cite{Denicol:2014tha,Denicol:2014xca,Florkowski:2013lza,Florkowski:2013lya,Florkowski:2014sfa,Martinez:2017ibh}. The common approaches to finding approximate solutions of the Boltzmann equation are the Chapman-Enskog approximation ~\cite{chapman1990mathematical} and Grad's moment method~\cite{Grad,Groot:1980}. In the Chapman-Enskog method, the distribution function is expanded as a power series in the Knudsen number around its equilibrium value, which eventually leads to a divergent series~\cite{santos,Denicol:2016bjh}. In Grad's moment method, the distribution function is expanded about the equilibrium state in terms of an orthogonal set of polynomials, and the coefficients of such expansion are average momentum moments~\cite{Grad}. The second approach, however, turns out to be more convenient when dealing with far-from-equilibrium backgrounds~\cite{Bazow:2013ifa,Molnar:2016vvu,Bluhm:2015raa,Martinez:2010sc,Martinez:2010sd}.  

In this work, we also adopt the latter method and study the full nonlinear aspects of the moments' evolution equations. We carry out our research plan by considering a weakly coupled relativistic system of massless particles which undergoes Bjorken expansion~\cite{Bjorken:1982qr} with vanishing chemical potential. We simplify our problem by assuming the Boltzmann equation within the relaxation time approximation (RTA-BE).    
   
For the Bjorken flow we use the Milne coordinates $x^\mu=(\tau,x,y,\varsigma)$ (with the longitudinal proper time $\tau=\sqrt{t^2-z^2}$ and pseudorapidity $\varsigma={\rm arctanh}\left(z/t\right)$), and the metric is $g_{\mu\nu}=\text{diag.}\left(-1,1,1,\tau^2\right)$. The time-like normal vector identified with the fluid velocity is taken to be $u^\mu=(1,0,0,0)$ (with $u^\mu u_\mu=-1$) and the spatial-like normal vector pointing along the $\varsigma$-direction is $l^\mu=(0,0,0,1)$ (with $l^\mu l_\mu=1$). The Bjorken flow symmetries reduce the RTA-BE to the following relaxation type equation~\cite{Baym:1984np}
\be
\label{eq:RTABoltzmann}
\partial_\tau f\left(\tau,p_T,p_\varsigma\right)=-\frac{1}{\tau_r(\tau)}\left[f\left(\tau,p_T,p_\varsigma\right)-f_{eq.}\left(-u\cdot p/T\right)\right]\,,
\ee
where $p_T$ is the transverse momentum, $p_\varsigma$ denotes the momentum component along $\varsigma$-direction, and $\tau_r$ represents the relaxation time scale. Also, $f_{eq.}$ is the local equilibrium J\"uttner distribution which without loss of generality is chosen to be of Maxwell–Boltzmann type, i.e., $f_{eq}(x)=e^{-x}$. We recall that $\tau_r$ sets the time scale at which the system relaxes to its thermal equilibrium. We shall consider models whose relaxation time is a power law in the effective temperature, namely
\be
\label{eq:rela-time}
\tau_r=\frac{\theta_0}{T^{1-\Delta}}\,.
\ee
For $\Delta=1$ means that the constant $\theta_0$ is dimensionful, while for $\Delta=0$ the theory is conformally invariant.\footnote{A slightly more general class of models for the relaxation time approximation have been studied in Ref.~\cite{Dusling:2009df}.} For pedagogical purposes we perform explicit calculations for the case of $\Delta=0$ in the main body of the paper, while the gist of results for the general case are discussed in Appendices ~\ref{app:general-Delta},~\ref{app:RTABE} and \ref{app:thermalizationDelta1}.  

The RTA-BE~\eqref{eq:RTABoltzmann} can be solved exactly~\cite{Baym:1984np}. In what follows, we will take a distinct approach in which the mathematical problem of solving the Boltzmann equation is recast into seeking solutions to a set of nonlinear ODEs for the moments. For the Bjorken flow we propose the following ansatz for the single-particle distribution function~\cite{Behtash:2018moe}
\be
\label{eq:ansfunc}
f\left(\tau, p_T,p_\varsigma\right)\,=f_{eq.}\left(\frac{p^\tau}{T}\right)\,\left[\sum_{n=0}^{N_n}\,\sum_{l=0}^{N_l}\,\cnl(\tau)\,\mc P_{2l}\left(\frac{p_\varsigma}{\tau\,p^\tau}\right)\,
\mc L_n^{(3)}\left(\frac{p^\tau}{T}\right)\,
\right]\,,
\ee
where $p^\tau=\sqrt{p_T^2+(p_\varsigma/\tau)^2}$ is the energy of the particle in the comoving frame, $\mc L_n^{(3)}$ and $P_{2l}$ denote the generalized Laguerre and the Legendre polynomials,  respectively. The ansatz~\eqref{eq:ansfunc} allows us to study hydrodynamization processes~\cite{Behtash:2018moe}. Furthermore, the nonlinear relaxation of the low energy ($n=0$) as well as the high energy tails ($n>0$) of the distribution function are better understood when $f(\tau,p_T,p_\varsigma)$ is expanded in terms of orthogonal polynomials~\cite{Bazow:2015dha,Bazow:2015cha}. 

The moments $\cnl$ are read directly from Eq.~\eqref{eq:ansfunc} as follows~\footnote{Blaizot and Li studied the time evolution of similar moments for a constant relaxation time ~\cite{Blaizot:2017lht} and a more general nonlinear collisional kernel in the small angle approximation~\cite{Blaizot:2017ucy}. In their case, the authors were interested in the details of the longitudinal momentum anisotropy which in our notation this corresponds to those moments with $n=0$. Up to some normalization factor, the main difference between the moments $\mc L_l$ (see Eq.~(2.8) in Ref.~\cite{Blaizot:2017lht}) and our moments $c_{0l}$ is that the former are dimensionful.}

\be
\label{eq:momnl}
\cnl(\tau)=\,2\pi^2\,\frac{(4l+1)}{T^4(\tau)}\,\frac{\Gamma(n+1)}{\Gamma(n+4)}\,
\biggl\langle\,\left(p^\tau\right)^2\,P_{2l}\left(\frac{p_\varsigma}{\tau\,p^\tau}\right)\,
\mc L_n^{(3)}\left(\frac{p^\tau}{T}\right)\,\biggr\rangle\,,
\ee
where $\Gamma(n)$ is the Gamma function and the momentum average of any observable $\mc O(x^\mu,p^\mu)$ weighted by an arbitrary distribution function $f_X$ is denoted as $\langle\,\mc O\,\rangle_X\equiv\int_{\bf p}\mc O(x^\mu,p^\mu) f_X(x^\mu,p_i)$  with $\int_{\bf p}\equiv \int d^2p_T dp_\varsigma/[(2\pi)^3\,\tau \,p^\tau]$. If $f_{eq.}(x)=e^{-x}$, then have that the hydrodynamic equilibrium (asymptotic IR fixed point) is given by $\cnl^{eq.}=\delta_{n0}\delta_{l0}$. 

For the Bjorken flow the energy momentum tensor $T^{\mu\nu}=\langle\,p^\mu\,p^\nu\,\rangle$ is~\cite{Molnar:2016vvu,Molnar:2016gwq,Huang:2009ue,Huang:2011dc,Gedalin1,Gedalin2} \footnote{It is customary to use the following tensor decomposition of the energy-momentum tensor
\begin{equation*}
T^{\mu\nu}=(\epsilon+P_0)\,u^\mu u^\nu\,+\,g^{\mu\nu}\,P_0\,+\,\pi^{\mu\nu}\,,
\end{equation*}
where $P_0$ is the equilibrium pressure and $\pi^{\mu\nu}$ is the shear viscous tensor. Nonetheless, for highly anisotropic systems, M\'olnar et. al~\cite{Molnar:2016vvu,Molnar:2016gwq} showed that using $T^{\mu\nu}$~\eqref{eq:tmn} is convenient. It should be noted that both formulations are equivalent~\cite{Molnar:2016vvu} and explicit examples for the Bjorken~\cite{Molnar:2016gwq} and Gubser flows~\cite{Martinez:2017ibh,Behtash:2017wqg} have already been discussed in the literature.}
\be
\label{eq:tmn}
\begin{split}
T^{\mu\nu}&=\epsilon\,u^\mu u^\nu\,+\,P_L\,l^\mu l^\mu\,+\,P_T\,\Xi^{\mu\nu}\,,
\end{split}
\ee
with the projector operator $\Xi^{\mu\nu}=g^{\mu\nu}+u^\mu u^\nu- l^\mu l^\nu$  which is orthogonal to both $u^\mu$ and $l^\mu$. The energy density $\epsilon$, the transverse and longitudinal pressures ($P_L$ and $P_T$ respectively) can be written in terms of the moments $\cnl$~\cite{Behtash:2018moe}
\bs
\label{eq:tmnmom}
\beal
\label{eq:enemom}
\epsilon=\langle \left(-u\cdot p\right)^2\rangle&=\,\frac{3}{\pi^2}\,c_{00}\,T^4\,,\\
\label{eq:ptmom}
P_T=\biggl\langle\,\frac{1}{2}\Xi^{\mu\nu}p_\mu p_\nu\,\biggr\rangle &=\epsilon\,\left(\frac{1}{3}-\frac{1}{15}c_{01}\,\right)\,,\\
\label{eq:plmom}
P_L=\biggl\langle\,\left(l\cdot p\right)^2\,\biggr\rangle &=\,\epsilon\,\left(\frac{1}{3}+\frac{2}{15}\,c_{01}
\,\right)\,.
\end{align}
\es
It follows that $\epsilon=2P_T+P_L$ from these expressions. We also mention the nonnegativity of pressure components, $P_T,P_L\ge0$, sets the physical range for $c_{01}$ as
\be
-2.5 \le c_{01} \le 5. \label{eq:physical_range}
\ee
These bounds are satisfied by the exact solution of the RTA-BE~\eqref{eq:RTABoltzmann} but it is not expected to be satisfied by a particular truncation scheme for the distribution function.

The energy-momentum conservation  together with the Landau matching condition for the energy density imply $c_{00}\equiv 1$. The only independent (normalized) shear viscous component $\bar{\pi}\equiv\tau^2\,\pi^{\varsigma\varsigma}/\epsilon$ for the Bjorken flow is proportional to the moment $c_{01}$~\cite{Behtash:2018moe} 
\be
\label{eq:pressanis}
\bar{\pi}&=\frac{2}{3}\left(\frac{P_L-P_T}{\epsilon}\right)\,=\,\frac{2}{15}\,c_{01}.
\ee

In a previous work of some of us~\cite{Behtash:2018moe}, it was shown that the time evolution of the temperature $T$ and the moments $\cnl$ with $n \ge 0$ and $l\geq 1$ is described by an infinite number of coupled nonlinear ODEs as follows
\bs
\label{eq:ODEsBjor}
\beal
\label{eq:Tevol}
&\frac{1}{T}\frac{d T}{d\tau}+\frac{1}{3\tau}=-\frac{c_{01}}{30\,\tau}\,,\\
\label{eq:cnlevol}
&\frac{d\cnl}{d\tau} +\frac{1}{\tau}\left[\alpha_{nl}\,c_{nl+1}\,+\beta_{nl}\,\cnl+\gamma_{nl}\,c_{nl-1}\,-\,n\left(\rho_{l}\,c_{n-1l+1}\,+\,\psi_{l}\,c_{n-1l}\,+\,\phi_{l}\,c_{n-1l-1}\right)\right]
+\frac{1}{\tau_r(\tau)}\left(c_{nl}-\delta_{n,0}\delta_{l,0}\right)
=0\,,
\end{align}
\es 
where the coefficients are given by
\be
\label{eq:coefevolcnl}
\begin{split}
&\alpha_{nl}= \frac{(2 + 2l)(1 + 2 l) (n+1 - 2l)}{(4l+3)(4l+5)}\,,\hspace{1cm}
\beta_{nl}=\frac{2l(2l+1)(5+2n)}{3(4l+3)(4l-1)}-\frac{(4+n)}{30}c_{01}\,,
\hspace{1cm} \phi_{l}= \frac{(2l)(2l-1)}{(4l-3)(4l-1)}\,,
\,,\\
&\gamma_{nl}= (2l+2+n)\frac{(2l)(2l-1)}{(4l-3)(4l-1)}\,\,,\hspace{1.5cm}
\psi_{l}= \frac{1}{3}\left(\frac{4l(2l+1)}{(4l+3)(4l-1)}\right)\,-\,\frac{c_{01}}{30}\,,\hspace{1cm}
 \rho_{    l}=\frac{(2l+1)(2l+2)}{(4l+3)(4l+5)}\,,
\end{split}
\ee
The hierarchy of equations in~\eqref{eq:ODEsBjor} constitute a dynamical system where  moments of different degree $n$ and $l$ mix amongselves in a nontrivial way. The nonlinear nature of the RTA-BE is manifest in the set of ODEs~\eqref{eq:cnlevol} by the {\it mode-to-mode} coupling term $\sim c_{nl}\,c_{01}$. Nonetheless, one observes that the low energy modes $c_{0l}$ decouple entirely from the high energy ones $\cnl$ with $n > 0$. As a result, the time evolution of the energy momentum tensor is fully reconstructed from the solutions of the temperature and the modes $c_{0l}$. Notice that one cannot deduce the same thing about the high energy modes whose evolution receive major contribution from the lower energy modes.  Furthermore, the high energy moments shall play a role in the stability of the system and its convergence to its asymptotic thermal state, which is the subject of discussion in Sect.~\ref{sec:hydrodyn}.

\section{Transseries solutions and Costin's formula}
\label{sec:coustin}
In this section, we construct the transseries solution to the dynamical system in~\eqref{eq:ODEsBjor}.
Once one reduces to the Boltzmann equation to a dynamical system, next immediate step would be to construct an exact form of the transseries around each asymptotic fixed point and obtain all the coefficients recursively from the evolution equations. In what follows, we will attempt to build the general form of the exact transseries solution by slightly modifying the original set of ODEs.
For technical reasons, we also start with a truncation of the dynamical system at $0 \le l \le L $ and $0 \le n \le N $, hence the original Boltzmann distribution will be reproduced by taking the limit $N,L \rightarrow \infty$.
Since we are interested in building the solutions starting at late times, we will assume the convergence at infinity, i.e., $c(\tau) \rightarrow 0$ for $\tau \rightarrow \infty$.

Let us prepare the set of ODEs in the dynamical system \eqref{eq:ODEsBjor} in the following asymptotically linearized form:
\be
&& \frac{d {\bf c}}{d w} \, = \, {\mathbf f}(w,{\bf c}),  \nl
&& {\mathbf f}(w,{\bf c}) \,=\, - \left[ \hat{\Lambda} {\bf c} + \frac{1}{w} \left( {\mathfrak B} {\bf c} + {\bf A} \right)   \right] + \mathcal{O}( {\bf c}^2, \, {\bf c}/w^2), \qquad (w \rightarrow \infty, \, {\bf c} \rightarrow 0) \label{eq:dcdw_o}
\ee
where $\hat{\Lambda}$ and ${\frak B}$ are constant matrices, 
${\bf c}$ is an $((N+1) \times (L+1) -1)$-dimensional vector given by
\be
&&   {\bf c} \, = \, (c_{01},\,\dots,\ c_{0L} ,\,c_{10},\,c_{11},\,\dots, \, c_{1L},\, \dots\,, c_{N0},\,\dots\, , \, c_{NL}),
\ee
and ${\bf A}$ is a constant vector.
Here, we have defined a new time coordinate as $w=\tau T(\tau)$ which behaves asymptotically like $w \sim \tau^{2/3}$ at late times.
We remind that ${\rm rank} (\hat{\Lambda}) = {\rm rank}({\frak B})= (N+1) \times (L+1) -1 =: I$. Furthermore, we assume $\hat{\Lambda}$ is a diagonal matrix proportial to the unit matrix. To construct the transseries solution, we suitably diagonalize ${\mathfrak B}$ by defining an invertible matrix $U$ such that
\be \label{eq:transformations}
\tilde{\bf c}(w) = U {\bf c}(w), \quad 
\hat{\mathfrak B} = U  {\mathfrak B}  U^{-1}={\rm diag}\, (b_1 ,\dots , b_{L}) \in {\mathbb C}^{L}, \quad \tilde{\bf A} = U{\bf A},\quad \hat{\Lambda} = U \hat{\Lambda} U^{-1}.
\ee
These transformations cast Eqs.~\eqref{eq:dcdw_o} in the following form
\be
&& \frac{d \tilde{\bf c}}{d w} \, = \, \tilde{{\mathbf f}}(w,\tilde{\bf c}), \label{eq:dcdw_tilde}\nl
&&\tilde{\mathbf f}(w,{\bf c}) \,=\, - \left[ \hat{\Lambda} \tilde{{\bf c}} + \frac{1}{w} \left( \hat{{\mathfrak B}} {\bf c} + \tilde{{\bf A}} \right)   \right] + \mathcal{O}( \tilde{\bf c}^2, \, \tilde{\bf c}/w^2). \qquad (w \rightarrow \infty, \, \tilde{\bf c} \rightarrow 0) \label{eq:dcdw_o_tilde}
\ee
We hereby call the components of the vectors $\tilde{\bf c}$ {\it pseudomodes} due to being generally complex-valued, thus not physical. This makes the components of the matrix $U$ complex-valued as well. But the inverse transformations of $\tilde{\bf c}$ achieved by the action of $U^{-1}$ always yield real vectors that eventually contribute to the observables in our theory.

The classical asymptotics beyond naive asymptotic power series expansion can be carried out with the help of a \textit{transseries} ansatz first put forward by O. Costin in~Ref. \cite{costin1998}. In that seminal work, the author proves that the set of ODEs written in the prepared form  \eqref{eq:dcdw_o_tilde} has the exact transseries solution 
\be
&& \tilde{c}_{i}(w) = \sum_{|{\bf m}| \ge 0}^{\infty} \sum_{k=0}^{\infty} \tilde{u}^{({\bf m})}_{i,k}  E^{({\bf m})}_k(w), \label{eq:ansatz_c} \\
&& E^{({\bf m})}_k(w) = \bm{\sigma}^{{\bf m}} \bm{\zeta}^{{\bf m}}(w) w^{-k},\label{eq:ansatz_c_basis} \\
&& \bm{\zeta}^{{\bf m}}(w) = e^{-({\bf m}  \cdot {\bf S}) w} w^{ {\bf m} \cdot \tilde{\bf b}} = \prod_{i=1}^{I} \left[ \zeta_{i}(w) \right]^{m_i}, \\
&& \zeta_i(w) =  e^{-S_i w} w^{\tilde{b}_i} \label{eq:IRdata} , \\
&& \bm{\sigma}^{\bf m} = \prod_{i=1}^{I} \sigma^{m_i}_{i} ,
\ee
where $I = {\rm dim} \, ({\bf c})$, ${\bf m} \in {\mathbb N}_0^{I}$ is an integer vector, and the dot denotes the inner product between any two vectors. The real numbers $\sigma_i$ are going be referred to as ``integration constants'' throughout this work as they would really symbolize the constants to be obtained if we were to integrate the ODEs in Eqs. \eqref{eq:dcdw_o_tilde}. Here, for simplicity we have defined $E^{({\bf m})}_{k}(w)$ that stand for the basis of transmonomials (i.e., the exponential factors and the fractional powers $w^{-1}$).

The transseries {\it data} such as the coefficients $\tilde{u}^{({\bf n})}_{i,k}$, the Lyapunov exponents $S_i$ and the {\it anomalous dimensions} $\tilde{b}_i$ can be recursively determined by the evolution equations up to normalization of $\sigma_i$.
Without loss of generality, we pick the normalization fixed by $u_{j,0}^{({\bf m})} = \delta_{ij}$ for $m_{j}= \delta_{ij}$. This leaves no room for ambiguity in determining the coefficients $\tilde{u}_{i,k}^{({\bf m})}$.
It is noteworthy that the transseries solution of $c_i(w)$ can be reproduced by the inverse transformations of \eqref{eq:transformations}, and one can find that $c_i(w)$ has essentially the same transseries as $\tilde{c}_i(w)$ due to the fact that the matrix $U$ acts on the index $i$ (mode number) in $\tilde{u}^{({\bf m})}_{i,k}$ only.

Although the transseries of $\tilde{c}_i(w)$ given by (\ref{eq:ansatz_c}) is generally complex-valued due to $\tilde{b}_i$ and $\tilde{u}^{({\bf 0})}_{ik}$ taking values in complex numbers, we can still recover a real transseries solution for the moments $c_i(w)$.
The important fact is that if $\tilde{b}_i$ for some $i$ is complex, it is always accompanied by its complex conjugate counterpart, i.e., $\tilde{b}_{i} = \tilde{b}^*_{j}$ for some $j$, where the associated coefficients satisfy $\tilde{u}^{({\bf m})}_{i,k} = \tilde{u}^{({\bf m})*}_{j, k}$ under a certain normalization of the eigenvectors of $\hat{\frak B}$ in such a way that $U^{-1}_{Ii}=1$ for every $i$ where again $I$ is the rank of matrix $\hat{\frak B}$.
Therefore, the reality of $c_i(w)$ is guaranteed if the following conditions are satisfied:
\be
 \sigma_{i} = \sigma^{*}_{j},&& \qquad \mbox{if \, $ \tilde{b}_{i} = \tilde{b}^*_j$}\\
 \sigma_{i} \in {\mathbb R}, && \qquad  \mbox{if \, $\tilde{b}_{i} \in {\mathbb R}$}\,.
\ee
\subsection{Evolution equations: $N=0$ case}
\label{subsubsec:n0}
In terms of $w$, Eqs.~\eqref{eq:ODEsBjor} can be reduced to the following non-autonomous dynamical system:
\bs
\label{eq:weqs}
\beal
\label{eq:atractor}
\frac{d\log T}{d\log\tau} &=-\frac{1}{3}\left(\frac{c_{01}}{10}+1\right)\,,\\
\label{eq:momweq}
\frac{d c_{0l}}{d w} \,&=-\frac{1}{1-\tfrac{1}{20}c_{01}}\left[\frac{3}{2w}\left(\aol\,c_{0l+1}\,+\,\beta_{0l}\,c_{0l}\,+\,\gol
\,c_{0l-1}\right)+\,\,\frac{3c_{0l}}{2\theta_0}\right]\,.
\end{align}
\es
The second equation is valid for any $l>0$. Since temperature $T$ is now sort of washed away from Eq.~\eqref{eq:momweq}, we shall solve these ODEs for $c_{nl}$ by the mathematical techniques discussed in Refs.~\cite{Iwano1982,costin1998}. These tools are well known in the context of resurgence theory. A curious reader is invited to check e.g., Ref.~\cite{costin} for technical details, Ref.~\cite{Dorigoni:2014hea} for a nice introduction to the subject, and Ref.~\cite{Dunne:2015eaa} for the summary of its applications to quantum mechanics and quantum field theories). We will afterwards substitute $c_{01}$
in Eq.~\eqref{eq:atractor} to solve for $T$.

It is convenient to cast Eqs.~\eqref{eq:atractor}-\eqref{eq:momweq} in the following familiar form
\bea
&& \frac{d {\bf c}}{dw} = {\bf f}(w,{\bf c}), \label{eq:dif_LH}\\
&& {\bf f}(w,{\bf c}) = -\frac{1}{1-\frac{c_{1}}{20}}
\left[ \hat{\Lambda} {\bf c} + \frac{1}{w} \left( {\frak B}  {\bf c} - \frac{c_{1}}{5 }   {\bf c} +  {\bf A} \right)
  \right], \label{eq:dif_RH}
\eea
where
\bea
&&   {\bf c}  = (c_{01},c_{02}, \dots ,c_{0L-1}, c_{0L})^{\top},\\
&& {\bf A}= \frac{3}{2}\left( \gamma_{01},0,\dots, 0 \right)^{\top}, \\
&& \hat{\Lambda} = {\rm diag} \left( \frac{3}{2 \theta_0}, \dots, \frac{3}{2 \theta_0} \right), \\ 
&& {\frak B} = \frac{3}{2}
\begin{pmatrix}
 \frac{2}{3} \Omega_{1}  & \alpha_{01} & & & & \\
 \gamma_{02}& \frac{2}{3}\Omega_{2}   & \alpha_{02} & & & \\
 &   \gamma_{03}& \frac{2}{3}\Omega_{3}   & \alpha_{03} & & \\
 && \ddots &\ddots & \ddots & \\
 & & &   \gamma_{0L-1}& \frac{2}{3}\Omega_{L-1}   & \alpha_{0L-1}   \\
&  & & &   \gamma_{0L}& \frac{2}{3}\Omega_{L}   
\end{pmatrix}, \\
&& \Omega_{l} = \frac{5l(2l+1)}{(4l+3)(4l-1)}.
\eea
Here, the index of vectors and matrices runs over $i = 1, \dots, L$.
To directly apply Costin's formula, one has to diagonalize ${\frak B}$ in Eq.~\eqref{eq:dif_LH} using the transformations in \eqref{eq:transformations}
to get the equation
\bea
&& \frac{d \tilde{{\bf c}}}{dw} = \tilde{\bf f}(w, \tilde{{\bf c}}), \label{eq:diff_til} \\
&& \tilde{\bf f}(w, \tilde{{\bf c}}) = -\frac{1}{1-\frac{c_{1}}{20}} \left[ \hat{\Lambda} \tilde{{\bf c}} + \frac{1}{w} \left( \hat{\frak B} \tilde{{\bf c}} - \frac{c_{1}}{5} \tilde{\bf c} +  \tilde{\bf A} \right) \right], \label{eq:f_til}
\eea
We can easily see that $c_{0i}$ can be reproduced by $\tilde{\bf c}$ and $U$ as
\bea \label{eq:c_i}
c_{i} = \sum_{l^{\prime}=1}^{L} U^{-1}_{i i^{\prime}}  \tilde{c}_{i^{\prime}}.
\eea
As is seen in Eq.~\eqref{eq:f_til}, from now on whenever $N=0$, we may drop the index $0$ in $c_{0l}$ for ease of notation. Notice that the asymptotic expansion of $\tilde{c}_{i},c_{i}$ both start at order $\mathcal{O}(w^{-1})$ asymptotically, and due to this we have $\tilde{u}^{(0)}_{i,0}=0$.

Before trying to solve the dynamical system in \eqref{eq:dif_LH}, it would be instructive to first linearize the r.h.s. of it to obtain the so-called {IR data} $S_i$, $\tilde{b}_i$ in Eq.~\eqref{eq:IRdata} of the transseries ansatz. These are going to be the input data obtained uniquely from the profile of the solution of the linarized equation around the asymptotic fixed point of the dynamical system at $w\rightarrow\infty$. To do so, we expand the factor $\frac{1}{1-c_{1}/20}$ to get
\bea
\label{eq:rhsseries}
&& {\bf f}(w,{\bf c}) = 
- \sum_{n=0}^{\infty} \left(\frac{c_{1}}{20} \right)^n
\left[ \hat{\Lambda} {\bf c} + \frac{1}{w} \left( {\frak B}  {\bf c}  - \frac{c_{1}}{5 }   {\bf c} + {\bf A} \right) \right] \nl
&&  \qquad \quad \, = -\frac{1}{w} {\bf A}
- \left[ \hat{\Lambda} {\bf c} + \left( \frac{1}{w}  {\frak B} + \frac{c_{1}}{20} \hat{\Lambda} \right) {\bf c} + \frac{3 \gamma_{01}}{40w} {\bf c}_1 
  \right]  + \mathcal{O}(c_i^2 w^{-1},c_{i}^3),
\eea
where ${\bf c}_{1}(w)= \left( c_{1}(w),0,\dots, 0 \right)^{\top}$.
The coefficients of the lowest power of $w^{-1}$, i.e., $u^{({\bf 0})}_{i,k}$, can be found from this equation, and one can readily see that $c_{i=l}=\mathcal{O}(w^{-l})$.
In particular,
\bea
\label{eq:c01_NS}
u^{(0)}_{11} = -\gamma_{01} \theta_0 = -\frac{8 \theta_0}{3}.
\eea
Now, we may proceed to linearize the equation \eqref{eq:rhsseries} by substituting $c_{i} \rightarrow \bar{c}_{i} + \delta c_{i}$ with $\bar{c}_{i}= \sum_{k} u^{({\bf 0})}_{i,k}w^{-k}$, and expanding to first order in $\delta c_{i}$
\bea
&&  \frac{d \delta c_{i}}{dw} = \left. \sum_{i^\prime=1}^{L} \frac{\pde f_i(w,{\bf c})}{\pde c_{i^{\prime}}} \right|_{{\bf c} \rightarrow \bar{\bf c}} \delta c_{i^\prime} \nl
&\Rightarrow \quad & \frac{d \delta {\bf c}}{dw} = -\left[ \hat{\Lambda}  + \frac{1}{w} \left( {\frak B} - \frac{1}{5} {\bf 1}_{L} \right) \right] \delta {\bf c} + \mathcal{O}(w^{-2} \delta c_{i}).
\eea
Using the matrix $U$, the solution to the linearized equation is found to be given by
\bea
\delta \tilde{c}_{i}(w) = \sigma_i \frac{e^{- \frac{3}{2\theta_0} w}}{w^{b_i-1/5}}, \label{eq:sol_linear}
\eea
where $\sigma_i$ is the integration constant and $\delta \tilde{\bf c} = U \delta {\bf c}$.
Therefore, we have the IR data (Lyapunov exponent and anomalous dimension) as
\bea
\label{eq:IRDATA}
S_{i} = \frac{3}{2 \theta_0} , \qquad \tilde{b}_{i} = - \left( b_{i} - \frac{1}{5} \right).
\eea
Finally, we substitute the ansatz \eqref{eq:ansatz_c} in Eq.~\eqref{eq:c_i} and then insert the resulting transseries for $\bf c$ in Eqs.\eqref{eq:diff_til}-\eqref{eq:f_til} to find the recursive relation for the transseries coefficients as
\bea
&& 20 \left[\left( {\bf m} \cdot \tilde{\bf b}  + b_{i}   -k \right) \tilde{u}^{({\bf m})}_{i,k} + \left( \frac{3}{2 \theta_0} - {\bf m} \cdot {\bf S} \right) \tilde{u}^{({\bf m})}_{i,k+1}  \right]  + 20\tilde{A}_i \,  \delta_{k,0} \delta_{{\bf m},{\bf 0}} \nl
&& - \sum_{|{\bf m}^\prime| \ge {\bf 0}}^{\bf m} \left[ \sum_{k^\prime=0}^{k} \left(
   {\bf m}^\prime \cdot \tilde{\bf b}  + 4 -k^\prime \right) \, u^{({\bf m}-{\bf m}^\prime)}_{1,k-k^\prime} \tilde{u}^{({\bf m}^\prime)}_{i,k^\prime} 
  - {\bf m}^{\prime} \cdot {\bf S}
  \sum_{k^\prime=0}^{k+1} u^{({\bf m}-{\bf m}^\prime)}_{1,k-k^\prime+1} \tilde{u}^{({\bf m}^\prime)}_{i,k^\prime} \right] = 0,  \label{eq:evo_nk}
\eea
where $ u^{({\bf m}^\prime)}_{1,k^\prime} = \sum_{i=1}^{L} U^{-1}_{1i} \tilde{u}^{({\bf n}^\prime)}_{i,k^\prime}.$
Fixing $\tilde{u}^{({\bf m})}_{i,0}=1$ for $m_{i^\prime} =\delta_{i,i^\prime}$, we can solve this equation order by order to
get the transseries data $u^{({\bf n})}_{1,k}$, $S_i$ and $\tilde{b}_i$ without any (imaginary) ambiguity. Note that the
IR data should match the numbers found using linearization around the asymptotic fixed point in Eq.~\eqref{eq:IRDATA}.

\subsection{Evolution equation: $N\neq 0$ case}
\label{subsubsec:ngen}
The technique advocated in the $N=0$ case above can be extended to include higher order moments $c_{nl}$. The energy dependence of hydrodynamization is only materialized by $c_{nl}~ (n=1,\dots,N)$ even if $c_{0l}$ have been for the most part sidelined in the literature. To get a complete and correct picture, however, all the moments ought to be taken into consideration. In principle, this means that one has to eventually 
solve an infinite-dimensional dynamical system for getting the full hydrodynamization process in both IR and UV regimes, which is obviously ambitious. As a result, a practical approach is to consider a truncated dynamical system whose proposed transseries solution has of course the ability to be generalized to the exact result. So our starting point in this section will be Eq.~\eqref{eq:momweq},
\be
\begin{split}
\frac{d c_{nl} }{d w } &= - \frac{1}{ 1 - \frac{c_{01} }{20} } \left[\frac{3}{2 w} \left( \alpha_{nl}  c_{nl+1} + \beta_{nl} c_{nl} + \gamma_{nl} c_{nl-1} - n \left(\rho_{l}  c_{n-1l+1} + \psi_{l} c_{n-1l} + \phi_{l} c_{n-1l-1} \right) + \frac{3c_{nl}}{2\theta_0} \right)\right].
\end{split}
\label{eq:evolutioneq_cnl}
\ee 

The equation \eqref{eq:evolutioneq_cnl} can be written in a concise matrix form,
\bea
&& \frac{d {\bf c}}{dw} = {\bf f}(w,{\bf c}), \label{eq:dif_LH_cnl}\\
&& {\bf f}(w,{\bf c}) = -\frac{1}{1-\frac{c_{01}}{20}}
\left[
  \hat{\Lambda} {\bf c} + \frac{1}{w} \left( {\frak B} {\bf c} +  c_{01}{\frak D}{\bf c} + {\bf A} \right)
  \right], \label{eq:dif_RH_cnl}
\eea
where the quantities ${\bf c}$, $\hat{\Lambda}$, ${\bf A}$ are explicitly given by
\bea
&& {\bf c} \, = \, (\underbrace{c_{01},\,\dots,\ c_{0L}}_{L} ,\, \underbrace{c_{10},\,c_{11},\,\dots, \, c_{1L}}_{L+1},\, \dots\,, c_{N0},\,\dots\, , \, c_{NL})^\top , \\
&& \hat \Lambda = \rm{diag} \left( \frac{3}{2\theta_0},\dots,\frac{3}{2\theta_0} \right), \\
&& {\bf A} = \frac{3}{2} ( \underbrace{\gamma_{01},0,\dots,0}_{L},0,\phi_{1},0,\dots,0 )^{\top},
\ee
and we have defined the block matrices ${\frak B}$ and ${\frak D}$ as
\be
&& {\frak B} = \frac{3}{2}
\begin{pmatrix}
  \bar{\frak B}_{00} & & & & \\
  \bar{\frak B}_{10} &    \bar{\frak B}_{11} & & & \\
  &  \bar{\frak B}_{21} &    \bar{\frak B}_{22} &  & \\
  & & \ddots & \ddots & \\
&    &  & \bar{\frak B}_{NN-1} &    \bar{\frak B}_{NN}   \\
\end{pmatrix}, \qquad
 {\frak D} = 
\begin{pmatrix}
  \bar{\frak D}_{00} & & & & \\
  \bar{\frak D}_{10} &    \bar{\frak D}_{11} & & & \\
  &  \bar{\frak D}_{21} &    \bar{\frak D}_{22} &  & \\
  & & \ddots & \ddots & \\
&    &  & \bar{\frak D}_{NN-1} &    \bar{\frak D}_{NN}   \\
\end{pmatrix},
\ee
with 
\be
&& \bar{\frak B}_{00} = 
\begin{pmatrix}
\frac{2}{3}\Omega_{01} & \alpha_{01} &  &  & \\ 
\gamma_{02} & \frac{2}{3}\Omega_{02} & \alpha_{02} &   &\\ 
 &  \ddots & \ddots &  \ddots & \\ 
&   & \gamma_{0L-1} & \frac{2}{3} \Omega_{0L-1} & \alpha_{0L-1} \\
 & &  & \gamma_{0L} & \frac{2}{3}\Omega_{0L}  \\ 
\end{pmatrix}, \quad 
  \bar{\frak B}_{nn\,(n>0)} = \begin{pmatrix}
\frac{2}{3}\Omega_{n0} & \alpha_{n0} &  &  & \\ 
\gamma_{n1} & \frac{2}{3}\Omega_{n1} & \alpha_{n1} &   &\\ 
 &  \ddots & \ddots &  \ddots & \\ 
&   & \gamma_{nL-1} & \frac{2}{3} \Omega_{nL-1} & \alpha_{nL-1} \\
 & &  & \gamma_{nL} & \frac{2}{3}\Omega_{nL}  \\ 
  \end{pmatrix},  \nl
&& \bar{\frak B}_{10} = - \begin{pmatrix}
  \rho_{0} - \frac{1}{30}&  &  & & \\ 
  \Psi_{1} & \rho_{1} &   & &\\
 \phi_2  & \Psi_{2} & \rho_{2}    & &\\ 
 &  \ddots & \ddots &  \ddots & \\ 
 &  & \phi_{L-1} &  \Psi_{L-1} & \rho_{L-1} \\
 & &  & \phi_{L} &  \Psi_{L}  \\ 
\end{pmatrix}, \quad 
   \bar{\frak B}_{nn-1\, (n>1)} = -n \begin{pmatrix}
 \Psi_0 & \rho_{0} &  &  & \\ 
\phi_{1} &  \Psi_{1} & \rho_{1} &   &\\ 
 &  \ddots & \ddots &  \ddots & \\ 
&   & \phi_{L-1} &  \Psi_{L-1} & \rho_{L-1} \\
 & &  & \phi_{L} &  \Psi_{L}  \\ 
   \end{pmatrix}, \\ \nl
   && \bar{\frak D}_{nn} = - {\rm diag} \left( \frac{4+n}{20}, \cdots ,\frac{4+n}{20} \right), \quad  \bar{\frak D}_{10} = \frac{1}{20} \begin{pmatrix}
   &  &  & \\ 
  1 &      & &\\
    & \ddots &    \\ 
  &  &  1  \\
 \end{pmatrix}, \quad 
   \bar{\frak D}_{nn-1 \, (n>1)} =  {\rm diag} \left(  \frac{n}{20}, \dots ,\frac{n}{20} \right), \\ \nl
   && \Omega_{nl}=\frac{l(2l+1)(5+2n)}{(4l+3)(4l-1)}, \qquad \Psi_{l}=  \frac{4l(2l+1)}{3(4l+3)(4l-1)} .
\ee
Here, the blocks $\bar{\frak B}_{00}$, $\bar{\frak D}_{00}$ are both $L\times L$ matrices, 
$\bar{\frak B}_{10}$,$\bar{\frak D}_{10}$ are $(L+1)\times L$ matrices, and the remaining blocks are $(L+1)\times(L+1)$ matrices.
As in the $N=0$ case, we will be defining the matrix $U$ to diagonalize $ {\frak B}$,
\be
&& \tilde{\bf c} = U {\bf c}, \quad  \tilde{\bf A} = U {\bf A}, \\
&& \hat{\frak B} = U {\frak B} U^{-1} = {\rm diag}\, (b_1 ,\dots , b_{I}) \in {\mathbb C}^{I}.
\ee
We finally employ the transseries ansatz in the diagonalized form of Eqs.~\eqref{eq:dif_LH_cnl}-\eqref{eq:dif_RH_cnl} as before to obtain the recursive relations involving the transseries data, namely
\bea
&& 20 \left[\left( {\bf m} \cdot \tilde{\bf b}  + b_{i}   - k \right) \tilde{u}^{({\bf m})}_{i,k} + \left( \frac{3}{2 \theta_0} - {\bf m} \cdot {\bf S} \right) \tilde{u}^{({\bf m})}_{i,k+1}  \right]  + 20\tilde{A}_i \,  \delta_{k,0} \delta_{{\bf m},{\bf 0}} \nl
&& - \sum_{|{\bf m}^\prime| \ge {\bf 0}}^{\bf m} \left[ \sum_{k^\prime=0}^{k} \left(
   {\bf m}^\prime \cdot \tilde{\bf b}  -k^\prime \right) \, u^{({\bf m}-{\bf m}^\prime)}_{1,k-k^\prime} \tilde{u}^{({\bf m}^\prime)}_{i,k^\prime} -20 \sum_{i^\prime=1}^I \sum_{k^\prime=0}^k u^{({\bf m}-{\bf m}^\prime)}_{1,k-k^\prime} \tilde{\frak D}_{ii^\prime}\tilde{u}^{({\bf m}^\prime)}_{i^\prime,k^\prime} 
  - {\bf m}^{\prime} \cdot {\bf S}
  \sum_{k^\prime=0}^{k+1} u^{({\bf m}-{\bf m}^\prime)}_{1,k-k^\prime+1} \tilde{u}^{({\bf m}^\prime)}_{i,k^\prime} \right] = 0,  \label{eq:evo_nk_ngen}
\eea
where $\tilde{\frak D}=U {\frak D}U^{-1}$, $\tilde{c}_{i}(w) = 
\sum_{|{\bf m}| \ge 0}^{\infty} \sum_{k=0}^{\infty} \tilde{u}^{({\bf m})}_{i,k}  E^{({\bf m})}_k(w)$, and $\bf m$ is the index of each non-perturbative sector of the pseudomodes, $\tilde{c}_{i}(w)$.
Eq.~\eqref{eq:evo_nk_ngen} can be solved recursively. The higher non-perturbative sectors are connected to the lower ones through mode-to-mode couplings of the form $c_{01} c_{nl}$, whereas operations in charge of promoting the asymptotic order are differentiation and multiplication by a $1/w$ factor. Once the perturbative order is fixed, the 1st-order non-perturbative sector can be constructed explicitly. 

For pedagogic purposes, we show how to compute the leading order contribution in the perturbative sector of the transseries corresponding to $\bm m=\bm 0$ (also known as the leading order in the asymptotic expansion). As an example, first, we determine which moments behave like $\mathcal{O}(1/w)$. This is effectively done by taking $k=0$ in \eqref{eq:evo_nk_ngen}. Since the asymptotic fixed point has all the moments other than $c_{00}$ go to $0$ at $w\rightarrow\infty$, the coefficients $\tilde u^{\bm 0}_{i,0}$ have to vanish. By default, all the other coefficients promoting the asymptotic order also vanish. For the lowest asymptotic behavior for each $n$ and $l$, Eq.~\eqref{eq:dif_LH_cnl} gives
\be
&&  
   \frac{d c_{nl}}{dw} + \frac{3}{2 \theta_0} c_{nl} + \frac{3 c_{01}}{40 \theta_0}    {c}_{nl} + \frac{1}{w} \left[ \left( {\frak B} {\bf c}\right)_{nl}  + A_{nl}  \right] + {\cal O}(c_{01}^2 {\bf c}, c_{01} {\bf c}/w, c_{01}{\bf A}/w) = 0.
  \ee
Here, we relabeled the index $i$ by $n,l$ for convenience. Since $c_{nl}\sim{\cal O}(1/w)$ and $A_{nl} \neq 0$ only when $(n,l)\in\{(0,1),(1,1)\}$, we find that both $c_{01}$ and $c_{11}$ have a non-zero coefficient for the 1st-order asymptotics given by 
\be
 c_{nl}   + \frac{2\theta_0}{3 w}  A_{nl} + {\cal O}(1/w^2) = 0 \quad \mbox{ for \, $(n,l)\in\{(0,1),(1,1)\}$}.
 \label{eq:c11-c01asy}
\ee
Since $\bm A$ is constructed by $\gamma_{01}$ and $\phi_1$, only these two moments survive at the 1st asymptotic order, that is, $c_{01}$ (cf. Eq. \eqref{eq:c01_NS} aka Navier-Stokes) and $c_{11}\sim -\phi_{1} \theta_0/w$. For the other moments, the leading asymptotic order is dictated by their neighbors $c_{n^\prime l^\prime}$ alongside mode-to-mode coupling. The equation responsible for identifying the leading order of $c_{nl}\sim \mathcal{O}(1/w^k)\,\,(k>1)$ reads
\be
   c_{nl} +  \frac{c_{01}}{20} c_{nl} +  \frac{2 \theta_0}{3w } \left( {\frak B} {\bf c}\right)_{nl} 
   + {\cal O}(c_{01}^2  {\bf c}, c_{01} {\bf c}/w) = 0 \quad \mbox{ for \, $(n,l)\notin\{(0,1),(1,1)\}$}.
\ee
All the other operations promoting the asymptotic order vanish, such as differentiation. For example, the leading asymptotic order of $c_{2l} ~ (l=0,1,2)$ is identified by
\bea
&& c_{10} \sim -\frac{\theta_0}{w} \left(\alpha_{10}c_{11} - \rho_{0} c_{01} + \frac{c_{01}}{30}\right) \sim \mathcal{O}(1/w^2)\,, \label{eq:c10LeadOrder}\\
&& c_{20} \sim \frac{2\theta_0}{w} \rho_{0} c_{11} \sim  \mathcal{O}(1/w^2), 
\label{eq:c20leadOrder}\\
&& c_{21} \sim \frac{2\theta_0}{w} \Psi_{1} c_{11} \sim  \mathcal{O}(1/w^2), \\
&& c_{22} \sim \frac{2\theta_0}{w} \phi_{2} c_{11} \sim  \mathcal{O}(1/w^2).
\eea
As can be seen, they both have the lowest asymptotic order $1/w^2$ because of $c_{11}$. By virtue of Eq.~\eqref{eq:evo_nk_ngen}, the leading order asymptotics of higher moments is determined to be $c_{nl}\sim \mathcal{O}(1/w^k)$ where $k=\max(n,l)$ for $n,l\geq 2$. 
The higher modes, therefore, decay faster in general as all the Lyapunov exponents are equal due to the RTA-BE having a single scale which in this case is $\theta_0$, and the suppression in the IR is just determined by the leading asymptotic order. 
All the modes satisfying $n,l\leq 1$, however, are exempt from this rule as $c_{10}\sim 1/w^2$, and as mentioned earlier, the two remaining
lower modes $c_{01}$ and $c_{11}$ are the slowest modes of all in the IR. 

\begin{figure}[htpb]
\begin{center}
\includegraphics[width=0.8\linewidth]{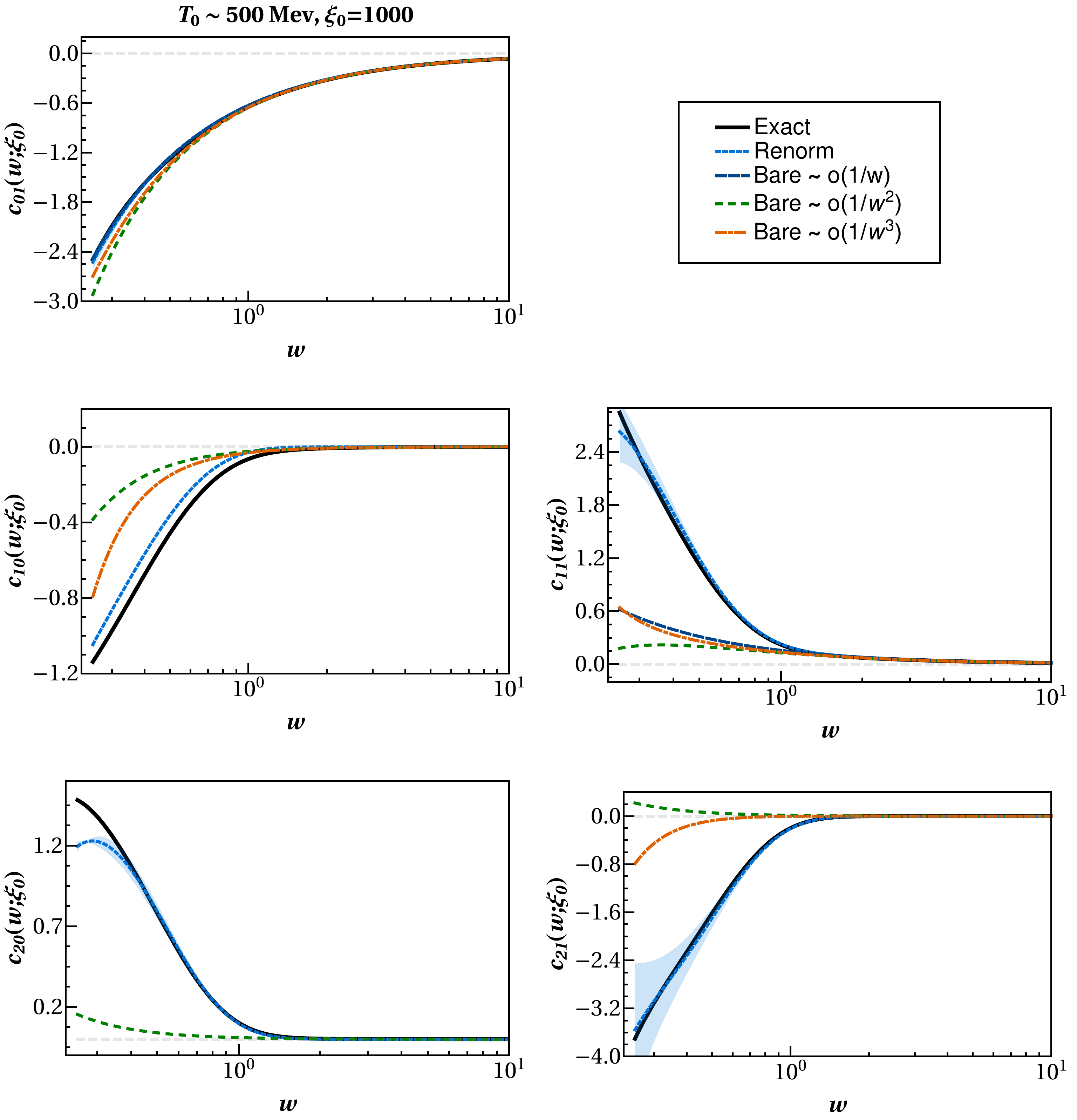}
\caption{Five lowest non-hydrodynamic modes computed by transseries, leading bare asymptotic expansion and exact numerics. The blue shaded area depicts the variation in the transseries solution due to the standard deviation of $\sigma_i$. For each moment, the transseries is truncated and renormalized by transasymptotic matching including all the transmonomials up to $15$th order. For example $c_{01}^{Remorn}=C_{01,1}/w$, where $C_{01,1}$ is constructed by including up to 15th-order transmonomials.}
\label{fig:transseries_Delta0}
\end{center}
\end{figure}

Once the coefficients of the bare asymptotic series for each $c_{nl}$ are obtained, the transseries coefficients can be achieved by solving the equation including higher ${\bf m} > 0$ non-perturbative sectors. As an example, we show the transseries and leading order bare asymptotics of five moments $c_{01},\dots, c_{21}$ of the truncated dynamical system $N=2,~L=1$ in Fig. \ref{fig:transseries_Delta0}
~\footnote{In this work the initial condition for the distribution function is given by the RS ansatz of the distribution function~\cite{Romatschke:2003ms}, i.e.,
\begin{equation*}
    f_0(\tau,p_T,p_\varsigma)=\exp\left[-\frac{\sqrt{p_T^2+(1+\xi_0)\left(\frac{p_\varsigma}{\tau}\right)^2}}{\Lambda_0}\right]\,.
\end{equation*}
where $\Lambda_0$ and $-1<\xi_0<\infty$ are the initial effective temperature and momentum anisotropy along the $\varsigma$ direction. Thus the set of initial conditions for the moments $\cnl$ depend on the initial time $\tau_0$ and initial anisotropy parameter $\xi_0$. See App.~\ref{app:RTABE} for further details.
}. Each moment is renormalized by the inclusion of non-perturbative/perturbative sectors available using the transasymptotics as $C_{nl,k}/w^k + \mathcal{O}(1/w^k)$, where the transasymptotic matching is approximated up to $15$th-order transmonomials. On the numerical front, the integration constants $\bm \sigma$ are determined by employing {\it simultaneous} least-squares method, which aims to minimize the difference of the exact and transseries solutions of all the moments involved in 
a given truncated dynamical system simultaneously. So the overall average error will be reduced across all the moments. In this figure, five integration constants $\sigma_{1},\dots,\sigma_{5}$ are given by their optimized values along with individual standard deviations reported in the plots. Also, the light blue dashed lines stand for the transseries solution with the best optimized $\sigma_i$ while the blue shaded areas highlighting the possible variation of transseries due to standard deviation. Furthermore, bare asymptotic expansion at $1/w,~ 1/w^2,$ and $1/w^3$ orders 
for each moment is plotted as a comparison. These plots show that continuing to a higher asymptotic order will not necessarily lead to a better result, as opposed to adding more transmonomials at each order, which results in a better approximation of the transasymptotic matching overall (see next section).
\subsection{Transasymptotic matching}
\label{subsec:transasymp}
In this section, we construct the transasymptotic matching condition responsible for the full-blown form of the time evolution of the transport coefficients including all the effects of non-perturbative sectors, which is a 1st-order PDE. In doing so, we  first redefine $\tilde{c}_{i,k}(w)$ as
\be
\tilde{c}_{i}(w)  = \sum_{k \ge 0}^{\infty} \tilde{C}_{i,k}(\bm{\sigma \zeta}(w)) w^{-k} 
\ee
and sum over ${\bf m}$ in Eq.~\eqref{eq:evo_nk_ngen}. Then the transasymptotic matching condition turns out to have the following formal form
\be
&& 20 \left[ \left(   (\tilde{\bf b} \cdot   \hat{\bm{\zeta}}   -k ) + b_{i}   \right) \tilde{C}_{i,k} -  {\bf S} \cdot \hat{\bm{\zeta}}  \tilde{C}_{i,k+1}  +  \frac{3}{2 \theta_0} \tilde{C}_{i,k+1}  \right] + 20 \tilde{A}_i   \delta_{k,0} \nl
&& -  \sum_{k^\prime=0}^{k} C_{1,k-k^\prime} \left( \tilde{\bf b} \cdot \hat{\bm{\zeta}}  -k^\prime \right) \tilde{C}_{i,k^\prime}
+ 20 \sum_{i^\prime=1}^I \sum_{k^\prime=0}^{k} C_{1,k-k^\prime} \tilde{\frak D}_{ii^\prime} \tilde{C}_{i^\prime,k^\prime}
+   \sum_{k^\prime=0}^{k+1} C_{1,k-k^\prime+1}     {\bf S} \cdot \hat{\bm{\zeta}} \tilde{C}_{i,k^\prime}= 0, \label{eq:transasymptotic}
\ee
where $\hat{\zeta}_i := \partial /\partial \log ( \sigma_i \zeta_i)$.
Here, $\tilde{C}_{i,k}$ is supposed to be only a function of $\sigma_i\zeta_i$. 
We again  note that Eq.~\eqref{eq:transasymptotic} is a 1st-order PDE, whose solution yields the time evolution of the transport coefficients with all the $\sigma_i \zeta_i(w)$ included. Moreover, the integration constants may be determined from the transseries coefficients $\tilde{u}^{({\bf m})}_{i,k}$.

Due to its partial differential nature, it is technically not easy to solve \eqref{eq:transasymptotic} for the general case even though for $L=1,N=0$, it can be exactly solved. For instance, the first three coefficients are given by solving \eqref{eq:transasymptotic} in this case as \cite{Behtash:2018moe}
\be
\label{eq:transasym}
&& C_{1,0}( \sigma \zeta) = - 20 W_{\zeta}, \\
&& C_{1,1}(\sigma \zeta) =  -\frac{8 \theta_0 \left( 50 W_\zeta^3 + 105 W^2_{\zeta} + 36 W_{\zeta}+5 \right)}{15\left( W_{\zeta}+1 \right)}, \\
&& C_{1,2}(\sigma \zeta) = -\frac{8 \theta^2_{0}}{7875(W_{\zeta}+1)} 
\left[ \frac{25 \left( 700 W_{\zeta}^4 + 2195 W^{3}_{\zeta} + 966 W_{\zeta}^2 + 20  \right)}{W_\zeta} + \frac{4032}{(W_{\zeta}+1)^2} + 3685
\right],
\ee
where $W_{\zeta} := W(- \sigma \zeta/20)$ is Lambert $W$-function.

For a system consisting of a larger number of moments, the above equation becomes a PDE, which renders the matching condition hard to solve. 
However, the transasymptotic matching, in this case, is at least effectively consolidated by summing over a large number of transmonomials 
just in the same manner as in the renormalization group equation for a running coupling constant, where loop diagram contributions are added to the r.h.s. in order to capture a more realistic RG flow. As a special example, we guide the readers to take a look at App. \ref{sec:transasymptotic_Delta1} 
 in which the transasymptotic matching for $N=0,~ L=1,~ \Delta=1$ system yields polynomials of finite degree, which are the exact solutions to the dynamical system. In Fig. \ref{fig:transseries_Delta0}, the transmonomials are taken up to 15th order to get as close as possible to the exact solution of the truncated system. We again mention that the renormalized moment is meant to be a solution compatible with the full-fledged transasymptotics at its leading asymptotic order.

\subsection{Comment on resurgence and cancellation of the imaginary ambiguities}
\label{subsubsec:cancel}
Once one has an ODE of the form \eqref{eq:dcdw_o}, we can realize imaginary ambiguity cancellation in a systematic way.
In general cases, it is hard to prove the cancellation of imaginary ambiguities. However, for ODEs of the type \eqref{eq:dcdw_o} 
there are mathematically rigorous proofs for the ambiguity cancellations due to O. Costin (cf. \cite{costin1998} and references within
for more technical details). In this section, we consider an easy example involving only one mode $i=1$, and therefore we omit the index for simplicity.

Let us begin with a formal transseries ansatz given by
\be
&& F(w;\sigma) = \sum_{n=0}^\infty \sigma^n F^{(n)}(w), \quad  F^{(n)}(w)=e^{-nSw} w^{n \beta} \hat{\Phi}(w) , \quad \hat{\Phi}^{(n)}(w) = \sum_{k=0}^{\infty} a^{(n)}_{k} w^{-k}. \label{eq:hatPhi}
\ee
Suppose that $F(w)$ solves \eqref{eq:dcdw_o} and let us only focus on the $0$th non-perturbative sector $a^{(0)}_k$.
By calculating the radius of convergence one can see that the power series \eqref{eq:hatPhi} is divergent. The approximate form of the 
growing upper bound is 
\be
|a^{(0)}_k| \le M S^{-k}  k! \qquad \mbox{as \, $k \rightarrow \infty$},
\ee
with some positive real constants, $M$ and $S$.
This means that $\hat{\Phi}^{(0)}(w)$ is asymptotically of {\it Gevrey-1 class}~\footnote{The Gevrey-$n$ class means that there exists a smooth function $f$ on $\mathbb{R}^d$ such that on every compact subset $C$, there are constants $p,q$ such that $|D^{\alpha}f(x)|\leq p q^{k}(k!)^{n}$. Here, $D^\alpha$ is a differential operator of some multi-indices $\alpha$ such that $|\alpha|=k$.}, hence the Borel transform is just enough for convergence purposes, defined by
\be
&& {\cal B} [w^{-(k+1)}] := \frac{\xi^k}{\Gamma(k+1)}, \\
&& \Phi^{(0)}(\xi) := {\cal B} [ \hat{\Phi}(w) ] = \sum_{k=0}^{\infty} \frac{a^{(0)}_{k+1}}{\Gamma(k+1)} \xi^k.
\ee
Note that $\Phi^{(0)}(\xi)$ has a finite convergence radius and $r_c=S$.
Since $\hat{\Phi}^{(0)}(w)$ is a divergent series, there should exist a branch-cut on the positive real axis with a branch-point $S$ on the Borel plane. The position of the first singularity corresponds to the Lyapunov exponent of the flow lines approaching the asymptotically stable fixed point in the dynamical system. 
Notice that the appearance of a branch-cut is a reminder that the series expansions are coupled to higher order transmonomials such as $\exp(-S w)/w$ which independently satisfy the same ODE, i.e. Eq.~\eqref{eq:dcdw_o}.

The original divergent series \eqref{eq:hatPhi} can be reproduced through the Laplace integral,
\be
\tilde{\Phi}^{(0)}_{\theta}(w) :=  {\cal L}^{\theta} [\Phi^{(0)}(\xi)] = \int_0^{\infty e^{i \theta}} e^{- \xi w} d\xi \, \Phi^{(0)}(\xi), \label{eq:laplace_int}
\ee
and subsequently taking the asymptotic limit
\be
\hat{\Phi}^{(0)} (w) \sim \tilde{\Phi}_{+0}^{(0)} (w) \sim \tilde{\Phi}_{-0}^{(0)} (w)  \qquad \mbox{as \, $w \rightarrow \infty$}.
\ee
It is noteworthy that the imaginary ambiguity depending on the contour of integration is nothing but an \textbf{artifact} of
going to the Borel plane and it is exponentially suppressed in the large-$w$ limit.
However, when one observes a singularity on the Borel plane, one can build a relationship among non-perturbative sectors by taking the Hankel contour going around the singularity in the Laplace integral,
\be
   {\cal L}^{+0} [\Phi(\xi)] - {\cal L}^{-0} [\Phi(\xi)] =    \int_\gamma e^{- \xi w} d\xi \, \Phi(\xi).
\ee
The relations obtained in this way are called \textit{resurgence relations}.

We now are in position to introduce the {\it Stokes} automorphism $\underline{{\frak S}}^{\theta}$. Formally speaking, the integration constants
in the transseries may jump once the singularity (Stokes) rays on the Borel plane are crossed. This is associated with the existence of an automorphism that takes one 
integration constant to another plus some new contribution, defined by
\be
&& {\cal S}^\theta := {\cal L}^{\theta} \circ {\cal B}, \\
&& {\cal S}^{\theta^+} = {\cal S}^{\theta^-} \circ \underline{{\frak S}}^{\theta} = {\cal S}^{\theta^-} \circ ({\bf 1} - {\rm Disc}^{\theta})
, \\
&& \underline{{\frak S}}^{\theta} := \exp \left( \sum_{\rho \in \{\rho_{\theta}\}} \underline{\dot{\Delta}}_{\rho}  \right), \qquad 
\underline{\dot{\Delta}}_{\rho} := e^{-\rho w} \Delta_{\rho},
\ee
where $\{\rho_\theta \}$ denotes the set of singular points along the Stokes ray with the contour enclosing the singularities 
being controlled by an angle parameter $\theta$. Here, $\Delta_\rho$ represents an abstract derivative operator known as {\it alien derivative} which can be understood as an infinitesimal change in the asymptotic behavior due to a singularity in the direction parametrized by $\theta$.
Note that
\be
( {\cal S}^{+\theta} - {\cal S}^{-\theta} ) \hat{\Phi}(w) = 0 ,
\ee
if $\hat{\Phi}^{(0)}(w)$ is not a divergent series along $\theta$ direction. However, when a singularity appears for a particular angle $\theta$,
we can make the resurgence relations among different sectors. In our case, the set of singularities on the real positive axis can be given by
 $\{\rho_0\} = \{n S \, | \, n \in {\mathbb N}\}$, where again $n=1$ shows the Lyapunov exponent of the flow lines flowing to the asymptotic stable
 fixed point of the dynamical system.
By taking the Hankel contour with a singularity at $nS$, one can obtain the so-called {\it bridge equation}
\be
\underline{\dot{\Delta}}_{nS} F(w;\sigma) = A_n(\sigma) \frac{\partial F(w;\sigma)}{\partial \sigma},
\ee
and the relationships among different sectors are given by
\be
\left( {\cal S}^{+0} - {\cal S}^{-0}\right) F^{(n)}(w) = \sum_{l=1}^{\infty}
\begin{pmatrix}
  n+l \\
  n
\end{pmatrix}
A^l {\cal S}^{-0} F^{(n+l)}(w), \label{eq:SSF}
\ee 
where $A \in i \, {\mathbb R}$ the famous {\it Stokes constant}. This equation means that information of some sector is carried over to 
higher sectors once Stokes rays are crossed.

Although the construction of resurgence relations is generally an independent issue from the imaginary ambiguity cancellation, we can systematically
find such a cancellation mechanism for the transseries solution of the ODE in \eqref{eq:dcdw_o} without the need to resort to the discussion of this section. The significantly important fact is that a nonzero r.h.s. Eq.~\eqref{eq:SSF},
satisfies the same ODE. Therefore, the Borel-summed transseries along the $\theta=0$ ray may well be expressed by the arbitrariness in the integration constant, and the fact that nothing prevents one from extending $\sigma \in {\mathbb R}$ to $\sigma \in {\mathbb C}$.
The relationship between two integration constants across a Stokes ray is then given by
\be
\sigma(\theta) =
\begin{cases}
  \sigma_- = \sigma(-0) & \mbox{for \, $\theta < 0$} \\
  \sigma_0 = \sigma(-0) + \frac{A}{2}& \mbox{for \, $\theta = 0$} \\
  \sigma_+ = \sigma(-0) + A & \mbox{for \, $\theta > 0$}
\end{cases}.
\ee
Therefore, the imaginary ambiguity can be cancelled by shifting the integration constant in such a way that
\be
{\cal S}^{+0}F(w;\sigma+A/2) = {\cal S}^{-0}F(w;\sigma-A/2).
\ee

Let us demonstrate the imaginary ambiguity cancellation for simple cases. For $L=1$ and $N=0$,
the leading large-order growth of the asymptotic expansion is given by \cite{Basar:2015ava}
\be
\tilde{u}^{(0)}_{1,k} &\sim& C(\theta_0) \frac{\Gamma(k+\tilde{b}_1)}{2 \pi i S_1^{k+\tilde{b}_1}} \left( \tilde{u}_{1,0}^{(1)} + \frac{S_1}{k+\tilde{b}_1 -1} \tilde{u}_{1,1}^{(1)} + \cdots \right) + \cdots \nl
&\rightarrow& C(\theta_0) \frac{\Gamma(k+\tilde{b}_1)}{2 \pi i S_1^{k+\tilde{b}_1}}, \qquad \mbox{as \, $k \rightarrow \infty$} \label{eq:asym_u}
\ee
where $S_1 = \frac{3}{2 \theta_0}, \tilde{b}_1 = -18/35$, and $C(\theta_0)$ is an overall factor depending on $\theta_0$.
Here, we have used $\tilde{u}_{1,0}^{(1)}=1$ in the second line. Since $w$ always appears with a constant factor $1/\theta_0$ in $\tilde{c}_l(w)$, one finds that
\be
C(\theta) = C_0 \theta_0^{-\tilde{b}_1},
\ee
where $C_0$ is a constant.
We measured $C_0$ by numerical fitting with the a prepared form \eqref{eq:asym_u} and obtained
\be
C_0 \approx 0.4898.
\ee
Therefore, the Stokes constant $A_1$ is related to $C(\theta)$ in the following way
\be
\label{eq:Stokes}
&& A_{1} = 2 \pi i C(\theta_0),
\ee
leading to the imaginary ambiguity cancellation once we set
\be
{\rm Im}\,  \sigma_1 = -\pi C(\theta_0).
\ee
For general $L$ and $N$, $\tilde{u}^{(0)}_{l,k}$ are complex-valued so are the integration constants $\sigma_l$ for the reality condition to hold.
For $L=2$, the overall factor in \eqref{eq:asym_u} is found to be
\be
C_1(\theta_0) = C_0 \theta_0^{-\tilde{b}_1} = [C_2(\theta_{0})]^*,
\ee
in which $\tilde{b}_1 = (75+i \sqrt{10655})/110 = \tilde{b}_2^*$. By fitting numerically we can estimate the complex constant $C_0$ as
\be
C_0 \approx 0.8270 -  0.4060 i.
\ee
Therefore, the Stokes constant as a function of $\theta_0$ can be fully evaluated from Eq.~\eqref{eq:Stokes}. Finally, the imaginary ambiguity cancellation follows by just shifting the integration constant as
\be
\sigma_1 \rightarrow \sigma_1 -  i \pi C_1(\theta_0).
\ee
\section{Global dynamics and asymptotic stability analysis}
\label{sec:global}
In this section we will explore the generic properties of the Bjorken flow 
from the perspective of dynamical systems with some useful definitions provided in App.~\ref{app:dynsys} for readability.
\subsection{Phase space and asymptotic UV/IR fixed points}
\label{subsec:fixedpoint}
The explicit time-dependence of \eqref{eq:ODEsBjor} 
means that the system is {\it non-autonomous} and the phase space is larger than 
that of time-independent (autonomous) system with the same number of variables. Let us assume that 
the total number of Legendre and Laguerre terms in the expansion of the distribution function is
$L+1$ and $N+1$, respectively. Then one can write down the phase space of this truncated dynamical system as
\be
\mathcal{M} = \mathbb{R}^{(L+1)(N+1)-1}\times  \mathbb{R}^+\times\mathbb{R}^+ = (M_{0},M_{1},M_{2},\dots,M_{N}, \mathbb{T}, \mathbb{t}) \equiv (X, \mathbb{t}) \label{eq:phase_space}
\ee
where each subspace $M_i$ consists of $L$ dimensions, and we excluded the constant component $c_{00}=1$ from the
phase space. So \eqref{eq:phase_space} is just the hyperspace $c_{00}=1$ of the full phase space. Eventually, the limit 
$N,L\rightarrow\infty$ has to be taken, meaning that the true physical dynamical system at its full glory is
{\bf infinite-dimensional} \footnote{This is indeed expected by noting that the distribution function $f$ solving the Boltzmann equation has an exponential kernel which admits infinitely many terms in its series expansion, leading to an infinite number of
moments.}. The tuple $(M_0,\dots)$ is just short for the product of all the entries, and the time manifold $\mathbb{t}$ is taken to be the set of positive real numbers $\mathbb{R}^+$. This choice is made due to the fact that Eq.~\eqref{eq:Tevol} is not regular at $\tau= 0$ so to keep the continuity, $\tau<0$ is not allowed. Since temperature $T$
has its own equation in \eqref{eq:ODEsBjor}, we also add its manifold $\mathbb{T}$ to the crowd and let it take values over $\mathbb{T}=\mathbb{R}^+$. In total we are then left with $(L+1)(N+1) + 1 + 1 -1 = (L+1)(N+1)+1$ dimensions for $\mathcal{M}$.

By looking at the case where $N=0$, the dynamical system reduces to $L+2$ dimensional subspace
$(M_0, \mathbb{T}, \mathbb{t})$ which is denoted by $\mathcal{M}_0\subset \mathcal{M}$. The special facet of this subspace is that it indeed encompasses the {\it invariant manifold} \footnote{This manifold is basically an embedding in an $L+2$ dimensional phase space. Because
    the exact solution of RTA-BE has only the parameters $\xi_0$, $T_0$, and $\tau_0$ controlling the shape of flow lines, the invariant manifold has to be a three dimensional object. For the argument in the $w$ coordinate, cf. Sect. \ref{subsec:initial}.} because its equations are completely 
decoupled from the rest $(n>0)$ in case we started with a larger system involving more moments, i.e., $N>0$. Therefore, any solution of this sub-system entirely lies inside $\mathcal{M}_0$ at all times $\tau>0$. The second observation is that the mode-to-mode coupling terms in $\eqref{eq:ODEsBjor}$ 
for $n>0$ all depend on the elements of the subspace $\mathcal{M}_0$. So it is crucial to understand the dynamics
happening in this subspace by first locating the asymptotic fixed points and analyzing their stability, and then
obtaining the flow lines connecting these fixed points to find the qualitative shape of the invariant manifold.

In terms of the variable $w=\tau T(\tau)$, the non-autonomous dynamical system is described by
\be
\mathcal{W} = \mathbb{R}^{(L+1)(N+1)-1}\times\mathbb{R}_+ = (W_{0},W_{1},W_{2},\dots,W_{N}, \mathbb{w})\equiv (Y,\mathbb{w}), \label{eq:phase_space_w}
\ee
where $\mathbb{w}=\mathbb{R}_+:= \mathbb{R}^+ \cup \{0\}$. Note that $w=0$ has been allowed
since the combination of $\tau T(\tau)$ {\it does} essentially have a well-defined zero limit. The invariant manifold now is inside
$W_0$ parametrized by ${c}_{0l}$. The structure of this invariant manifold will be the subject of next subsection.

\begin{figure}[!htpb]
    \begin{center}
        \includegraphics[scale=.58]{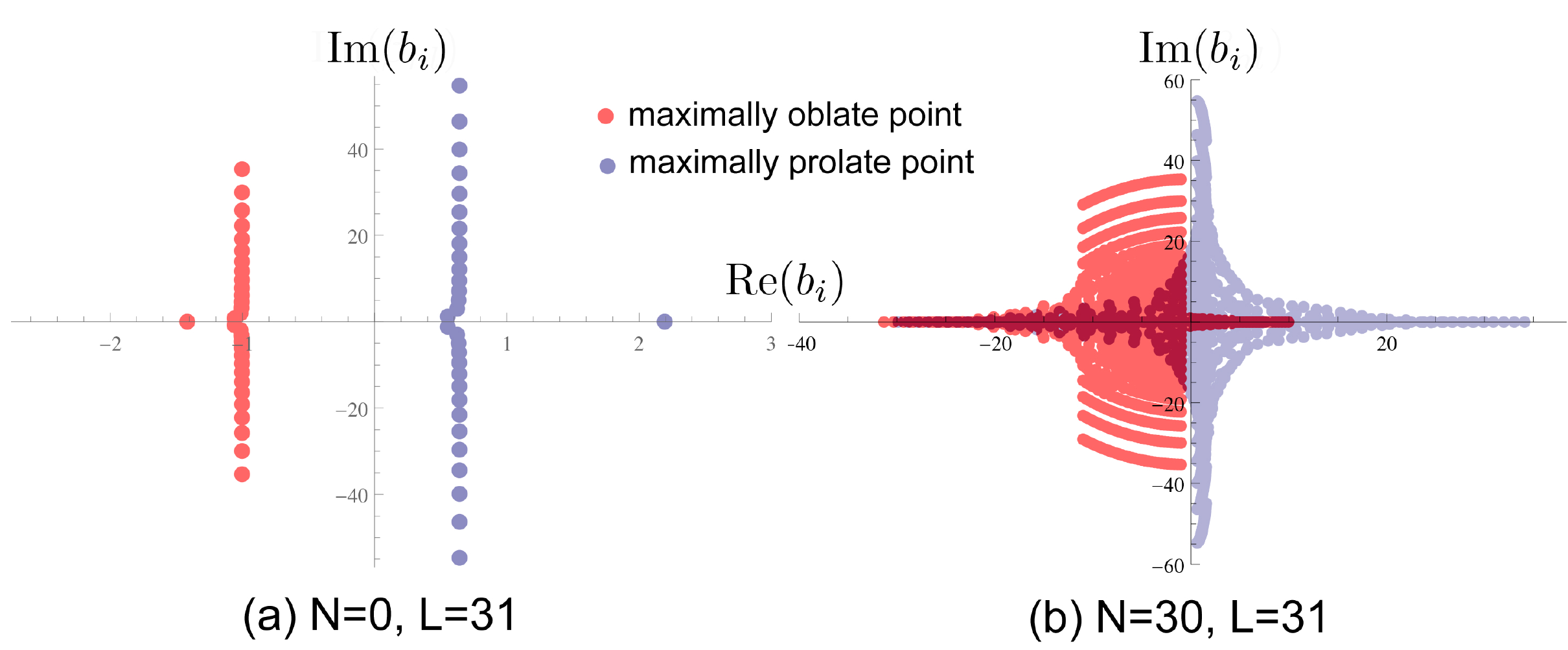} 
                \caption{
                Eigenvalues of the Jacobian matrix around asymptotic UV fixed points in the truncated system. Red and blue points denote eigenvalues of the maximally oblate point and the spiral source, respectively.
                In (a), maximally oblate point behaves like a spiral sink because $\Re (b_i)$ are all negative. For the maximally prolate point, however, $\Re (b_i)$ are all positive, a property that classifies it as a spiral source.
                In (b), both of these points are shown to be able to have positive and negative $\Re(b_i)$ at the same time, hence they are in general spiral saddle points. From the perspective of transseries, upon choosing those integration constants $\sigma_i=0$ associated with the monomials with $b_i$ satisfying $\Re(b_i)>0$, we can make the stability of the UV fixed points change to a source in the general case $N>0$.
                }
                        \label{fig:stability_UV}
    \end{center}
\end{figure}

\begin{figure}[!htpb]
\begin{center}
\includegraphics[scale=.9]{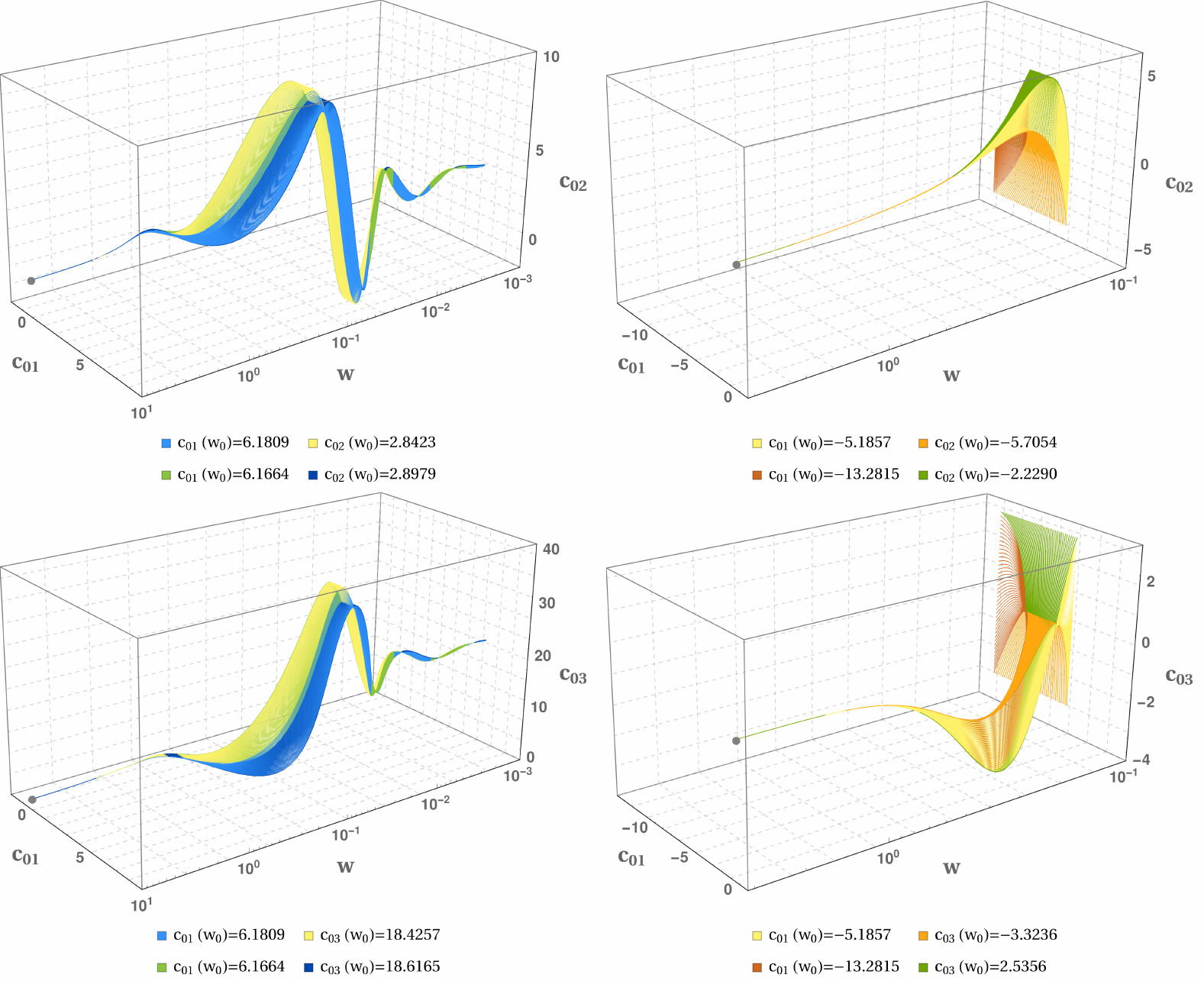}
\caption{The phase portraits for the system truncated at $N=0,\, L=3$. The initial value of the flow is taken to be $c_{nl}(w_0)=c_{nl}(w_0;\xi_0)$. In the UV (early time), the flows are coming out of a spiral source as seen on the left where anistropy parameters are close to $-1$ that corresponds to $c_{01}\sim 6.6$ \cite{Behtash:2018moe}. In the range closer to the maximally oblate point, that is $10<\xi_0<1000$, the flow lines do not converge since they are initiated outside of the basin of attraction (of the invariant manifold). The maximally oblate point is exactly located on the boundary of the basin. Note that the flow lines are shown to be spiraling in the UV as $w\rightarrow 0$, a symptom of being in the vicinity of a spiral saddle. We can see the forward attractor in the IR for the flows initiated in the basin of attraction.}
\label{fig:flows_N=0}
\end{center}
\end{figure}

\begin{figure}[!htpb]
    \begin{center}
        \includegraphics[scale=.9]{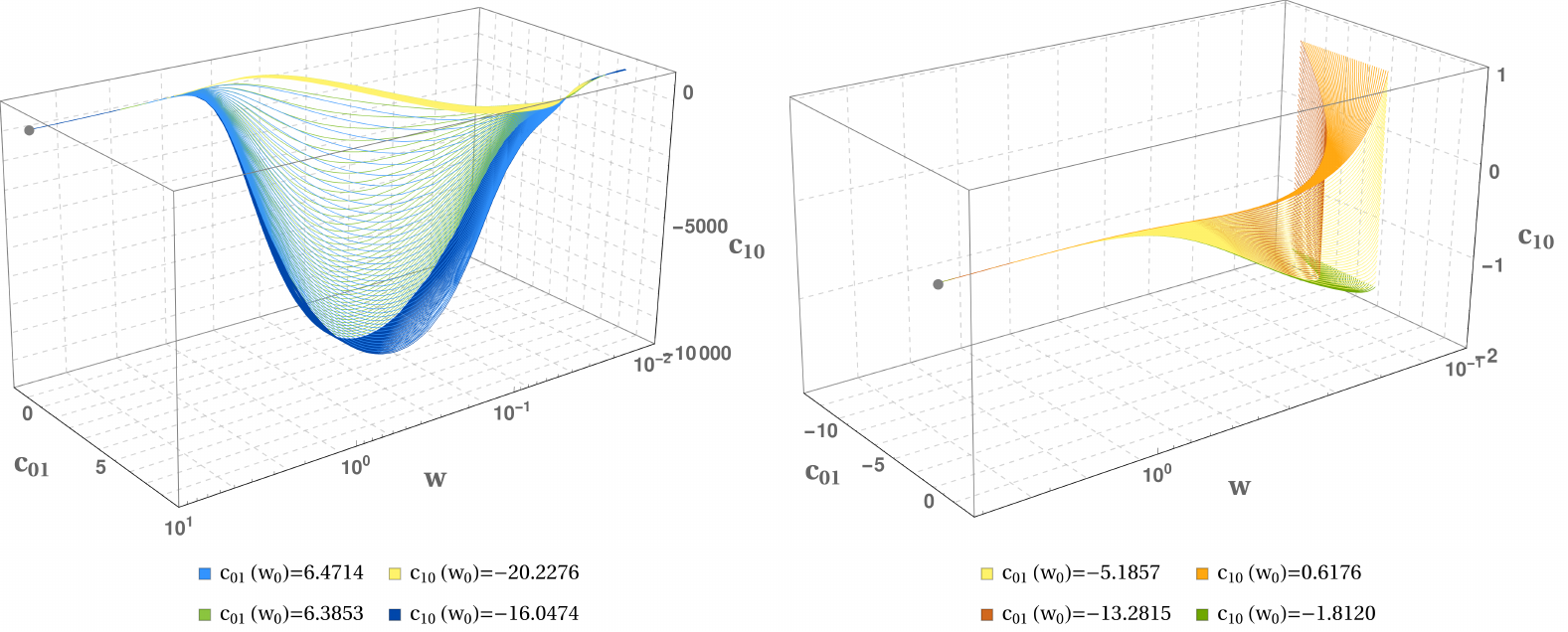}
        \includegraphics[scale=1]{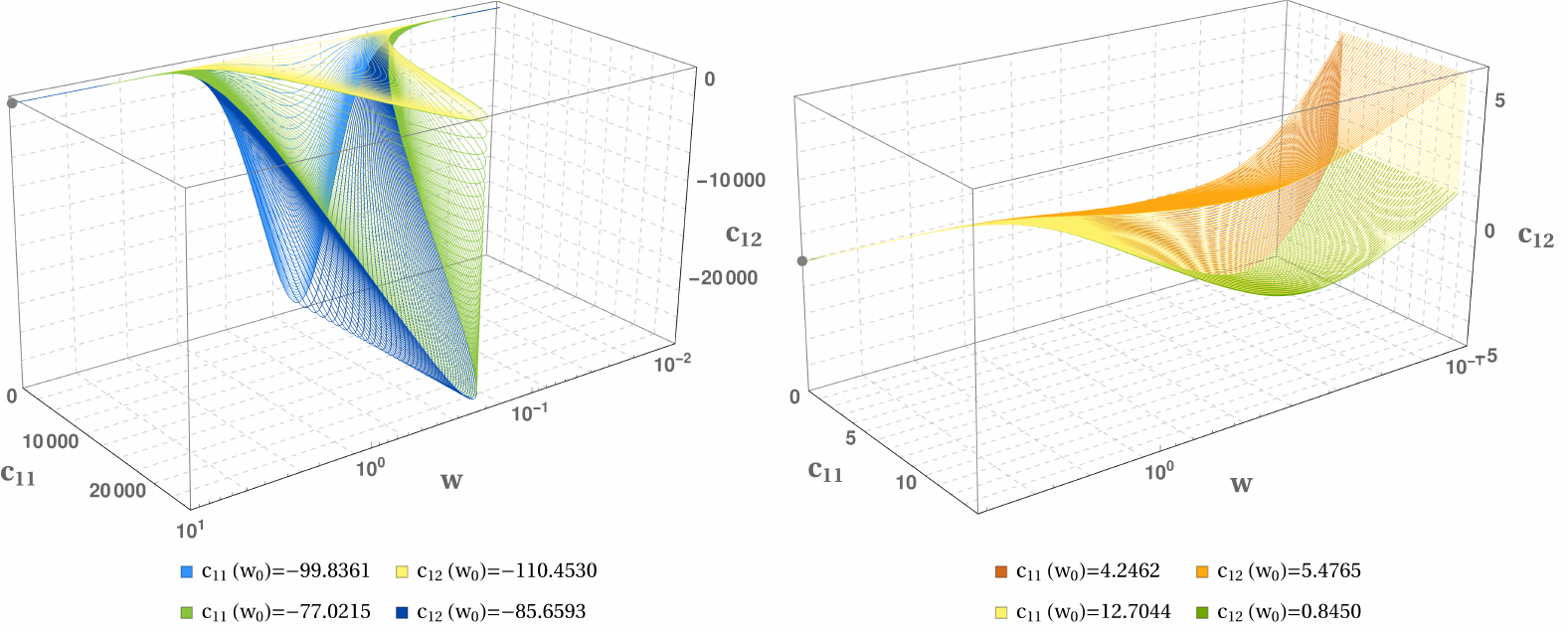} 
        \caption{The phase space portraits for the system truncated at $N=0,\, L=3$. In the UV (early time), the flows are coming out of a spiral source as seen on the left where anisotropy parameters are close to $-1$. In the range closer to the maximally oblate point, that is $10<\xi_0<1000$, the flow lines are shown to be spiraling without any convergence happening in the UV as $w\rightarrow 0$, a symptom of being in the vicinity of a spiral saddle. We can see the forward attractor in the IR for the flows initiated in either range of anisotropy parameters. }
    \label{fig:moreflow}
\end{center}
\end{figure}

There are however three major differences between the two parametrizations of the phase space of the Bjorken flow. Namely,
\begin{itemize}
\item[1-] In terms of $\tau$, the early-time $\tau<1$ or UV limit of the theory does have a pair of fixed points for the truncation order $ L,N=\infty$; one being the {\it maximally oblate point} at which the longitudinal pressure $p_L$ vanishes and transverse pressure 
$p_T$ is maximized, and one more fixed point, at which $p_L$ is maximum and transverse pressure is vanishing, to be called {\it maximally prolate point}. The early-time shape of flow lines in $\mathcal{W}$ looks completely different by allowing a continuous flow from the maximally prolate point, while in $\mathcal{M}$ there is no such flow in general since the flows cannot reach $T=0$ line on which both these points are located. Either one of the two parametrizations, nonetheless, leads to the same IR
structure for flow lines in both phase spaces $\mathcal W$ and $\mathcal{M}$. 
\item[2-] A nonphysical singularity hypersurface $c_{01}=20$ emerges in $\mathcal{W}$ that again affects the UV physics of the flow lines initiated in the basin of attraction of the invariant manifold admitting the range given in Eq. \eqref{eq:physical_range}. 
\item[3-] In the original version of the dynamical system, Eqs. \eqref{eq:ODEsBjor}, the temperature does not converge in the UV so practically speaking it is impossible to connect continuously to the maximally oblate fixed point at $T=0$ by running the evolution equations backward in time, say, from somewhere in the IR. It is also not possible to find a complete flow connecting to another UV fixed point present in the Bjorken flow phase space $\mathcal{M}$ \cite{Behtash:2018moe}. In terms of $w$, the UV structure of the theory is altered through a change in the stability of fixed points at $w \rightarrow 0$ even though the IR will remain intact as mentioned above. It is now possible to search for flow lines starting at the maximally prolate point (see Fig~\ref{fig:flows_N=0} for instance), but still it will not be feasible to have a flow that begins its journey at exactly the maximally oblate fixed point. This latter solution, if existed, would be a {\it critical line} due to the extreme fine-tuning required around a saddle point to initiate a {\it structurally unstable} flow on the boundary of the invariant manifold that connects to the IR fixed point $c_{nl}=\delta_{n0}\delta_{l0}$ at $w\rightarrow \infty$. This is explained below.

If we consider the full theory with $L=\infty$ and $N=0$ in the phase space  $\mathcal M$ of the Bjorken flow, there are two
UV fixed points in general whose coordinates are given by
\be
\label{eq:maxob}
&&{\rm maximally\,\,oblate\,\,point}:\quad \left(0,0,\left(-2.5,\left\{{(-1)^l (4 l+1) \binom{2 l}{l}}{4^{-l}}\right\}_{l=2}^\infty\right)\right),\\
&&{\rm maximally\,\,prolate\,\,point}:\quad \left(0,0,\left(5,\left\{4 l+1\right\}_{l=2}^\infty\right)\right),
\label{eq:maxpro}
\ee
where we have adopted the notation $(\tau,T,\{c_{0l},c_{02},\dots\})$.
In Fig.~\ref{fig:stability_UV} the stability analysis of these two fixed points in the truncated system $N=0,L=31$ as well as $N=30,L=31$ is shown. At a general truncation order, the stability analysis is summarized as follows for any relaxation time $\tau_r=\frac{\theta_0}{T^{1-\Delta}}$ ($0\leq \Delta\leq 1$):
\begin{enumerate}
    \item[(a)] $N=0,~L=\text{odd}$: The system has 2 fixed points in the UV. The maximally prolate fixed point is a source while the maximally oblate fixed point is a saddle point with one repelling direction, i.e., index $1$ fixed point);
    \item[(b)] $N=0,~L=\text{even}$: The system has only an index $1$ maximally oblate point. The total index is still conserved since the other fixed point becomes a complex saddle;
    \item[(c)] $N>0,~L=\text{odd}$: The maximally prolate fixed point becomes an $\text{index} (I-N+1)$ saddle point whereas the maximally oblate fixed point is an $\text{index} (N+1)$ saddle point;
    \item[(d)] $N>0,~L=\text{even}$: The maximally oblate fixed point is a saddle of $\text{index} (N+1)$.
\end{enumerate}
Here, we have again used the notation $I=(L+1)(N+1)-1$.
In $L={\rm odd}$ case the total UV index is always $I+2$. Considering that an asymptotically stable hydrodynamic fixed point has zero index, then the total Morse index is $I+2$. A comparison between the numerical solutions of the flow lines in $L={\rm odd}$ system and the exact RTA-BE shows that $I+2$ is the correct topological invariant of
the real phase portrait of the RTA-BE. Note that the temperature is always an attractive direction in both UV and IR, and therefore it does not contribute to the index.
Furthermore, due to the lack of one real fixed point (maximally prolate point) in the UV when $L=\rm {even}$, this kind of truncation will not be physically sensible and will be hereby omitted from our discussions.\footnote{Again, the past of the dynamical system in $w$ coordinate is altered from the original case in a way that $w=0$ does not capture the physics of original theory, which is an argument related to the stability of individual fixed points. Nonetheless, the conclusion about the index is true across the board.} The phase portraits of different 3d sectors of the dynamical system in \eqref{eq:evolutioneq_cnl} at odd truncation orders are plotted in Figs.~\ref{fig:flows_N=0} and \ref{fig:moreflow}.
\end{itemize}

An interesting question that comes to mind concerns the existence of a flow line that connects the maximally oblate fixed point to the asymptotic hydrodynamic fixed point in the phase portrait of the Bjorken dynamical system using $w$ coordinate. There are two things that one might keep in mind here. First, the invariant manifold is in the subspace $W_0\subset\mathcal{W}$ or simply the subspace of the phase space with coordinates $c_{0l}$. Since the maximally oblate point for $N>0$ is a saddle point of index $N+1$ (see (c) above), it must be located on the boundary of the invariant manifold. So at least in the  $N>0,~L=\text{odd}$ system, there cannot be a critical line since it will always flow outside of the basin of attraction of the invariant manifold due to the repelling directions being in the subspace $\bigcup_{n=0}^NW_n$. Second, for $N=0$, the maximally oblate point is essentially a sink (time is the only repelling direction). As a result, the $c_{0l}$ set initially to be the coordinates of this fixed point at $w=0$ will change its position with time, while being a fixed point at any moment $w>0$ until eventually it becomes the hydrodynamic fixed point at $w\rightarrow\infty$. So there will not be a flow line (process) to begin with, proving that the critical line does not exist in the $w$ parametrization of the Bjorken flow just like it could not exist in the $\tau$ coordinate but for a completely different reason.\footnote{Another way to phrase this is that the r.h.s. of the dynamical system in Eq.~\eqref{eq:dif_LH} is obviously zero at the maximally oblate fixed point, but it does remain zero at all times $w>0$ even if the position of this fixed point changes. This means that $dc_{0l}(w)/dw = 0$ so a flow will not happen to exist at any later time if ${\bf c}_0$ are equal to the values given in \eqref{eq:maxob}.} 

\subsection{Initial value problem for transseries and invariant manifold }
\label{subsec:initial}
Although the transseries contains a large amount of information about a given dynamical system, fixing the integration constants is an important, yet challenging, problem. On the one hand, since our (truncated) dynamical system has $I$ variables-thus $I$ first-order ODEs, it consequently has  $I$ integration constants needed to be tuned in order for the basin of attraction (of the invariant space) to be determined. On the other hand, the initial condition of the exact integral solution of the Boltzmann equation has only three parameters, namely $\tau_0, T_0$, and $\xi_0$.

By redefining the ODEs using $w$ as a time coordinate,  $T_0$ can be completely separated from the rest of the dynamical system which now has the form \eqref{eq:atractor}. The new initial time $w_0$ determines a point on a flow (more precisely a process), hence $\sigma_i$ {\it do} only depend on the anisotropy parameter $\xi_0$. We could behold from the numerical results of the integral solution of the Boltzmann equation that every flow is closed on an open subset of the $\mathcal{W}$ as long as $\xi_{0}$ is chosen from the domain $[-1,+\infty)$. Moreover, we recall that if a flow is initiated in the subspace $W_0\subset \mathcal{W}$ parametrized by $c_{0l}$, it always stays in that subspace, meaning that the invariant manifold has to be an embedding in the space $(W_0,\mathbb{w})$. But how does transseries know about the invariant manifold? Since $\xi_0$ determines the basin of attraction of the isolated invariant space of the Boltzmann equation, there has to be a map from the $\xi_0$-space to $W_0$ such that
\be
 {\bf \sigma}:  \,   [-1, + \infty)   &\rightarrow& (W_0,\mathbb w)\, \simeq \, {\cal F}   , \quad w_0 \in {\mathbb w}  \label{eq:sigmamap} \\
\xi_0 &\mapsto& \bm{\sigma} (\xi_0), \label{eq:ximapssigma}
\ee
where ${\mathbb w} = [0,+\infty)$ is the time manifold and ${\cal F}$ is the space of integral curves given by the $n=0$ ODEs in the dynamical system.
Because of the uniqueness theorem, the map in \eqref{eq:sigmamap} is a bijection only when $L,N=\infty$, and consequently one can find that the invariant manifold is a two dimensional surface embedded in the space $\cal F$.

It is a quite challenging problem to figure out the correspondence defined by \eqref{eq:ximapssigma}, but we can narrow  ${\mathbb R}^{I}$ down to a subspace formed by those $\bm{\sigma}(\xi_0)$ which yield flow lines starting in the basin of attraction of the invariant manifold. This means that $\bm{\sigma}({\xi_0)}$ is able to give the invariant manifold once the stability analysis of each fixed point in the UV is performed and the approximate boundary of this space is determined by finding possible critical lines.  To facilitate the search for the invariant manifold, one can construct the transseries around every fixed point individually at $w \rightarrow 0$, which would be something for the form 
\be
\tilde{c}_{i} (w) = \sum_{|{\bf m}| \ge 0}^{\infty} \sum_{k=0}^{\infty}   \bm{\sigma}^{\bf m} w^{{\bf m} \cdot \tilde{\bf b} } \tilde{u}^{({\bf m})}_{i,k} w^{k},
\ee
where the anomalous dimensions $\tilde{b}_i$ of the pseudomodes are now the eigenvalues of the linearization matrix about the corresponding fixed point.\footnote{Since pseudomodes are not physical, $\tilde{b}_i$ can take complex values in general.}
Choosing $\sigma_i = 0$ for every $i$ for which $\Re (\tilde{b}_i) < 0$, we get a maximally prolate point that is a source (that is otherwise a spiral) fixed point. We will discuss the stability of the UV fixed points in Sect.~\ref{subsec:fixedpoint}.

An immediate question that comes to mind is whether one can construct a complete flow line and/or a critical line from the transseries solutions around individual fixed points. One part of the answer involves topological arguments to be discussed in an upcoming work from the perspective of dynamical systems and equivariant cohomology Conley index \cite{Behtash:Topology}. Another piece comes from transferring the information from one transseries to another by tuning $\sigma_i$ to a particular set of numbers from both UV and IR sides, provided that the flow line achieved this way remains on the invariant manifold at all times.

As a final remark, let us mention that one can analytically continue the transseries by rotating in the time axis $w\rightarrow e^{i\alpha} w$ such that transseries solution around the asymptotic hydrodynamic fixed point is now exactly a Fourier decomposition at $\alpha=\pi/2$. Neglecting an overall imaginary factor at each non-perturbative order, it is easy to see that $\sigma_i$ serve as amplitudes of the individual wave components in the Fourier decomposition. The significant advantage of analytic continuation is that the attractor entails oscillatory components which were exponentially suppressed in real time, thus fitting the transseries to the complex numerical solutions of the analytically continued Boltzmann equation is much more tractable.
\subsection{Dynamical renormalization of 2nd order transport coefficients}
\label{subsec:dynrenor}
In this section, we aim to construct an RG equation for the transport coefficients in analogy with the gradient descent approach to the RG flows in the context of quantum field theories using the language of dynamical systems (cf. Ref. \cite{Gukov:2016tnp}). 

The variable $w$ so far has been playing the role of flow `time' in our system of ODEs, i.e., Eq. \eqref{eq:momweq}. But in this section, we want to interpret it as playing the role of `energy scale' for the renormalization scheme arising from resuming all the non-perturbative fluctuations around the asymptotic expansion of non-hydrodynamic modes ${\bf c}:=\{c_{nl}\}$. The reason is simple: the Knudsen number $Kn=\tau_r(\tau)|D_\mu u^\mu|\sim w^{-1}$ in Bjorken flow provides the only parameter by which the system evolves from the UV regime all the way to IR. Since the transseries coefficients of individual moments include transport coefficients with exponentially suppressed factors in $w$ following \cite{Behtash:2018moe}, the far-from-equilibrium dynamics of the moments can then be associated to the running of these coefficients as $w$ changes. Hence,
preparing a time-dependent (non-autonomous) dynamical system for $\bf c$ automatically amounts to having a renormalization group equation on the phase space of moments $\bf c$ and $w$, together denoted by $\mathcal{W}$, the so-called phase space of the Bjorken flow in $w$ parametrization as in \eqref{eq:phase_space_w}. 

A flow (process) on the phase space is described by a continuous map 
\begin{equation}
{\phi}^{w,w_0}(\mathbf{c}_0): \mathcal{W}\times\mathbb{w}\rightarrow \mathcal{W}
\end{equation}
such that ${\phi}^{w,w_0}(\mathbf{c}_0) = \mathbf{c}(w)$, and ${\phi}^{w,w_0}({\phi}^{s,w_0}(\mathbf{c})) = {\phi}^{w + s,w_0}(\mathbf{c})$
where $w\in \mathbb{R}^+$ is the RG time. The IR regime of the theory is captured by the behavior of the flow lines approaching to an {\it asymptotic} stable fixed point aka hydrodynamic fixed point satisfying $\cl^{eq} = \delta_{0l}\in \mathcal{M}$ for which ${\phi}^{w,w_0}(\cl^{eq}) = \cl^{eq}$ for 
all $w\in\mathbb{R}^+$. We alternatively refer to $\cl^{eq}$ as the asymptotic IR fixed point since it is reached at $w\rightarrow \infty$.

Let us formulate our dynamical system as 
    \bea
    \frac{d  {\bf c}}{dw } = {\bf f}(w,{\bf c}) &\quad {\rm or} \quad&  w\, \frac{d  {\bf c}}{dw} = w\, {\bf f}(w,{\bf c})  \quad {\rm or} \quad \frac{d  {\bf c}}{d\log\,\mu} = w\, {\bf f}(w,{\bf c}).
    \eea
    where $\mu:=w/\theta$ is the RG scale and $\log(\mu)\in \mathbb{R}$. As $\mu$ varies, the moments ${\bf c}$ are mapped to themselves in a self-similar way, and any initial RG scale $\mu_0$ can initiate a process to a later $\mu$ by the renormalization group action. Since the r.h.s. only depends on ${\bf c}$ and $\mu$, one can write
    \be
    \frac{d  {\bf c}}{d \log \mu} = \bm{\beta}(\mu,{\bf c})
    \equiv -\frac{1}{1-\frac{c_{1}}{20}} \left[ \frac{3 \mu}{2} {\bf c} + {\cal B}{\bf c} - \frac{c_{1}}{5}{\bf c} + \frac{3}{2} \bm{\gamma} \right].
    \ee
    This is nothing but an RG equation in the space of moments $\bf c$ and the vector $\bm{\beta}(\mu,{\bf c})$ consists of $\beta$-functions, each encoding the dependence of  every non-hydrodynamic mode on the renormalization scale, $\mu$, in the process of equilibration.
    In this expression, it is convenient to redefine our notations as
    \be
    \begin{split}
    & \tilde{u}_{l,k} \rightarrow \tilde{u}_{l,k} \theta^{k}, \quad F_{l,k} \rightarrow F_{l,k} \theta^{k}, \\
    & \lambda \rightarrow \lambda \theta^{-1}, \quad S_l \rightarrow S_l \theta^{-1} , \quad \zeta_l \rightarrow \zeta_l \theta^{\tilde{b}_l}.
    \end{split}
    \ee
    Note that the ordinary derivative along an RG flow can be expressed by two partial derivatives as
    \be
    \frac{d}{d \log \mu} =  \sum_{l=1}^{L}  \left(\tilde{b}_{l} -S_{l} \mu \right) \frac{\partial}{\partial \log\zeta_l} +\frac{\partial }{\partial \log \mu}.
    \ee

In the formalism we have been seeking to build so far, one prominent assumption is that an observable ${\cal O}$ is able to be expressed in terms of $c_i(w)$ (or $\tilde{c}_i(w)$), namely ${\cal O}={\cal O}({\bf c}(w))$. Therefore by solving the RG equation via transseries, we can achieve a renormalized form of the observable, say a transport coefficient, up to an arbitrary order of our choosing.

It is straightforward to define the RG equation by starting with the ODE in \eqref{eq:dif_LH_cnl}. To make things more tractable, we again go back to the RG time $w$ and compute the derivative of an observable ${\cal O}({\bf c}(w))$ w.r.t. $\log w$
    \be
    \label{eq:ren_O_b}
    \frac{d  {\cal O} ({\bf c}(w))}{d \log w} 
    &=& \sum_{i=1}^I \left[ \sum_{k=0}^{\infty} \left\{ \left(\tilde{\bf b} - {\bf S} w \right) \cdot  \hat{\bm{\zeta}} \tilde{C}_{i,k}  - k \tilde{C}_{i,k} \right\} w^{-k} \right]  \cdot  \frac{\partial  {\cal O}}{\partial \tilde{c}_i} \nl
    &=& \sum_{i=1}^I \tilde \beta_i  \cdot \frac{\partial  {\cal O} }{\partial \tilde{c}_i}, 
    \ee
    where $\tilde{\bf f} = U {\bf f}$ and $\tilde{\bm{\beta}}:=w \tilde{\bf f}$. 
    Here, use was made of $d/d \log w = (\tilde{\bf b}-{\bf S} w) \cdot \hat{\zeta} + \partial/\partial \log w$ and the fact that $\tilde{c}_i(w)=\sum_{k=1}^{\infty} \tilde{C}_{i,k}(\bm{\sigma \zeta}(w)) w^{-k}$.
    We should also point out that $\hat{\zeta}_i:= \partial/\partial \log (\sigma_i \zeta_i)$.    The first line describes the scaling behavior of ${\cal O}$ in terms of $w$, and the second line contains information of the dynamical system through $\bm{\beta}$.
    Hence, solving the RG equation determines a renormalization of the observable ${\cal O}$.
    Since we want to consider the RG equation for the transport coefficients, we are slightly better off with the definition of ${\cal O}$ changed to ${\cal O}={\cal O} ({\bf C}_k(w))$.
    Roughly speaking, this implies that ${\cal O}$ is a function of the transport coefficients and depends only on $\bm{\zeta}(w)$.
    Consequently, the scaling behavior could be derived in a fashion similar to what we did to obtain \eqref{eq:ren_O_b}, so
    \be
    \label{eq:ren_O_C}
    \frac{d  {\cal O} ({\bf C}_k(w))}{d \log w} 
    &=& \sum_{i=1}^I \sum_{k=0}^\infty \left[ \left(\tilde{\bf b} - {\bf S} w \right) \cdot \hat{\bm{\zeta}} \tilde{C}_{i,k} \right]  \cdot  \frac{\partial  {\cal O}}{\partial \tilde{C}_{i,k}}.
    \ee
    However, the link  between the scaling behavior and the dynamical information hidden in $\bm{\beta}$ is very non-trivial.
    Yet, one can proceed to calculate the RG equation for the $L=1,\,N=0$ system.
    
    To do so, we employ the scaling behavior in \eqref{eq:ren_O_C} which gives
    \bea
    \sum_{k=0}^{\infty} \left\{ \left(\tilde{b}_1 - S_1 w \right)   \hat{{\zeta}}_1 \tilde{C}_{1,k}  - k \tilde{C}_{1,k} \right\} w^{-k}    =\tilde {\beta}_1. \label{eq:sc_b}
    \eea
    Suppose now that $\zeta_1$ and $w$ are independent of each other, i.e., $\tilde{c}_1 = \tilde{c}_1(\sigma_1 \zeta_1,w)$ and $\tilde{C}_{1,k} = \tilde{C}_{1,k}(\sigma_1 \zeta_1)$. In this case we have
    \bea
    \sum_{k=0}^{\infty}  \hat{{\zeta}}_1 \tilde{C}_{1,k}(\sigma_1 \zeta_1) w^{-k}  = \frac{\sum_{k=0}^{\infty} k \tilde{C}_{1,k} (\sigma_1 \zeta_1)  w^{-k} + \tilde {\beta}_1(w,\tilde{c}_1(\sigma_1 \zeta_1,w))}{(\tilde{b}_1 - S_1 w )}. \label{eq:zetaCw}
    \eea
    To remove $w$ from the l.h.s. of Eq.~\eqref{eq:zetaCw}, we take a contour integration around the origin after multiplying by $w^{k-1}$, which then gives
    \bea 
    \hat{{\zeta}}_1 \tilde{C}_{1,k}(\sigma_1 \zeta_1)   = \frac{1}{2 \pi i} \oint_{|w| \ll 1} dw \, \frac{\sum_{k^\prime=0}^{\infty} k^\prime \tilde{C}_{1,k^\prime} (\sigma_1 \zeta_1)  w^{-k^\prime} + \tilde {\beta}_1(w,\tilde{c}_1(\sigma_1 \zeta_1,w))}{w^{1-k} (\tilde{b}_1 - S_1 w )}. \label{eq:hatC_int}
    \eea
        Hence, Eq.~\eqref{eq:ren_O_C} for $L=1$ and $N=0$ reads
    \bea
    \label{eq:RG}
    \frac{d  {\cal O} (C_{1,k}(w))}{d \log w} 
    &=& 
    \sum_{k=0}^\infty \left( \tilde{b}_1 - S_1 w \right) \cdot \left[  \hat{\zeta}_1 \tilde{C}_{1,k}(\sigma_1 \zeta_1)  \right]_{\zeta_1 = \sigma_1 e^{-S w} w^{\tilde{b}_1}} \cdot  \frac{\partial  {\cal O}}{\partial  \tilde{C}_{1,k}} \label{eq:ren_O_C2},
    \eea
    where $\hat{\zeta}_1 \tilde{C}_{1,k}(\sigma_1 \zeta_1)$ is given by Eq.~\eqref{eq:hatC_int}.
    
    The definition in \eqref{eq:ren_O_C2} has a direct connection with the dynamical information encoded in $\beta_1$, so that \eqref{eq:ren_O_C2} is regarded as the RG equation of the transport coefficients by setting ${\cal O}(C_{1,k}(w))=C_{1,k}(w)$.
    In the $L=1,N=0$ system, choosing $k=1$ in Eq.~\eqref{eq:ren_O_C2} and solving for the renormalized quantity $C_{1,1}$ gives that it is
      proportional to the shear-to-entropy ratio at the equilibrium \cite{Behtash:2018moe}. Plugging $C_{1,1}$ back in the expansion of the non-hydrodynamic moment $c_{1}$ obtained using the linear response theory (see App.~\ref{app:asymp}) automatically promotes the 1st-order transport coefficient to the non-equilibrium case compatible with the transasymptotic matching along an RG flow approaching the IR fixed point asymptotically. This essentially gives a dynamically renormalized $\eta/s$ \cite{Behtash:2018moe} as follows
    \bea
    c^{\rm 1st}_{1}(w\rightarrow \infty) &=& - \frac{8}{3} \frac{\theta_0}{w} = - \frac{40}{3} \frac{\left(\eta/s\right)_0}{w},\\
    \left(\frac{\eta}{s}\right)_{\rm reno} &=& - \frac{3}{40} C_{1,1}(w),
    \eea
    where $-\frac{3}{40} C_{1,1}(w\rightarrow \infty) \equiv (\eta/s)_0$~\cite{Behtash:2018moe}.
    Likewise, the 2nd order renormalized transport coefficients are given by 
    \bea
    c^{\rm 2nd}_{1}(w\rightarrow \infty) =  - \frac{32}{63} \frac{\theta_0^2}{w^2} &=& -\frac{40}{9}T\left[\tau_\pi \left(\frac{\eta}{s}\right)_0 - \left(\frac{\lambda_1
    }{s}\right)_0\right] \frac{1}{w^2},\\
     T\left[\tau_\pi \frac{\eta}{s} - \frac{\lambda_1}{s}\right]_{\rm reno}&=&  \frac{9}{40}C_{1,2}(w),
    \eea
    where again $\frac{9}{40} C_{1,2}(w\rightarrow \infty) \equiv T\left[\tau_\pi \left(\frac{\eta}{s}\right)_0 - \left(\frac{\lambda_1
    }{s}\right)_0\right]$.
    
    We conclude this section by discussing whether there is some sort of a Lyapunov function that would identify the RG flows from a global dynamical standpoint. It is well-known that one can formally define an object analogous to a \textit{dynamical potential} from the Newtonian mechanics. For simplicity, we assume that $\beta$ is independent of $w$ (a scenario 
    that would hold once the limit $\theta_0 \rightarrow \infty$ is taken).\footnote{When we keep $w$ in the $\beta$-function, we have to promote the non-autonomous system to an autonomous system of one dimension higher by introducing an ODE for $w$ in terms of a new flow time $\rho$ as follows
        \be 
        \frac{d{\bf c}(\rho)}{d \log \rho} = \bm{\beta}({\bf c}(\rho),w(\rho)), \quad \frac{dw(\rho)}{d \log \rho} = \beta_w(w(\rho)).
        \ee
    }
    Now, we define a positive-definite differentiable function $V$
    \bea
    && \frac{d c_i(w)}{d \log w} = \beta_i ({\bf c}(w)), \qquad  \beta_i({\bf c}) =- \frac{\partial V({\bf c})}{\partial c_i}.
    \eea
    We can easily show that $V$ is a monotonically decreasing function in terms of $w$ as
    \bea
    \frac{dV({\bf c}(w))}{d \log w} &=& \sum_{i=1}^I \frac{dc_i(w)}{d \log w} \cdot \frac{\partial V({\bf c}(w))}{\partial c_i(w)} \nl
    &=& - \sum_{i=1}^I |\beta_i({\bf c}(w))|^2 \le 0.
    \eea
    thus a candidate for a global Lyapunov function satisfying the conditions in Eq.~\eqref{eq:lyapunov}.\footnote{In case $V$ is not positive
        definite, one can always add a positive constant to it such that it remains positive at all times $w<\infty$.}
\section{Hydrodynamization of soft and hard modes}
\label{sec:hydrodyn}

In previous sections, we demonstrated that the moments $c_{nl}$ solving the ODEs in the dynamical system of \eqref{eq:momweq} turn out to be of the multi-parameter transseries form. In this section, we continue our quest for what type of information one can get from these solutions on a more physical ground by analyzing the late-time behavior of different momentum and energy sectors of the distribution function. In doing so, we stumble upon something interesting: a flow line in the space of moments initiated at an arbitrary state in the basin of attraction of the asymptotically stable fixed point, is controlled at late times by not only the known non-hydrodynamic moment $c_{01}$ but also the mode $c_{11}$. The latter also happens to possess the same perturbative decay as $c_{01}$. We recall that this immediately leads to the statement that the distribution function and any observable projecting onto/or involving at least the $l=1$ sector of the distribution function would receive {\it two} major IR contributions. In this section, we shall analyze in great details the impact of the slowest ($=\mathcal{O}(1/(\tau T))$) non-hydrodynamic modes in the IR. 

Let us start rewrite our ansatz as
\be
f\left(\tau,p_T,p_\varsigma\right)= f_{eq.}\,+\,f_{s}\,+\,f_{sh}\,+f_{h}\,.
\ee
where the soft (s), semi-hard (sh), and hard (h) sectors of the distribution function are respectively defined by 
\bs
\label{eq:sectors}
\beal
f_s&:=\,f_{eq.}\left(\frac{p^\tau}{T}\right)\,\left[\sum_{l=1}^{N_l}\,c_{0l}(\tau)\,\mc P_{2l}\left(\frac{p_\varsigma}{\tau\,p^\tau}\right)\,\,\right]\,,\\
f_{sh}&:=\,\,f_{eq.}\left(\frac{p^\tau}{T}\right)\,\left[\sum_{n=1}^{N_n}\sum_{l=1}^{N_l}\,\cnl(\tau)\,\mc P_{2l}\left(\frac{p_\varsigma}{\tau\,p^\tau}\right)\,\mc L^{(3)}_{n}\left(\frac{p^\tau}{T}\right)\,\right]\,,\\
f_{h}&:=\,\,f_{eq.}\left(\frac{p^\tau}{T}\right)\,\left[\sum_{n=1}^{N_n}\,c_{n0}(\tau)\,\mc L^{(3)}_{n}\left(\frac{p^\tau}{T}\right)\,\right]\,.
\end{align}
\es
In order to assess which sector of the distribution function hydrodynamizes faster, we focus on the decay of the following normalized moments~\cite{Denicol:2016bjh,Strickland:2018ayk} \footnote{In this work we follow the notation of Strickland~\cite{Strickland:2018ayk}.}
\be
\label{eq:mnm}
\bar M^{nm} (\tau)  =\frac{\langle\, \left({p^\tau}\right)^n \left(p^\varsigma/\tau\,\right)^{2m}\,\rangle}{\langle\, \left({p^\tau}\right)^n \left(p^\varsigma/\tau\,\right)^{2m}\,\rangle_{eq}}\,.
\ee
$\bar M^{nm}$ are then certainly sensitive to the energy and momentum tails of the distribution function. To this explicitly, first note that $\bar M^{nm}\geq 0$ if and only if $f(\tau,p_T,p_\varsigma)\geq 0$. This condition holds for the exact solution of the RTA-BE, but not necessarily for the approximate distribution function. Second, the Landau matching condition for the energy density implies $\bar M^{20}\equiv 1$. Last, some of the moments in~\eqref{eq:mnm} such as $\bar{M}^{10}=n/n_{eq.}$ (,  where $n$ and $n_{eq.}$ are the non-equilibrium and thermodynamic particle densities, respectively) and $\bar{M}^{01}=P_L/P_0$ have a direct interpretation in terms of the usual macroscopic variables. 

After taking the time $\bar M^{nm}$ in Eq.~\eqref{eq:mnm}, we end up with the following equation
\be
\label{eq:Mnmeq}
\partial_\tau \bar M^{nm}  
+\left(\frac{2(3m+1)-n}{4\,\tau}\right)\,\bar M^{nm} \,-\,\left(\frac{n+2(m+1)}{12\,\tau}\right)\,\bar M^{nm}\bar M^{01}\,+\,\frac{(n-1)(1+2m)}{(1+2(m+1))}\,\frac{\bar M^{n-2,m+1}}{\tau}   & 
= -\frac{1}{\tau_r}\left( \bar M^{nm} -1 \right)\,.
\ee
The solutions to this system indicate that different momentum tails of the distribution function reach values at equilibrium state asymptotically, that is the asymptotic hydrodynamic fixed point. In order to put the problem into the framework of dynamical systems, we write $\bar M^{nm}$~\eqref{eq:mnm} as a linear combination of the moments $c_{nl}$ using the ansatz \eqref{eq:ansfunc} of the distribution function, i.e.\footnote{In order to get this expression, we utilize the following identities~\cite{gradshteyn2007,abramowitz_stegun64}
\bs
\begin{align*}
& \int_0^\infty\,dx\,e^{-x}\,x^{\gamma-1}\,\mc L^{(\mu)}_n(t)=\frac{\Gamma\left(\gamma\right)\,\Gamma\left(1+\mu+n-\gamma\right)}{n!\,\Gamma\left(1+\mu-\gamma\right)}\,\,,\hspace{1cm}\text{Re}[\gamma]>0\,,\\
& \int_0^1\,dx\,x^\rho\,\mc P_{m}(x)=\frac{\pi^{1/2}}{2^{\rho+1}}\,
\frac{\Gamma\left(\rho+1\right)}{\Gamma\left(1+(\rho-m)/2\right)\,\Gamma\left(m/2+3/2+\rho/2\right)}\,,\hspace{1cm} \text{Re}[\rho]>-1\,.
\end{align*}
\es
}
\begin{figure}[!htpb]
    \centering
    \includegraphics[width=.92\linewidth]{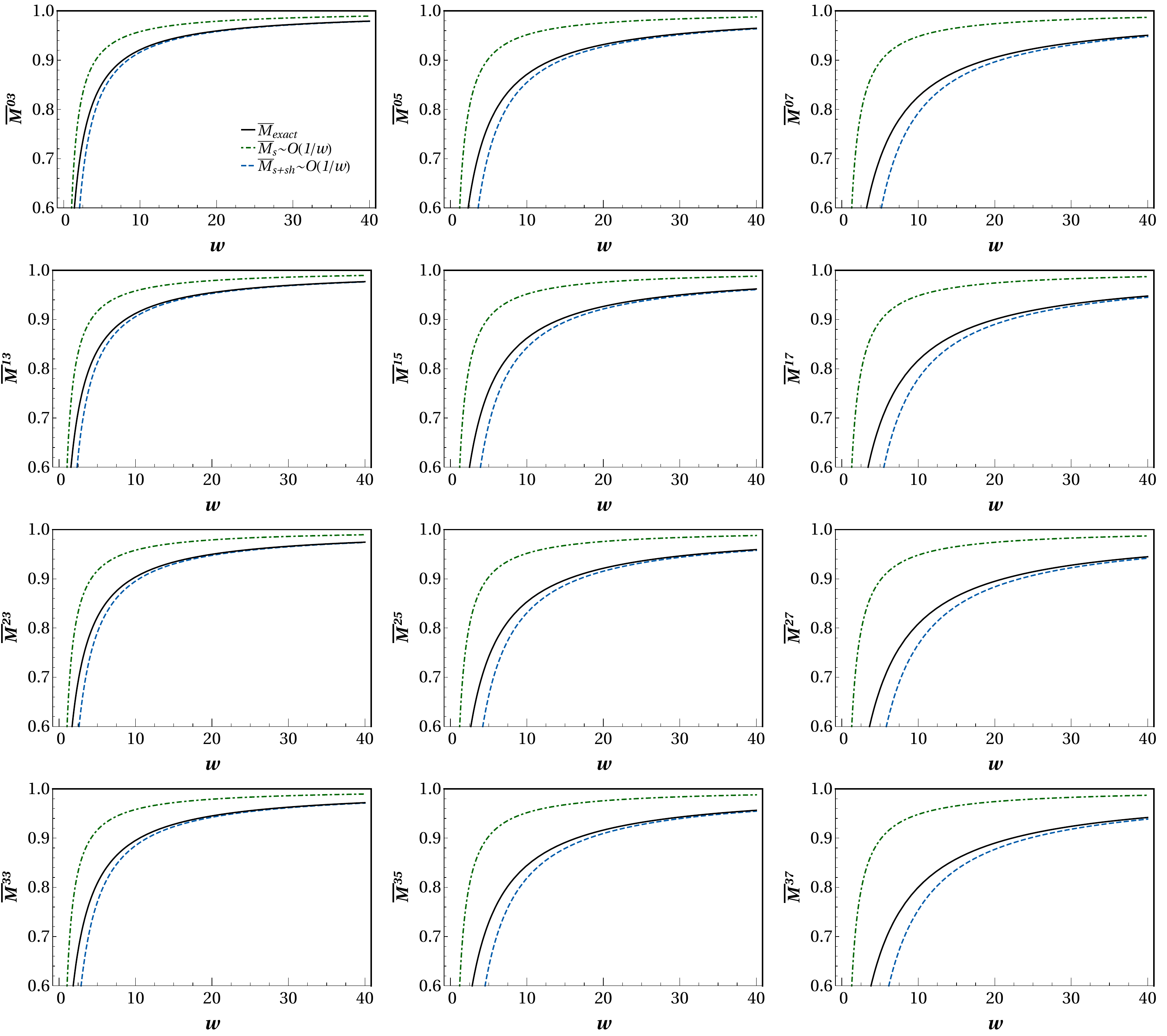}
\caption{Evolution of several normalized moments $\bar M^{nm}$ as a function of $w$. We compare the exact numerical expressions for the normalized moments obtained by solving RTA-BE~\eqref{eq:mnmexact} (black lines) alongside the soft limit~\eqref{eq:mnm-NS} (green dash-dotted lines) and the soft+semi-hard limit~\eqref{eq:mnm-c11} (blue dashed lines). For the initial conditions of the exact RTA-BE solutions, $\tau_0=0.25$ fm/c is chosen, and the initial temperature is set to $T_0=600$ MeV, with initial anisotropy parameter being $\xi_0=10$.}
\label{fig:thermalization}
\end{figure}
\begin{figure}[!htpb]
\begin{center}
\includegraphics[width=.8\linewidth]{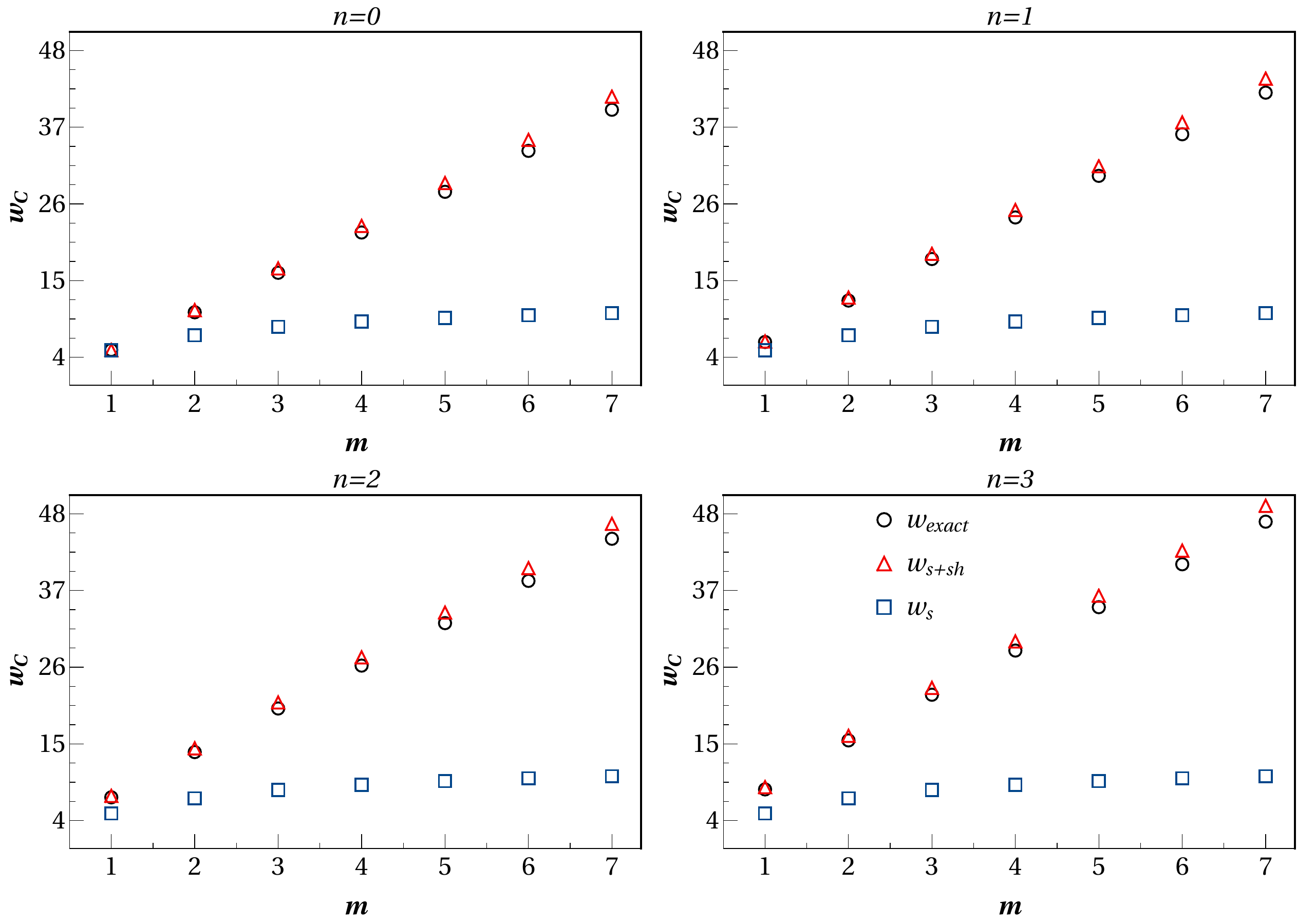}\\
\caption{Saturation time $w_c$ vs. $m$ for different values of $n=\{0,1,2,3\}$ (top right, top left, bottom right and bottom left panels, respectively). The black dots are computed using the exact RTA-BE solver; the blue squares and red triangles represent the values obtained in the soft and soft+semi-hard regimes. We also see that the normalized moments $M^{nm}_{exact}$ equilibrate later as both $n$ and $m$ are increased. }
\label{fig:thermtime}
\end{center}
\end{figure}
\be
\bar M^{nm} (\tau)= 1+ \frac{2m+1}{2} \sum_{\substack{k,l=0 \\ k+ l>0}} \,\frac{\Gamma(\frac{1}{2}+m) \Gamma(1+m)}{\Gamma(\frac{3}{2}+m+l)\Gamma(1+m-l)} \binom{1+k-n-2m}{k}\,c_{kl}(\tau) \,.
\label{eq:mnmcnl}
\ee
where $\Gamma(x)$ is the Gamma function and $\binom{a}{b}$ is the binomial coefficient. 
%
Therefore the solutions to the dynamical system in~\eqref{eq:Mnmeq} are also written in the form of multi-parameter transseries.\footnote{Alternatively one can also find a compact form of the moment $\bar M^{nm}$  from the exact RTA-BE solution of the distribution function (see Eq.~\eqref{eq:distf} in App. \ref{app:thermalizationDelta1}).} These solutions flow to the stable fixed point at late times once the initial state is chosen in the basin of attraction of the invariant manifold (of the truncated system), which indicates that each momentum sector is guaranteed to equilibrate eventually. In other words, $\bar M^{nm}\to 1$ since $\cnl^{eq.}\to\delta_{n0}\delta_{l0}$ asymptotically. 

In Sect.~\ref{subsubsec:ngen}, (cf. Eq.~\eqref{eq:c11-c01asy}), we showed that the slowest non-hydrodynamic moments decay perturbatively like $c_{01},c_{11}\sim 1/w$ and $c_{10},c_{20}\sim 1/w^2$ at large $w$. Thus, the asymptotic behavior of $\bar M^{nm}$ is dominated by the slowest moments appearing in each sector of the distribution function. 

We study first the normalized moments $\bar M^{nm}$ for $n,l>0$ where there are contributions coming only from both {\it soft} and {\it semi-hard} sectors the distribution function whose slowest non-hydrodynamic modes are $c_{01}$ and $c_{11}$, respectively. In this case, the impact of these non-hydrodynamic modes is determined by taking the following asymptotic limits in Eq.~\eqref{eq:mnmcnl}:
\begin{itemize}
    \item \textit{Soft (s) regime} in which there is only the leading order asymptotic contribution of $c_{01}\sim 1/w$, i.e.,\footnote{In Eq.~(4.6) of Ref.~\cite{Strickland:2018ayk}, the author assumes explicitly the 14-moment approximation when truncating the distribution function, that is
    \begin{equation*}
    \begin{split}
        f&=f_{eq.}\left(1+\frac{p_\mu p_\nu \pi^{\mu\nu}}{2(\epsilon+p)T^2}\right)\,,\\
        &=f_{eq.}\left(1+\frac{3}{16} \left[3\left(\frac{p_\varsigma}{\tau}\right)^2-\left(p^\tau\right)^2\right]\frac{\bar{\pi}}{T^2}\right)\,,\\
        &\approx\,f_{eq.}\left(1+\frac{1}{3T^2} \left[\left(p^\tau\right)^2-3\left(\frac{p_\varsigma}{\tau}\right)^2\right]\frac{\eta}{s}\,\frac{1}{T\,\tau}\,\right)\,.
    \end{split}
    \end{equation*}
    By substituting this expression in our definition~\eqref{eq:mnm}, one obtains the result derived by Strickland, Eq.(4.8)~\cite{Strickland:2018ayk}, which obviously differs from ours~\eqref{eq:mnm-NS}. This discrepancy arises from the truncation procedure. Here, we do not truncate the momentum and energy dependence of the distribution function and rather keep these in their exact form. Instead, we truncate the number of moments entering the distribution function at our discretion. We remind the reader of the work of Denicol et al.~\cite{Denicol:2012es}, where it was shown that the 14-moment approximation does not provide a unique set of equations of motion for the dissipative currents, an ambiguity which was resolved in Ref.~\cite{Denicol:2012cn}.
    }
    \be
    \label{eq:mnm-NS}
        \bar M^{nm}_{s} = 1 + \frac{2m}{2m+3} c_{01} \approx 1 - \frac{16m}{6m+9}\frac{\theta_0}{w} + \mathcal{O}(1/w^2), \quad {\rm for}\,\,\, m\geq 1.
    \ee
    \item \textit{Soft + semi-hard ($s+sh$) regime} which incorporates the leading order asymptotic contribution of {\it both} $c_{01}$ and $c_{11}$ up to $\mathcal{O}(w^{-1})$, namely 
    \be
    \label{eq:mnm-c11}
        \bar M^{nm}_{s+sh} = 1 + \frac{2m}{2m+3} c_{01} + \frac{2m(2-n-2m)}{2m+3} c_{11}\approx 1 - \left(\frac{4m\,(n+2m+2)}{6m+9}\right)\,\frac{\theta_0}{w} + \mathcal{O}(1/w^2), \quad {\rm for}\,\,\, m\geq 1.
    \ee
\end{itemize}
In Fig.~\ref{fig:thermalization}, a number of $\bar M^{nm}$ are plotted versus the variable $w$, obtained from the exact RTA-BE solution~\eqref{eq:mnmexact} (black line), the $s$~\eqref{eq:mnm-NS} (green dot-dashed line), and the $s+sh$~\eqref{eq:mnm-c11} (blue dashed line) regimes. We verify numerically that the IR behavior is not affected by the choice of initial conditions as long as they agree with the bounds set by the invariant manifold. In order to get a hold of how fast the normalized moments equilibrate, we set an arbitrary saturation bound at which a given moment falls within 5\% of its equilibrium value, that is to say $M^{nm}= 1\pm\delta$ with $\delta=0.05$, and the $+$ or $-$ is taken depending on whether the moment monotonically increases or decreases, respectively.  In this setting, as seen in Fig.~\ref{fig:thermalization}, it can be verified that independently of the truncation scheme, the normalized moments will asymptote to one, as expected. In the UV, however, the $s$ and $s+sh$ regimes disagree with the exact RTA-BE result. This should not come as a surprise since both limits are only valid when the distance from the hydrodynamic fixed point is small. But in the IR region, the moments $\bar M^{nm}_{s}$ first reach the saturation bound while $\bar M^{nm}_{s+sh}$ and $\bar M^{nm}_{exact}$ approach this value much later on, and above all else, {\it almost at the same time}. In the same limit, we behold a remarkable agreement between the exact and $s+sh$ solutions. Thus, the asymptotic behavior of the normalized moments $\bar{M}^{nm}$ ($n,l>0$) can only be given entirely by a linear combination of both non-hydrodynamic modes $c_{01}$ and $c_{11}$. 

In Fig.~\ref{fig:thermtime}, the value of $w=w_c$ is numerically evaluated. As is shown, if the indices for energy, $n$, and longitudinal momentum, $m$, are increased, $w_s$ for the exact and $s+sh$ moments marks a later time. The convergence of $\bar{M}^{nm}_{s}$ depends only on $m$, which is evident from Eq.~\eqref{eq:mnm-NS}. The numerical results in this figure attests that in fact the exact and $s+sh$ moments are in good agreement, and therefore $s+sh$ is the correct description of the IR regime of different energy and momentum sectors of the distribution function solely based on a linear combination of {\it both} $c_{01}$ and $c_{11}$ as written in Eq.~\eqref{eq:mnm-NS}~\footnote{A careful reader would notice a minor disagreement (less than 5$\%$) between the values of $w_c^{exact}$ and $w_c^{s+sh}$. This small mismatch is due to the difficulty in extracting numerically the large-time behavior of the exact RTA-BE solver used here. While concluding the draft of this paper, we were informed of an optimized algorithm that uses logarithmically-spaced grid in proper time and reduces the numerical error at late times. This new code is available in the ancillary files included in the arXiv version of Ref.~\cite{Strickland:2018ayk}.}. Given this agreement, one concludes that the larger the index $n>0$, the more relevant the inclusion of the new non-hydrodynamic mode, $c_{11}$. We want to mention that this is solely due to the power of dynamical systems since the $1/w$ decay of both $c_{01}$ and $c_{11}$ in the perturbation theory around the hydrodynamic fixed point is directly confirmed from the form of ODEs in \eqref{eq:Mnmeq}.
\begin{figure}[!htpb]
    \centering
    \includegraphics[width=.92\linewidth]{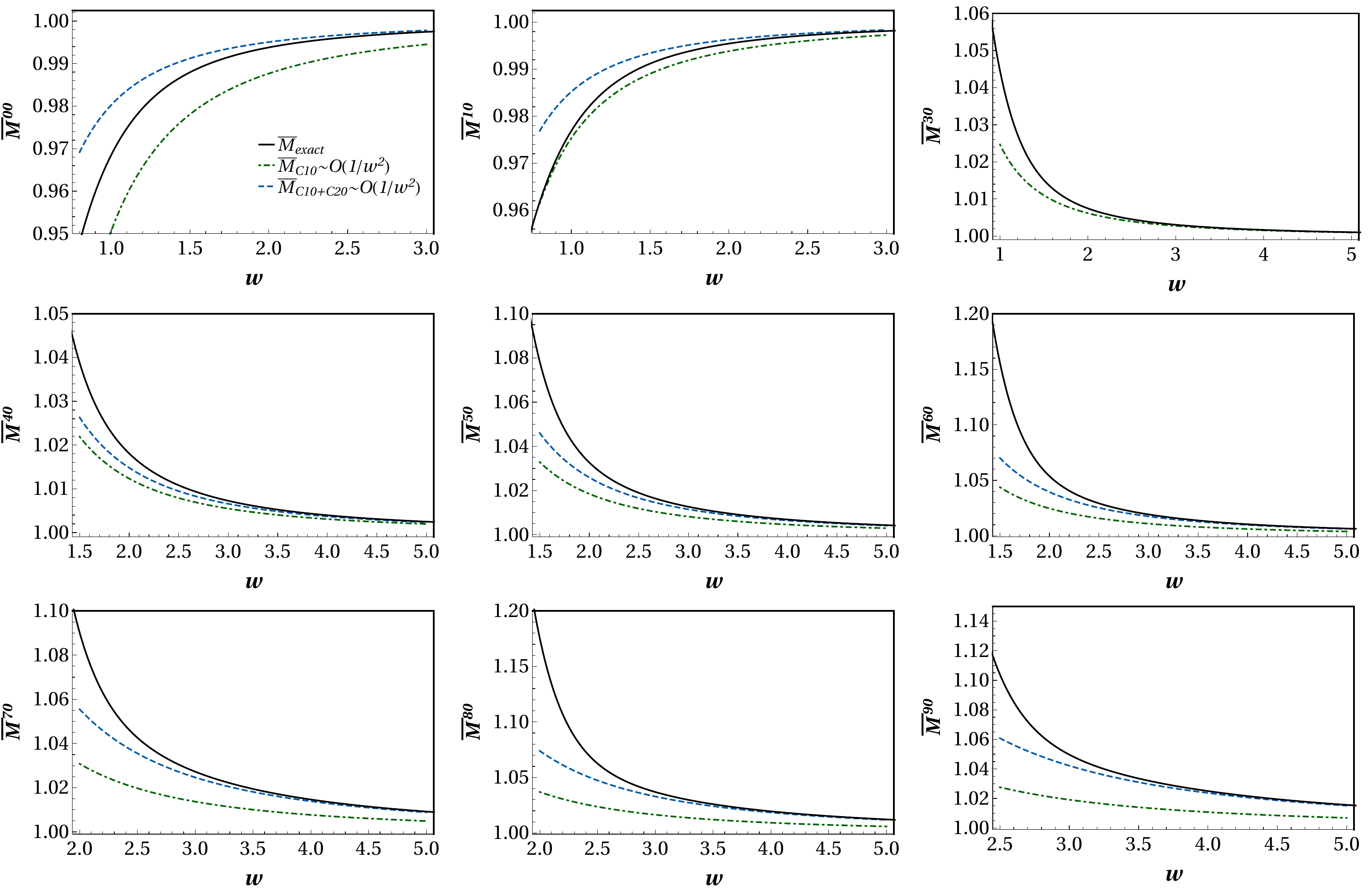}\\
\caption{Evolution of the normalized moments $\bar M^{n0}$ as a function of $w$. In this figure, the exact numerical expressions for the normalized moments obtained by solving RTA-BE~\eqref{eq:mnmexact} (black lines) are compared against Eqs.~\eqref{eq:mn0-cna} (green dot-dashed lines) and hard mode limit~\eqref{eq:mn0-cnb} (blue dashed lines). For the initial conditions of the exact RTA-BE solutions, $\tau_0=0.25$ fm/c is chosen, and the initial temperature is set to $T_0=600$ MeV, with anisotropy parameter being $\xi_0=10$.}
\label{fig:thermalization_Mn0}
\end{figure}
\begin{figure}[!htpb]
\begin{center}
\includegraphics[width=0.5\linewidth]{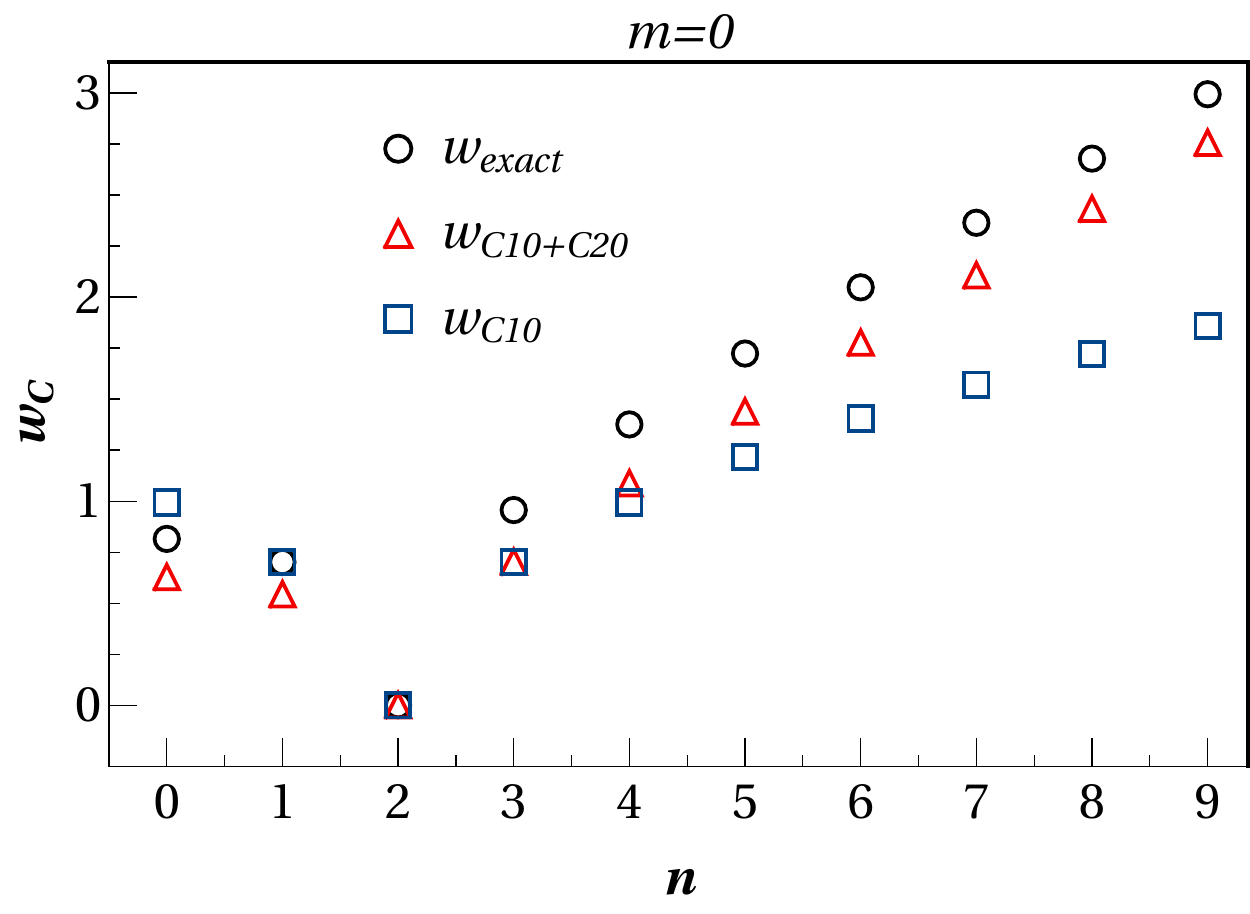}\\
\caption{Saturation time $w_c$ vs. $n$ for the moments $\bar M^{n0}$. The black circles are computed using the exact RTA-BE solutions; the blue squares and red triangles represent the asymptotic hard limits~\eqref{eq:mn0-cna} and~\eqref{eq:mn0-cnb}, respectively. The deviation between the exact and $c_{10}+c_{20}$ results is obviously due to truncation effects.}
\label{fig:thermtime_Mn0}
\end{center}
\end{figure}

We are now in position to analyze the normalized moments $\bar M^{nm}$ for $n>0,l=0$. In this case, only the hard sector of the distribution function contributes and the slowest modes playing a role in Eq.~\eqref{eq:mnm} are $c_{10}$ and $c_{20}$, both asymptotically decaying as $1/w^2$. As a double check, consider for instance the mode $\bar M^{10}=n/n_{eq.}$. For the Bjorken flow, the Chapman-Enskog expansion in the RTA approximation~\cite{Jaiswal:2015mxa} shows that the NS corrections to the particle density $n$ vanishes as chemical potential $\mu$ is turned off, and thus the leading order dissipative corrections are of the second order in the Knudsen number $\sim 1/w^2$. In general, it is known that second-order fluid dynamics theories, e.g. Israel-Stewart theory, do not reproduce properly the behavior of the heat flow and particle density as well as the dominant corrections arising from couplings between the particle density and the shear stress tensor~\cite{Jaiswal:2015mxa,Denicol:2012vq}.

Now, as done before, we study the impact of the non-hydrodynamic modes entering Eq.~\eqref{eq:mnm} by taking two truncation limits of the hard ($h$) regime
    \bs
    \label{eq:mn0}
    \beal
    \label{eq:mn0-cna}
    & \bar M^{n0}_{c_{10}+c_{20}~(h)} = 1 + (n^2-4)\frac{4 \theta_0^2}{45w^2}, \\
    \label{eq:mn0-cnb}
    & \bar M^{n0}_{c_{10}~(h)} = 1 - (2-n)\frac{4\theta_0^2}{9 w^2}.
    \end{align}
    \es

In Fig.~\ref{fig:thermalization_Mn0} the evolution of several normalized moments $\bar M^{n0}$ are shown. In the IR $\bar M^{n0}_{c_{10}+c_{20}~(h)}$ and $\bar M^{n0}_{exact}$ reach the asymptotic hydrodynamic fixed point faster than the moments $\bar M^{n0}_{c_{10}~(h)}$, supporting the compatibility  of the exact normalized moments with the former. Note that the asymptotic behavior of the normalized moments $M^{n0}$ is entirely determined by the linear combination of the non-hydrodynamic modes $c_{10}$ and $c_{20}$. This conclusion does not hold when $n\geq3$, where only the mode $c_{10}$ contributes. 

In Fig.~\ref{fig:thermtime_Mn0} the values of $w_c$ vs. $n$ for the moments $M^{n0}$ are plotted. It is seen that the saturation bound is approached as $n$ increases if $n\geq 3$, as opposed to when $n=0,1$ for which $w_c$ decreases. Now, when comparing with the results plotted in 
Fig.~\ref{fig:thermtime}, we observe that $\bar M^{nm}$ ($n>0,m\geq 1$) asymptote to their values at the equilibrium state later than the moments $\bar {M}^{n0}$ do. In terms of our orthogonal polynomial basis expansion, it is easy to understand this as the slowest non-hydrodynamic modes in the soft and semi-hard sectors decay like $\sim 1/w$, whereas in the hard sector, the slowest modes die away like $1/w^2$. On the other hand, for the given bound $\delta =0.05$, there is a disagreement between all the hard asymptotic limits~\eqref{eq:mn0} and the exact results due to the truncation limitations.
As we emphasized previously, close to the IR fixed point, the inclusion of both moodes $c_{10}$ and $c_{20}$ (Eq.~\eqref{eq:mn0-cnb}) is a must, which is also seen in Fig.~\ref{fig:thermtime_Mn0}. 

\begin{figure}[!htpb]
\begin{center}
\includegraphics[width=.8\linewidth]{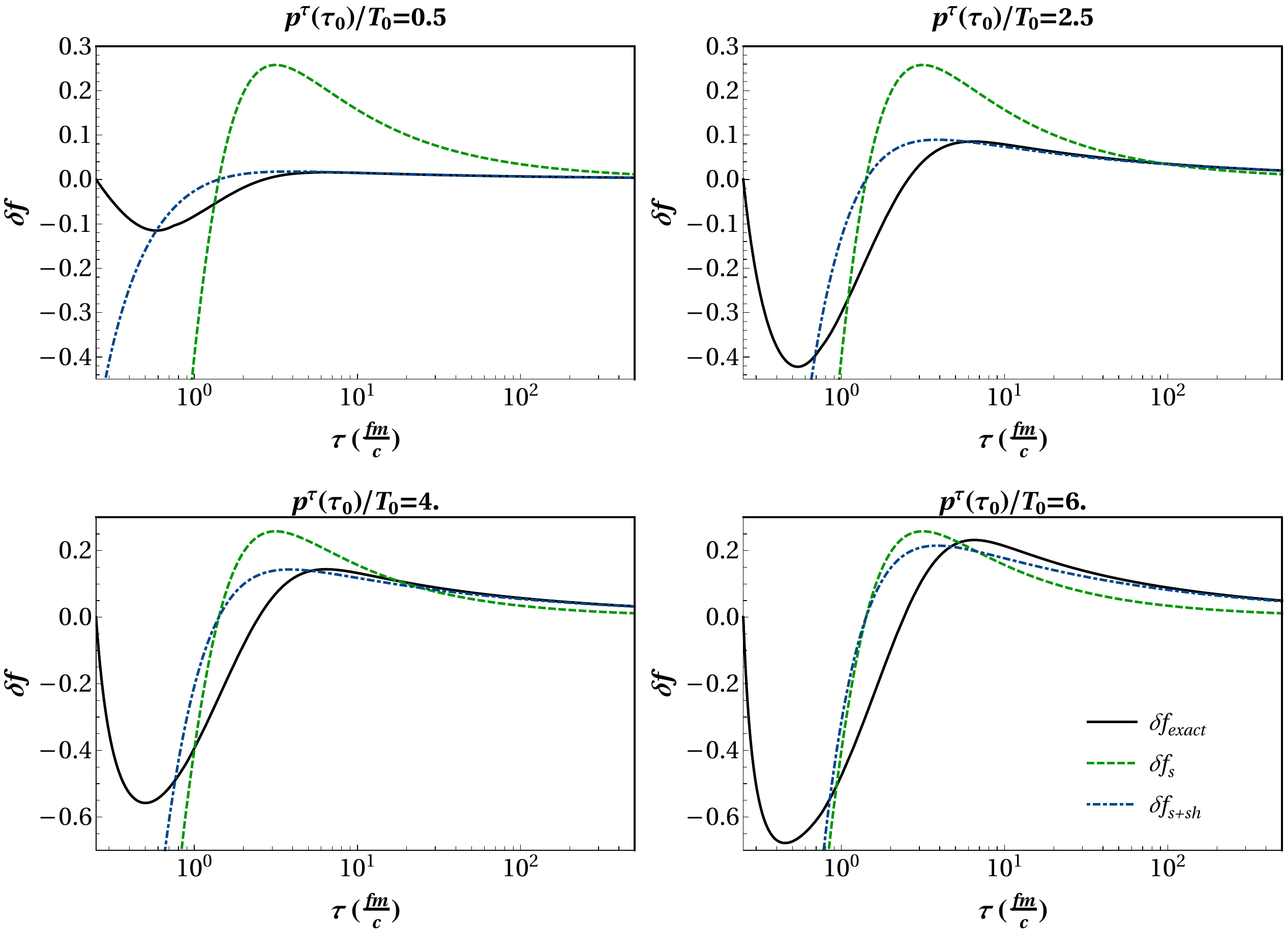}\\
\caption{Time evolution of the deviation from the equilibrium distribution function~\eqref{eq:deltafobs} for the exact RTA (black lines), the $s$ (green dot-dashed lines) and the $s+sh$ (blue dashed lines) regimes. For $\delta f_{exact}$, the initial ratio of $p^\tau_0/T_0$ are set to be $\{0.5,2.5,4,6\}$, and the initial temperature for the exact RTA-BE solver is set at $T_0=0.6 GeV$, $\tau_0=0.25$ fm/c. Finally, we initialize the solver with $\xi_0=0$.}
\label{fig:distributionf}
\end{center}
\end{figure}
The rest of this section is dedicated to showing how the presence of both $c_{01}$ and $c_{11}$ is a crucial factor in unveiling the asymptotic behavior of the exact distribution function in the IR. We calculate the deviations from the thermal equilibrium distribution function defined as
\be
\label{eq:deltafobs}
\delta f(\tau, p_T,p_\varsigma)=\frac{f(\tau, p_T,p_\varsigma)-f_{eq.}\left(p^\tau(\tau)/T(\tau)\right)}{f_{eq.}\left(p^\tau(\tau)/T(\tau)\right)}\,,
\ee
following the procedure outlined in Ref.~\cite{Heinz:2015cda}. In the $s$ and $s+sh$ regimes, $\delta f$ reads
\bs
\beal 
\label{eq:deltaf-cns}
\delta f_{s}&=\,c_{01}\,\mathcal{P}_{2}\left(\frac{p_\varsigma}{\tau p^\tau}\right)\approx\,-\frac{8}{3}\frac{\theta_0}{\tau T_{NS}}\,\mathcal{P}_{2}\left(\frac{p_\varsigma}{\tau p^\tau}\right)\,,\\
\label{eq:deltaf-ens}
\delta f_{s+sh}&=\,c_{01}\,\mathcal{P}_{2}\left(\frac{p_\varsigma}{\tau p^\tau}\right)\,+\,c_{11}\,
\mathcal{P}_{2}\left(\frac{p_\varsigma}{\tau p^\tau}\right)\,\mathcal{L}_{1}^{(3)}\left(\frac{p^\tau}{T_{NS}}\right)\,\nonumber\\
&\approx-\frac{8}{3}\frac{\theta_0}{\tau T_{NS}}\,\mathcal{P}_{2}\left(\frac{p_\varsigma}{\tau p^\tau}\right)\,+\,\frac{2}{3}\frac{\theta_0}{\tau T_{NS}}\,\mathcal{P}_{2}\left(\frac{p_\varsigma}{\tau p^\tau}\right)\,\mathcal{L}_{1}^{(3)}\left(\frac{p^\tau}{T_{NS}}\right)\,,
\end{align} 
\es
where $T_{NS}$ denotes the NS solution of the conservation law~\eqref{eq:Tevol}. The Fig.~\ref{fig:distributionf} outlines the time evolution of $\delta f$ for the exact and asymptotic limits. As a double check, other valid initial conditions are considered by varying the anisotropy parameter $\xi_0$. Two remarks are due here. First, no momentum and energy sector of the distribution function does in fact thermalize homogeneously, and second, the low energy particles ($p^\tau/T<1$) thermalize faster than the highly energetic ($p^\tau/T>1$) particles. This is not surprising since it is known that for weakly coupled systems, and in the absence of external fields, soft particles equilibrate faster~\cite{Bazow:2015dha,Bazow:2015cha,ERNST19811,PhysRevLett.36.1107}. On the one hand, the IR limit of $s$ truncation~\eqref{eq:deltaf-cns} does not capture the relaxation of the energy and momentum tails when compared against the exact RTA-BE solution. On the other hand, the exact RTA-BE result is perfectly compatible with the $s+sh$ limit~\eqref{eq:deltaf-ens}, which in turn confirms the {\it necessity} of including the new mode $c_{11}$. Therefore, the proper description of the IR in the Bjorken flow has to incorporate the dynamics of this non-hydrodynamic mode as well.

In sum, the numerical results presented in this section together with the previous analysis of the multi-parameter transseries solutions for the moments $\cnl$ lead to the true mechanism behind the nonlinear relaxation processes of different momentum sectors of the distribution function, which is completely governed by various mode-to-mode couplings among the moments in the framework of dynamical systems. Furthermore, the IR regime of the distribution function is determined uniquely by two non-hydrodynamic modes $c_{01}$ and $c_{11}$.  The conclusions derived in this section would still hold if one used other relaxation time models (cf. App.~\ref{app:thermalizationDelta1}).   

\section{Conclusions}
\label{sec:concl}
In this work, we have proposed a new dynamical renormalization scheme which allows us to study the nonlinear transient relaxation processes of a weakly-coupled plasma undergoing Bjorken expansion. The distribution function was expanded in terms of orthogonal polynomials. The nonlinear dynamics of the RTA-BE is then investigated through the kinetic equations of the coefficients entering this expansion, i.e. the average momentum moments $c_{nl}$ of the distribution function. We not only developed further our earlier findings~\cite{Behtash:2018moe} to include higher modes, but we also found new interesting results summarized in the following

\begin{itemize}
    \item Based on the seminal works of Costin~\cite{costin1998,Costin2001} we show that the coupled system of nonlinear ODEs for the moments $c_{nl}$ admit analytic multi-parameter transseries solutions. At every given order in the time-dependent perturbative asymptotic expansion of each mode, the summation over all the transient non-perturbative sectors appearing in the trans-series leads to renormalized transport coefficients. This presents a new description of the transport coefficients in the regimes far from equilibrium with an associated renormalization group equation, going beyond the usual linear response theory.
 \item As long as $0\le\Delta<1$, the $T=0$ limit of the Boltzmann equation is explicitly time-independent, meaning that the distribution function $f$ does not evolve with time. In the language of our dynamical system, this limit is equivalent to saying that the system is autonomous. The stability of the maximally oblate point shows in this case that it is a sink and consequently, there is no flow line that could connect it to the asymptotic hydrodynamic fixed point at $\tau\rightarrow\infty$. As soon as $T>0$, we have that the dynamical system explicitly depends on $\tau>0$ (i.e., it is non-autonomous) and continuously connected to the IR theory (hydrodynamics) at $\tau\rightarrow\infty$. But now the maximally oblate point is out of reach, hence a flow line (process) cannot be found with $\tau_0=0,T_0=0$ which leads to hydrodynamical behavior at late times. In other words, the actual phase space $\mathcal M$ of the Bjorken flow is a {\bf disjoint union} of hypersurfaces $T=0$ and $T\ne0$.
 \item The above conclusion does not hold true in the $w=\tau T$ parametrization. The phase space $\mathcal W$ is different in the UV in the sense that the system stays always non-autonomous for all $w\ge 0$. Also, notice that the temperature has been washed away from $\cal W$ at the cost of introducing a new singularity hypersurface $c_{01}=20$. However, the maximally prolate point is the only fixed point in the UV that {\it can} connect to the IR as shown on the l.h.s. of Fig.~\ref{fig:flows_N=0}. The maximally oblate point that in principle should correspond to the free-streaming limit, is a saddle point of index $N+1$. On the one hand, if $N=0$, it is basically a sink from the perspective of an observer sitting in the subspace $W_0\subset \mathcal{W}$ parametrized by $c_{0l}$. 
Therefore, if the initial conditions are set to be the coordinates of this fixed point, the observer only sees the maximally oblate fixed point moving to the IR as time passes. On the other hand, there is no bifurcation or change of stability along the way. Rather, the same fixed point at $w=0$ is just dislocated to become the  fixed point at $w\rightarrow\infty$.
As a result, there is {\it no} complete flow line connecting this fixed point to the hydrodynamic fixed point.

Also, if $N>0$, there appear to be repelling directions in the subspace $\bigcup_{n=1}^NW_n$. However, the invariant manifold is in $W_0$, and hence the search for a solution hits a snag again. If there was such a line, then we would have an example of a critical line. This critical flow line is often mistaken in the literature as a  {\it global attractor} as if the system was autonomous. In non-autonomous systems such as the Bjorken flow, the attractor is a concept encoding information of the past or future of the system of equations governing its dynamics independently of each other, as shown in  Fig.~\ref{fig:flows_N=0}. The relevant attractor is then either a forward or pullback attractor. In sum, given our study of the dynamical system for the Bjorken flow, we conclude that there is no critical line connecting the maximally oblate point to the hydrodynamic fixed point, which is otherwise known as ``attractor solution'' in the hydrodynamics literature.
\item For the Bjorken flow the asymptotic behavior of the kinetic equations unveils the existence of a new non-hydrodynamic mode whose decay rate is exactly the same as the usual NS shear viscous component, i.e., $\sim (\tau T)^{-1}$. The origin of the new mode's decay is due to the nonlinear mode-to-mode coupling $~c_{11}c_{01}$ which dominates close to the asymptotic hydrodynamic fixed point. This non-hydrodynamic mode is the slowest high energy mode and it determines quantitatively the late-time behavior of the transient high energy tails of the distribution function. 
\end{itemize}

In the present paper, we left out many interesting questions which certainly require appropriate answers. For instance, within our approach it is important to understand the analytic behavior of the retarded correlators together with the subtle interplay between branch cuts and poles as a function of the coupling~\cite{Romatschke:2015gic,Kurkela:2017xis}. It will be also important to investigate the possible phenomenological consequences of the new non-hydrodynamic mode $c_{11}$. For instance, one might wonder about the impact of this high energy mode in the modeling of a jet crossing an expanding QGP (quark-gluon plasma) using kinetic theory. Another possibility is to understand if the origin of azimuthal anisotropies at large momenta is connected to $c_{11}$. It has been argued by different authors that the azimuthal anisotropies at intermediate $p_T$ are related to non-hydrodynamic transport ~\cite{Romatschke:2018wgi,Borghini:2010hy,Kurkela:2018qeb}. On a more theoretical ground, it would be very interesting to come up with extending our analysis to the challenging case of nonlinear PDEs such as Israel-Stewart hydrodynamic equations for either $2+1$ or $3+1$ dimensions. Finally, our work can be generalized to studying nonlinear aspects of time-dependent systems of ODEs whose description is cast in the form of the properties of a dynamical system. We have left these exciting topics for future research projects.

\section*{Acknowledgements} 
We thank O. Costin for his patience, useful discussions, explanations and clarifications of some of the technical and conceptual matters in his groundbreaking work~\cite{costin1998}. We also thank G. Dunne, M. Spalinski, M. Heller, J.~Casalderrey-Solana, T. Schäfer and M. Ünsal for useful discussions on the subject of resurgence and non-equilibrium physics. We are grateful to T. Schäfer for bringing to our attention the possible relation between our calculations and non-Newtonian fluids. H. S. thanks C.~N.~Cruz-Camacho for his help and assistance with the numerical solver of the RTA Boltzmann equation. M. M. thanks L. Yan for clarifying the details of~\cite{Teaney:2013gca} and M. Strickland for his insistence on trying to understand the asymptotic behavior of the normalized moments $M^{n0}$. A. B., M. M., and S. K. are all supported in part by the US Department of Energy Grant No. DE-FG02-03ER41260. M. M. is partially supported by the BEST (Beam Energy Scan Theory) DOE Topical Collaboration.  A. B. was partially supported by the DOE grant DE-SC0013036 and the National Science Foundation under Grant No. NSF PHY-1125915. 

\appendix
\section{Asymptotic Chapman-Enskog expansion of the distribution function}
\label{app:asymp}

In this section, we briefly describe the asymptotic Chapman-Enskog expansion of the distribution function for the RTA Boltzmann equation. The Chapman-Enskog expansion has been calculated up to second order by different authors~\cite{Teaney:2013gca,Chattopadhyay:2014lya,Bhalerao:2013pza,Jaiswal:2013vta,Jaiswal:2013npa,Romatschke:2011qp,Yan:2012jb,Denicol:2012es}. We follow closely the derivation of Teaney and Yan~\cite{Teaney:2013gca} so any interested reader can take a look into their work for further details.

It is convenient to rewrite the general RTA-BE in the following suitable form~\cite{Teaney:2013gca}
\be
\label{eq:RTABoltzmann-2}
p^\mu\partial_\mu f_{\bm{p}}=-\frac{T^2}{C_{\bm{p}}}\left(f-f_{eq.}\right)\,,
\ee
where $C_{\bm{p}}=-T^2\tau_r/(u\cdot p)$ with $\tau_r=\theta_0/T^{1-\Delta}$. Next we expand the distribution function around $f_{eq.}$
\be
\label{eq:CEans}
f_{\bm{p}}=f_{eq.}\,+\,\alpha\,\delta f_{1}\,+\,\alpha^2\,\delta f_{2}\,+\,\ldots\,.
\ee
where $\alpha$ is an arbitrary parameter which keeps track of the order of the gradients.  One obtains the following set of coupled equations for $\delta f_1$ and $\delta f_2$ at first and second order in $\alpha$
\bs
\label{eq:deltaf}
\beal
\label{eq:f1}
\mathcal{O}(\alpha):&\hspace{0.5cm}\delta f_1=-\frac{C_{\bm{p}}}{T^2}\,p^\mu\partial_\mu f_{eq.}\,,
\\
\label{eq:f2}
\mathcal{O}(\alpha^2):&\hspace{0.5cm}\delta f_2=-\frac{C_{\bm{p}}}{T^2}\,p^\mu\partial_\mu\delta f_1\,,
\end{align}
\es
by substituting~\eqref{eq:CEans} in the RTA-BE~\eqref{eq:RTABoltzmann} and using the fact that the collisional kernel is  $\mc O(\alpha)$. In order to write the general form of $\delta f_1$ and $\delta f_2$, it is necessary to classify all the possible irreducible tensors invariant under rotations~\cite{Teaney:2013gca,Denicol:2011fa,DeGroot:1980dk}. For $d=4$ one gets after some algebra~\cite{Teaney:2013gca,Chattopadhyay:2014lya,Bhalerao:2013pza,Jaiswal:2013vta,Jaiswal:2013npa,Romatschke:2011qp}~\footnote{Here, we have used the Cauchy-Stokes decomposition of the fluid velocity 
\begin{equation*} 
\partial_\mu u_\nu=-u_\mu\left(u\cdot\partial u_\nu\right)+\Delta_{\mu\nu}(\partial_\lambda u^\lambda)/3+\frac{\sigma_{\mu\nu}}{2}+\omega_{\mu\nu}\,,
\end{equation*}
together with the (conformal) conservation laws to first order in gradients
\bs
\begin{align*}
D\log T&=-\frac{\partial_\mu u^\mu}{3}\,,\\
D u_\nu&=-\Delta_{\mu\nu}\partial^\nu\log T\,.
\end{align*}
\es
The use of conservation laws at first order in the gradients ensures that the value of $\delta f$ does not change the equilibrium energy density~\cite{York:2008rr}.} 
\bs
\label{eq:deltaf-2}
\beal
\label{eq:f1-2}
\delta f_1&=-
\frac{\sigma_{\mu\nu}p^\mu p^\nu}{T^3}\,\chi_{0\bm{p}}\left(\frac{u\cdot p}{T}\right)\,,\\
\label{eq:f2-2}
\delta f_2&=\chi_{1\bm{p}}\,\frac{p^\mu p^\nu p^\lambda p^\beta}{T^6}\,\sigma_{\langle\mu\nu}\sigma_{\lambda\beta\rangle}\,
+\,\chi_{2\bm{p}}\,\frac{p^\mu p^\nu p^\lambda}{T^5}\left[\nabla_{\langle\mu}\sigma_{\nu\lambda\rangle}-3\sigma_{\langle\mu\nu}\nabla_{\lambda\rangle}\log T\right]\nonumber\\
&+\left(\frac{4}{7}\frac{p^2}{T}\chi_{1\bm{p}}+(u\cdot p)\chi_{2\bm{p}}\right)\frac{p^\mu p^\nu}{T^5}\sigma^{\lambda}_{\hspace{.1cm}\langle\mu} \sigma_{\nu\rangle\lambda}\\
&\,-\,(u\cdot p)\frac{p^\mu p^\nu}{T^5}\chi_{2\bm{p}}\left[ D\sigma_{\langle\mu\nu\rangle}\,+\,\frac{\sigma_{\mu\nu}}{3}\theta-2\sigma^\lambda_{\langle\nu}\Omega_{\mu\rangle\lambda}\right]\nonumber\\
&+\chi_{3\bm{p}}\frac{p^\mu}{T^3}\left[-\nabla_{\langle\lambda}\sigma^\lambda_{\hspace{.1cm}\mu\rangle}-2\sigma_{\langle\mu\lambda}\nabla^{\lambda\rangle}\log T\right]+\frac{\chi_{4\bm{p}}}{T^2}\sigma^2\,,\nonumber
\end{align}
\es
where $A_{\langle\mu_1}\cdots B_{\mu_l\rangle}$ is a projector being traceless, symmetric and orthogonal to the fluid velocity $u^\mu$~\cite{DeGroot:1980dk}. In these expressions we have defined 
\bs
\label{eq:const}
\beal
\chi_{0\bm{p}}\left(\frac{u\cdot p}{T}\right)&=-\frac{1}{2}C_{\bm{p}}f'_{eq.}\,,\\
\chi_{1,\bm{p}}&=-C_{\bm{p}}\frac{\chi'_{0,\bm{p}}}{2}\,,\\
\chi_{2,\bm{p}}&= C_{\bm{p}}\,\chi_{0,\bm{p}}\,,\\
\chi_{4,\bm{p}}&= \chi_{1,\bm p}\frac{2 (p/T)^4}{(d-1)(d+1)} - \chi_{2,\bm p}\frac{(p \ T)^3}{d-1} - \chi_{0,\bm p} \frac{\eta}{s} \frac{( p/ T)^2}{d-1} + a_{E^\ast} n_p^\prime (p/T)\nonumber\\
 &=-\left( \frac{\tilde p^2}{6 T^2} + \frac{5 \tilde p}{6} + \frac{\tilde p^2}{6} + \frac{\tilde p^3}{6} + \frac{8 \tilde p^5}{105} - \frac{2 \tilde p^6}{105} \right) \chi_{\bm p} C_{\bm p}\,,\\
 \label{eq:ae}
a_{E^\ast} &= \frac{T\eta  \tau_\pi}{4 s} - \frac{d+3}{d-1} \frac{T \lambda_1}{4 s} + \frac{d+1}{2(d-1)} \left(\frac{\eta}{s}\right)^2\,,
\end{align}
\es
with $f_{eq.}'=d f_{eq.}(x)/dx$, $\theta=\nabla\cdot u$, $\tilde p:=p/T$ and tilde indicates that the quantity is dimensionless. The transport coefficients in $a_{E^\ast}$~\eqref{eq:ae} will be given below. 

For the Bjorken flow the gradient expansion of the distribution function reduces to the expression~\cite{Blaizot:2017lht,Blaizot:2017ucy}
\be
\label{eq:distf}
\begin{split}
\delta f &= \chi_p C_p \left[ - \tilde p^2 \frac{4}{9 T^4 \tau^2} - \tilde p \frac{20}{9 T^2\tau^2} - \tilde p^2 \frac{4}{9 T^2\tau^2} -  \tilde p^3 \frac{4}{9 T^2\tau^2} - \tilde p^5 \frac{64}{315 T^2\tau^2} + \tilde p^6 \frac{16}{315 T^2\tau^2}  + \cdots \right] P_0 (\cos\theta)\\
 & +\left[-\,\tilde\chi_p\tilde p^2 \left(\frac{2}{3\tau T}\right) + \tilde\chi_p^\prime\tilde C_p \tilde p^4\left(\frac{8}{63\tau^2 T^2}\right) -\tilde\chi_p\tilde C_p \tilde p^3\left(\frac{8}{9    \tau^2 T^2}\right) + \cdots \right] P_2(\cos\theta)\\
 & + \left[\tilde\chi_p^\prime \tilde C_p \tilde p^4\left( \frac{8}{35    \tau^2 T^2}\right) + \cdots \right]P_4(\cos\theta)
\end{split}
\ee
where $\tilde\chi_p=-C_pf^\prime_{eq}$. One immediately recognizes that the Chapman-Enskog expansion of the distribution function
 is an asymptotic series in terms of $1/(\tau T)$. Close to the thermal equilibrium $1/(\tau T)\approx\tau^{-2/3}$, which in the case of Bjorken flow, it 
 becomes proportional to the Knudsen and inverse Reynolds numbers. 

For the Bjorken flow using the second order Chapman-Enskog expansion for $\delta f$~\eqref{eq:distf} in Eq.~\eqref{eq:momnl} results in the asymptotic series expansion of the moments $c_{01}$ and $c_{02}$ as follows
\bs
\label{eq:asympc1c2}
\beal
c_{01} &= -\frac{8\tau_r}{5\pi^2\tau} - \frac{64\tau_r^2}{105\pi^2\tau^2} + \mathcal{O}(1/\tau^3) = -\frac{2\eta}{\tau T^4}+\frac{4}{3\tau^2 T^4}(\lambda_1-\eta\tau_\pi)+ \mathcal{O}(1/\tau^3)\,,\\
c_{02} &= \frac{32 \tau_r^2}{21\pi^2\tau^2} + \mathcal{O}(1/\tau^3) = \frac{4}{3\tau^2 T^4}(\lambda_1+\eta\tau_\pi) + \mathcal{O}(1/\tau^3)\,.
\end{align}
\es
The transport coefficients appearing in Eq.~\eqref{eq:asympc1c2} are given by \cite{Teaney:2013gca}
\bs
\label{eq:gexpcoeff}
\beal
\eta &= \frac{2}{15 T^3} \int_p p^4 \chi_{0,p} = \frac{4\tau_r T^4}{5 \pi ^2}, \\
\eta\tau_\pi &= \frac{2}{15 T^5} \int_p p^5 \chi_{2,p} = \frac{4 \tau_r ^2 T^4}{5 \pi ^2},\\
\lambda_1 &= \frac{8}{105 T^6} \int_p p^6 \chi_{1,p} - \eta\tau_\pi= \frac{12\tau_r ^2 T^4}{35 \pi ^2}.
\end{align}
\es

\section{A self-contained dictionary for the dynamical systems}
\label{app:dynsys}

In this section, we list a dictionary of words from the terminology of the mathematical field of dynamical systems used in Sect.~\ref{sec:global}. This is a self-contained and compact dictionary that 
could help the reader with some terminology of time-dependent dynamical systems.
\begin{itemize}
\item[1-] {\it Dynamical system}. A set of differential equations in terms of a state vector whose components are real numbers determined by a set of points in some suitable state space. Any small variation in the state of the system leads in turn to a change in the numbers. In short, the evolution of a dynamical system is based on a fixed deterministic rule that describes what future states follow from the current state. When explicitly time-dependent, however, equations of the system also evolve in form, and thus the rule changes as time passes. 
\item[2-] {\it Non-autonomous dynamical system}. A differential equation of the form
\be
\frac{d}{d\tau}x(\tau)=g(x(\tau),\tau) \quad ({\rm or} \,\, \frac{d}{dw}x(w)=h(x(w),w)) \label{eq:funct_g}
\ee
is explicitly time-dependent through the function $g$. This defines a non-autonomous dynamical system if the state function $x(\tau)$ is dynamically evolving with a flow time $\tau$. 
$T(\tau)$, and equivalent of $g$ in \eqref{eq:funct_g} is the r.h.s. of \eqref{eq:ODEsBjor}.
\item[3-] {\it (Asymptotic) fixed point}. The time-dependent solutions $x(\tau)$ of $g(x(\tau),\tau)=0$ are the {\bf moving} fixed points of the system \eqref{eq:funct_g}. For $\tau\rightarrow\infty$ these define a point(s) $x_*$ at which the state vector is in a steady equilibrium state for all times $t>0$ if $\tau = \infty+t$, referred to as fixed points of the non-autonomous system in \eqref{eq:funct_g}. Note that here the time $\tau$ takes values over $\mathbb{t}=\mathbb{R}^+$ \footnote{In math literature, a non-autonomous system is known to not admit any fixed point
by promoting it to an autonomous system of one dimension higher and observing that the r.h.s. of this
secondary system
\be
dx/d\tau =g(x),\quad d\tau/d\tau = 1,
\ee 
never vanishes. Nonetheless, our definition of a fixed point for a non-autonomous system is not for all times as in the autonomous case,
rather over just enough period of time beyond $\tau=\infty$ in order to distinguish it as an equilibrium state.
Studying topology of the flow lines around such fixed points is thus more complicated and requires
a lot of additional mathematical input \cite{Behtash:Topology}.}. If we reparametrize flow time to $\rho$ such that $\mathbb{t}=\mathbb{R}$, the fixed points can now be equally defined at $\rho\rightarrow -\infty$. Hence $\tau(\rho)$ turns into
a new independent variable in the dynamical system such that
\be
dx/d\rho = g (x,\tau),\quad d\tau/d\rho = h(\tau), \label{eq:auto_sys}
\ee
is autonomous. The fixed points are then defined as usual by finding those $(x_*,\tau_*)$ for which $g(x_*,\tau_*)=0=h(\tau_*)$. 
\item[4-] {\it Flow line (process)}. A flow line is a map defined by ${ \phi}^{\tau,\tau_0}({ x}_0):X\times\mathbb{t}\times \mathbb{t}\rightarrow X$ such that ${ \phi}^{\tau,\tau_0}({ x}_0)={ x}(\tau)$ solves \eqref{eq:funct_g} with some initial values $({ x}_0\equiv { x}(\tau_0),\tau_0)$. 
\item[5-] {\it Phase portrait}. It is a geometric representation of the flow lines for a dynamical system in its phase space given a set of initial conditions represented by $(x_i,\tau_i)$ with $i\in I$ where $I$ is simply an index set. 
\item[6-] {\it Invariant manifold}. As defined above, it is a subset of the phase portrait which contains
all the flow lines initiated inside it (or at its boundary) which keep being restricted in there at all times.
\item[7-] {\it Complete flow line}. Choosing a different parametrization of flow time $\tau$, say $w$,
may allow us to connect two fixed points at $(x_0,0)$ and $(x_\infty,\infty)$. Note again that the time manifold $\mathbb{w}$ now allows 0 to be a valid input. This flow line if existed, would be called complete. The set of fixed points of the dynamical system all belong to the boundary of the invariant manifold iff there is one complete flow line between every two fixed points.
\item[8-] {\it Stability of a fixed point}. Linearizing the system in \eqref{eq:auto_sys} about a fixed point $(x_*,\infty)$
leads to the understanding of how flow lines behave around that fixed point. Eigenvalues of the linearization matrix $J$ with components 
\be
J_{ij}:=\frac{\partial {G_i} }{\partial X_j},\quad { G} = (g(x,\tau),h(\tau)),\label{eq:lin_mat}
\ee
evaluated at $(x_*,\infty)$, determine the stability of $(x_*,\infty)$. Let us denote the eigenvalues of \eqref{eq:lin_mat}
by ${\lambda}_\infty$. Throughout this section we only consider {\it hyperbolic} fixed points for which $\Re(\lambda_\infty)\ne 0$. 
\item[9-] {\it The Hartman-Grobman  (HG) theorem}.  It states that near a hyperbolic fixed point $(x_*,\infty)$, the nonlinear system \eqref{eq:funct_g} has the same qualitative structure as the linear system 
\be
d x/d\tau = J(x_*,\infty)\,x. \label{eq:lin_sys}
\ee
Therefore, HG implies that two hyperbolic systems are locally topologically flow equivalent iff
their unstable manifolds have equal dimensions.
\item[10-] {\it Index of $(x_*,\infty)$}. The number of components of ${\lambda}_\infty$ whose real parts are positive is called
the (Morse) index of that fixed point, which is technically the dimension of the {\it unstable manifold} around every hyperbolic fixed point. So any nonzero index means that fixed point is basically unstable. This instability is severe for a larger index. For example, in a $d$ dimensional system $(x_1,\dots,x_{d-1},\tau)$, an index $n$ fixed point is
an unstable {\it saddle} whose stable manifold is $d-n$-dimensional. The instability of an index $n$ fixed point known as a {\it source} is far more pronounced than a saddle. Finally, a stable fixed point (so-called {\it sink}) has always a vanishing index \footnote{The imaginary part of ${\lambda}_\infty$ if present, means that the flow line is going to spiral its way in or out; thus the name
stable or unstable {\it spiral}, respectively. }.
\item {\it Invariant manifold}. Suppose $\phi$ is a flow on the phase space $(X,\mathbb{t})$ equipped with a
spatial metric $\rm dist$. A family ${\cal A}:=\bigcup_{\tau\in \mathbb{t}} A_\tau$ of nonempty subsets of $X$ is invariant
with respect to $\phi$ if for all $\tau \ge \tau_0$
\be
\phi^{\tau, \tau_0}(A_{\tau_0}) = A_\tau.
\ee
${\cal A}$ is called the invariant manifold. 

\item[12(a)-] {\it Forward attraction}. Let $\phi$ be a flow (process). A nonempty,
compact and invariant manifold ${\cal A}:=\bigcup_{\tau} A_\tau$ for compact $I$ is said to forward attract if 
\be
\lim_{\tau\rightarrow\infty}{\rm\,dist}\left[\phi(\tau, \tau_0, x_0),A_\tau\right]= 0, \label{eq:for_att}
\ee
for all $x_0 \in X$ and $\tau_0 \in \mathbb{t}$. The distance function between two sets $A,B$ is defined as usual on the metric space $X$ as ${\rm dist}(A,B) = \inf _{x\in A,y\in B} {\rm dist}(x,y)$.
\item[12(b)-] {\it Pullback attraction}. A nonempty,
compact and invariant non-autonomous set ${\cal A}:=\bigcup_{\tau} A_\tau$  is said to pullback attract if 
\be
\lim_{\tau_0\rightarrow-\infty}{\rm\,dist}\left[\phi(\tau, \tau_0, x_0),A_\tau\right]= 0, \label{eq:pull_att}
\ee
for all $x_0 \in X$ and $\tau \in \mathbb{t}$.
\item[13-] {\it Basin of attraction}. A  neighborhood of an attractor is called the basin of attraction $B$ if it consists of all points that evolve to the attractor  in the limit $\tau\rightarrow\infty$. In other words, $B$ is the set of all points $b\in B$ in the phase space provided that for any open neighborhood $N$ of the attractor, there is a positive constant $s$ such that $\phi^{\tau,\tau_0}(b) \in N$ for all real $\tau>s$.
\item[14-] {\it Lyapunov function}. Let us assume that the point $(x,\tau)$ at $\tau\rightarrow\infty$ is $(0,\infty)$ which defines an asymptotically stable fixed point
of a non-autonomous dynamical system. Then if there 
is a function $V(x,\tau)$ such that the Lyapunov stability conditions
\be
\label{eq:lyapunov}
V(x,\tau)>0,\,\,\, {\rm and}\,\,\, \frac{dV(x,\tau)}{d\tau}=\frac{dx(\tau)}{d\tau}\cdot \nabla V(x,\tau)<0,\quad  \forall x>0, \tau<\infty
\ee 
are satisfied, $V(x,\tau)$ is called a Lyapunov function.
\end{itemize}

It may be appropriate here to mention that in general, 11(a)-(b) are very different and independent concepts. Pullback attractor \eqref{eq:pull_att} contains information about the
past (early-time) of a non-autonomous dynamical system, whereas forward attractor \eqref{eq:for_att} makes use of information
about the future. It well may be the case that one exists while the other
one does not in a non-autonomous system. However, in the case of an autonomous system, they are equivalent and the resulting attraction is global with the set $\mathcal A$ referred to as {\it global attractor}. 
\section{Constructing the transseries for $T(\tau)$}
\label{app:T_trans}
\subsection{Algebraic properties of transseries}
Let us suppose the following transseries is given
\be
&& \Phi (z) =  \sum_{{\bf n} \ge 0;|{\bf n}| \ge 0}^{\infty}  \bm{\sigma}^{{\bf n}} \bm{\zeta}^{{\bf n}}(z) \sum_{k=0}^{\infty} \phi^{({\bf n})}_{k} z^{-k}\,, \\
&& \bm{\zeta}^{{\bf n}}(z) = e^{-({\bf n}  \cdot {\bf S}) z} z^{ {\bf n} \cdot \bm{\beta}} , \\
&& \bm{\sigma}^{\bf n} = \sigma^{n_1}_{1} \cdots \sigma^{n_L}_{L},
\ee
where $\phi^{({\bf n})}_{k}$ are complex-valued coefficients.
We define the basis of the transseries aka $E^{({\bf n})}_{k}(z)$ as
\be
&& \Phi(z) = \sum_{{\bf n} \ge 0;|{\bf n}| \ge 0}^{\infty}    \sum_{k=0}^{\infty} \Phi^{({\bf n})}_{k} (z) , \\
&& \Phi^{({\bf n})}_{k} (z) = \phi^{({\bf n})}_{k} E^{({\bf n})}_{k}(z), \\
&& E^{({\bf n})}_{k} (z) = \bm{\sigma}^{{\bf n}} \bm{\zeta}^{{\bf n}}(z) \, z^{-k}.
\ee
The transseries can be regarded as an element of a vector space defined as
\be
&& {\cal H} = \bigoplus_{{\bf n} \ge {\bf 0},k=0} {\cal H}^{({\bf n})}_{k}, \\
&& \Phi^{({\bf n})}_{k} (z) = \phi^{({\bf n})}_{k} E^{({\bf n})}_{k}(z) \quad \in {\cal H}^{({\bf n})}_{k}, \\
&& \Phi(z) = \sum_{{\bf n} \ge 0;|{\bf n}| \ge 0}^{\infty}  \sum_{k=0}^{\infty} \Phi^{({\bf n})}_{k} (z) \quad \in {\cal H}.
\ee
The operations on this vector space are briefly explained below. Note that the transseries is always closed under these operations.
\subsubsection*{Product}
The product operation $\times$ between two transseries is an additive operation on both transmonomial and asymptotic orders
given by
\be
\times &:& \,   {\cal H}^{({\bf n})}_{k} \, \times \, {\cal H}^{({\bf n}^\prime)}_{k^\prime}  \quad \rightarrow \quad {\cal H}^{({\bf n}+{\bf n}^\prime)}_{k+k^\prime}, \\
&&  ( \, \phi^{({\bf n})}_{k}  E^{({\bf n})}_{k} , \psi^{({\bf n}^{\prime})}_{k^\prime}   E^{({\bf n}^\prime)}_{k^\prime} \, )  \quad \mapsto \quad  \phi^{({\bf n})}_{k}   \psi^{({\bf n}^{\prime})}_{k^\prime} E^{({\bf n}+{\bf n}^\prime)}_{k+k^\prime}.
\ee
\subsubsection*{Exponential}
The exponential operation ${\rm Exp}$ is an extension of the product to the exponential terms defined by
\be
{\rm Exp}\, \Phi^{({\bf n})}_{k}  \, := \, \exp \Phi^{({\bf n})}_{k},
\ee
where $\Phi^{({\bf n})}_{k} \in {\cal H}^{({\bf n})}_{k}$.
Thus, the exponential is a linear mapping of the form
\be
\label{eq:Exp}
{\rm Exp} &:& \,   {\cal H}^{({\bf n})}_{k}  \quad  \rightarrow \quad \bigoplus_{l=0}^{\infty} {\cal H}^{(l{\bf n})}_{lk}, \\
&&  \phi^{({\bf n})}_{k}  E^{({\bf n})}_{k}   \quad \mapsto \quad  \sum_{l=0}^{\infty} \frac{ ( \phi^{({\bf n})}_{k})^l }{\Gamma(l+1)}  E^{(l{\bf n})}_{lk}. 
\ee
   
\subsubsection*{Integration}
We define an integration operation of the form
\be
\label{eq:Intop}
   {\rm Int}_{s} \, \Phi^{({\bf n})}_{k}  \, := \, \int \frac{d z}{z^s} \, \Phi^{({\bf n})}_{k}.
\ee
where $\Phi^{({\bf n})}_{k} \in {\cal H}^{({\bf n})}_{k}$ and $s \in {\mathbb N}_{0}$.
For realizing the closedness of transseries under the integration, we have to assume that
\be
\Phi^{({\bf 0})}_k = 0 \qquad \mbox{for any} \qquad  0 \le k \le 1-s.
\ee
Then after some algebra, Eq.~\eqref{eq:Intop} casts ${\rm Int}_s$ in the form of the following linear mapping 
\be
\label{eq:Int1}
   {\rm Int}_s &:& \,   {\cal H}^{({\bf n})}_{k} \quad \rightarrow \quad
   \begin{cases}
    \,  {\cal H}^{({\bf 0})}_{k+s-1} & \mbox{for }  {\bf n}={\bf 0}  \\
    \,  \bigoplus_{l=0}^{\infty}  {\cal H}^{({\bf n})}_{k+s+l}   & \mbox{otherwise}
     \end{cases}, \\
   &&  \,  \phi^{({\bf n})}_{k}  E^{({\bf n})}_{k}    \quad \mapsto \quad
   \begin{cases}
     -\frac{\phi^{({\bf 0})}_{k}}{k+s-1} E^{({\bf 0})}_{k+s-1} & \mbox{for }  {\bf n}={\bf 0}  \\
     - \sum_{l=0}^{\infty} \frac{ R_{l} (1+ \bm{\beta} \cdot {\bf n} -k-s) \phi^{({\bf n})}_{k}}{({\bf n} \cdot {\bf S})^{l+1}}   E^{({\bf n})}_{k+s+l} & \mbox{otherwise}\label{eq:Int2}
\end{cases},
\ee
where $R_l (a)$ is written in terms of the Pochhammer symbol $(a)_l$,
\be
&& R_l (a) = (-1)^l \, (1-a)_l, \\
&& (a)_l = \frac{\Gamma(a+l)}{\Gamma(a)}.
\ee
We recall that to get the expression in \eqref{eq:Int2}, use has been made of the asymptotic form of the upper incomplete Gamma function
for $z \rightarrow \infty$, namely
\be
\Gamma(a,z) = z^{a-1} e^{-z} \left( \sum_{k=0}^{n-1} \, R_k(a) z^{-k} + \mathcal{O}(z^{-n}) \right).
\ee

\subsection{Transseries solution for the temperature $T(\tau)$}
As we have seen in the previous section, $\hat{T}(\hat{\tau})$ can be expressed by the same transseries ansatz used for $c_l(\hat{\tau})$ due to the fact that $c_l(\hat{\tau}) = \mathcal{O}(\hat{\tau}^{-1})$. Using Exp and ${\rm Int}_s$ mappings defined in Eq.~\eqref{eq:Exp} and Eq.~\eqref{eq:Int1} respectively, we find that
\be
&& \hat{T} (\hat{\tau})  = {\rm Exp} \, C_T(\hat{\tau}), \label{eq:TT_exp} \\
&& C_T(\hat{\tau}) = -\frac{1}{20} \, {\rm Int}_1 \, c_1 (\hat{\tau}). \label{eq:CT_int} 
\ee
By assuming the ansatz for $c_{01}(\hat{\tau})$ as
\be
&& c_{01}(\hat{\tau}) = 
\sum_{|{\bf n}| \ge 0}^{\infty}  \bm{\sigma}^{{\bf n}} \bm{\zeta}^{{\bf n}}(\hat{\tau}) \sum_{k=0}^{\infty} u^{({\bf n})}_{01,k} \hat{\tau}^{-k}, \label{eq:ansatz_c1_tau} \\
&& C_{T}(\hat{\tau}) = 
 \sum_{|{\bf n}| \ge 0}^{\infty}  \bm{\sigma}^{{\bf n}} \bm{\zeta}^{{\bf n}}(\hat{\tau}) \sum_{k=0}^{\infty} u^{({\bf n})}_{T,k} \hat{\tau}^{-k}, \label{eq:ansatz_CT_tau} \\
&& \hat{T}(\hat{\tau}) = 
 \sum_{|{\bf n}| \ge 0}^{\infty}  \bm{\sigma}^{{\bf n}} \bm{\zeta}^{{\bf n}}(\hat{\tau}) \sum_{k=0}^{\infty} \hat{u}^{({\bf n})}_{T,k} \hat{\tau}^{-k}, \label{eq:ansatz_TT_tau}
\ee
we can get from Eq.~\eqref{eq:CT_int}
\be
 C_T(\hat{\tau})
&=& \frac{1}{20}  \sum_{k=1}^{\infty} \frac{u^{({\bf 0})}_{01,k}}{k} \hat{\tau}^{-k}   + \frac{1}{20}  \sum_{|{\bf n}| > 0}^{\infty}  \bm{\sigma}^{{\bf n}} \bm{\zeta}^{{\bf n}}(\hat{\tau})\sum_{k=0}^{\infty}  \sum_{p=0}^{k} 
 \frac{ R_{k-p} ( {\bf n} \cdot \bm{\beta} -p) u^{({\bf n})}_{01,p}}{({\bf n} \cdot {\bf S})^{k-p+1}}   \hat{\tau}^{-k-1}. \label{eq:ansatz_C_tau}
\ee
Hence, it is easy to read off the coefficients
\be
u^{({\bf n})}_{T,k} &=&
\begin{cases}
  0 &  \mbox{for any ${\bf n}$ and $k=0$} \\
  \frac{u_{1,k}^{({\bf 0})}}{20 k} &  \mbox{for ${\bf n}={\bf 0}$ and $k \ge 1$} \\
  \sum_{p=0}^{k-1}  \frac{ R_{k-p-1} ( {\bf n} \cdot \bm{\beta} -p) u^{({\bf n})}_{1,p}}{20({\bf n} \cdot {\bf S})^{k-p}}  &  \mbox{otherwise}
\end{cases}.
\ee
Therefore, inserting  Eq.~\eqref{eq:ansatz_C_tau} in \eqref{eq:TT_exp} results in
\be
&& \hat{T}(\hat{\tau}) =
\exp \left(  \sum_{|{\bf n}| \ge 0}^{\infty}  \bm{\sigma}^{{\bf n}} \bm{\zeta}^{{\bf n}}(\hat{\tau}) \sum_{k=1}^{\infty} u^{({\bf n})}_{T,k} \hat{\tau}^{-k} \right) =  1 + \Psi(\hat{\tau}), \\
&& \Psi(\hat{\tau}) = \sum_{m_1=1}^{\infty} \frac{ \left( \sum_{k=1}^{\infty}  u^{({\bf 0})}_{T,k} \hat{\tau}^{-k} \right)^{m_1}}{\Gamma(m_1+1)}  \left( 1+  \sum_{m_2=1}^{\infty} \frac{\left( \sum_{|{\bf n}| > 0}^{\infty}  \bm{\sigma}^{{\bf n}} \bm{\zeta}^{{{\bf n}}}(\hat{\tau}) \sum_{k=1}^{\infty}  u^{({\bf n})}_{T,k} \hat{\tau}^{-k} \right)^{m_2}}{\Gamma(m_2+1)} \right) \quad \in {\cal H},
\ee
where $\Psi(\hat{\tau})$ converges to zero at late times, that is $\lim_{\hat{\tau} \rightarrow \infty}\Psi(\hat{\tau})=0$.

\subsection{The ${\bf c}$-$T$ dynamical system}
\label{subsec:T_C}
In the analysis of the previous section, we used $w$ as a time variable, but something else that may be done is to consider the same analysis for the dynamical system including explicitly the temperature, which will require constructing transseries for $T(\tau)$ as well.
In this subsection, we seek to prove that this dynamical system entails the same structure as \eqref{eq:dcdw_o}, which is compatible with Costin's prepared form. 

The dynamical system in the $\tau$-coordinate takes the following form
\be
 \frac{d {\bf c} }{d \tau}
  &=& - \frac{1}{\tau} \left( {\frak B} {\bf c} +  c_{01} {\frak D} {\bf c} +  {\bf A}  \right) - \frac{T {\bf c}}{\theta_0} ,  \label{eq:diff_tau1} \\
 \frac{d T}{d \tau} &=& - \frac{T}{3\tau} \left( 1 + \frac{ c_{01}}{10} \right), \label{eq:diff_tau2}
 \ee
where ${\frak B}$ is a constant matrix, ${\frak D}$ is a diagonal constant matrix, and ${\bf A}$ is a constant vector.
From Eq.~\eqref{eq:diff_tau2}, one can obtain the formal solution of $T(\tau)$ as
\be
T(\tau) &=& C_0 \left( \frac{\tau_0}{\tau} \right)^{1/3} \exp \left[ - \frac{1}{30} \int^{ \tau}_{\tau_{0}} \frac{d\tau^\prime }{\tau^\prime} \, c_{01}(\tau^\prime) \right] \nl
&=& C_0 \left( \frac{\tau_0}{\tau} \right)^{1/3} e^{C_{T}(\tau) - C_{T}(\tau_{0})}, 
\ee
where $C_0$ is an integration constant, and $C_T(\tau)$ is defined as
\be
&& C_{T} (\tau) =  - \frac{1}{30} \int \frac{d\tau }{\tau} \, c_{01}(\tau) .
\ee
In the IR, or the late-time limit $\tau \rightarrow \infty$, $T(\tau)$ is observed to behave like
\be
\lim_{\tau \rightarrow \infty} T(\tau) = T_0 \left( \frac{\tau_0}{\tau} \right)^{1/3},
\ee
with a constant $T_0$. Hence, by redefining this constant, the temperature $T(\tau)$ takes the form 
\be
T(\tau) =
\frac{\hat{T}_0 e^{C_{T}(\tau) }}{\tau^{1/3}}, \qquad \hat{T}_{0}  = T_0 \tau_{0}^{1/3}  . \label{eq:hat_T0}
\ee
Now, by working \eqref{eq:hat_T0} into \eqref{eq:diff_tau1} and changing the parameters $\hat{\tau}$ and $\hat{\theta}_{0}$ as
\be
\hat{\tau} = \tau^{2/3}, \qquad \hat{\theta_0} = \frac{\theta_0}{\hat{T}_{0}}, \label{eq:hat_tau_theta}
\ee
the dynamical system becomes
\be 
\frac{d {\bf c}(\hat{\tau})}{d \hat{\tau}} 
&=& - \frac{1}{\hat{\tau}} \left( {\frak B} {\bf c}(\hat{\tau}) +  c_{01}(\hat{\tau}) {\frak D} {\bf c}(\hat{\tau}) +  {\bf A}  \right) - \frac{\hat{T}(\hat{\tau})}{\hat{\theta}_0} {\bf c}(\hat{\tau}) ,  \label{eq:diff_tau12T} 
\ee
where
\be
&& \hat{T}(\hat{\tau}) =  e^{C_{T}(\hat{\tau})}, \qquad  C_{T} (\hat{\tau}) =  - \frac{1}{20} \int \frac{d \hat{\tau}}{\hat{\tau}} \, c_{01}(\hat{\tau}).
\ee
Note that the temperature $T(\hat{\tau})$ can be reproduced by
\be
T(\hat{\tau}) = \hat{T}_{0} \frac{\hat{T}(\hat{\tau}) }{\hat{\tau}^{1/2}}.
\ee
As discussed in App.~\ref{app:T_trans}, we find that $\hat{T}(\hat{\tau})$ will be expressed by
\be
\hat{T}(\hat{\tau}) = 1 + \Psi(\hat{\tau}),
\ee
where $\Psi(\hat{\tau})$ is a continuous function with a transseries formula identical to that of $\tilde{\bf c}$, which converges
to zero asymptotically. Therefore, Eq.~\eqref{eq:diff_tau12T} bears the familiar prepared form as in Eq.~\eqref{eq:dcdw_o}
with only one significant difference from a physical standpoint: the temperature now is a {\it dimension} in the phase space of the
 dynamical system. In the UV limit, the divergence of $T(\tau)$ as $\tau\rightarrow 0$ in general (see Eq.~\eqref{eq:diff_tau2}) suggests that $T=0$ and $T\ne0$ theories are disconnected from each other, and there cannot be any flow line (process) that connects a UV fixed point to the hydrodynamic fixed point 
 achieved at $\tau\rightarrow\infty$ in the original kinetic model of the Bjorken flow. 
\section{Bjorken dynamical system and its trasseries solutions for general $\tau_r = \theta_0/T^{1-\Delta}$} \label{app:general-Delta}

\begin{figure}[htpb]
\begin{center}
\includegraphics[scale=0.34]{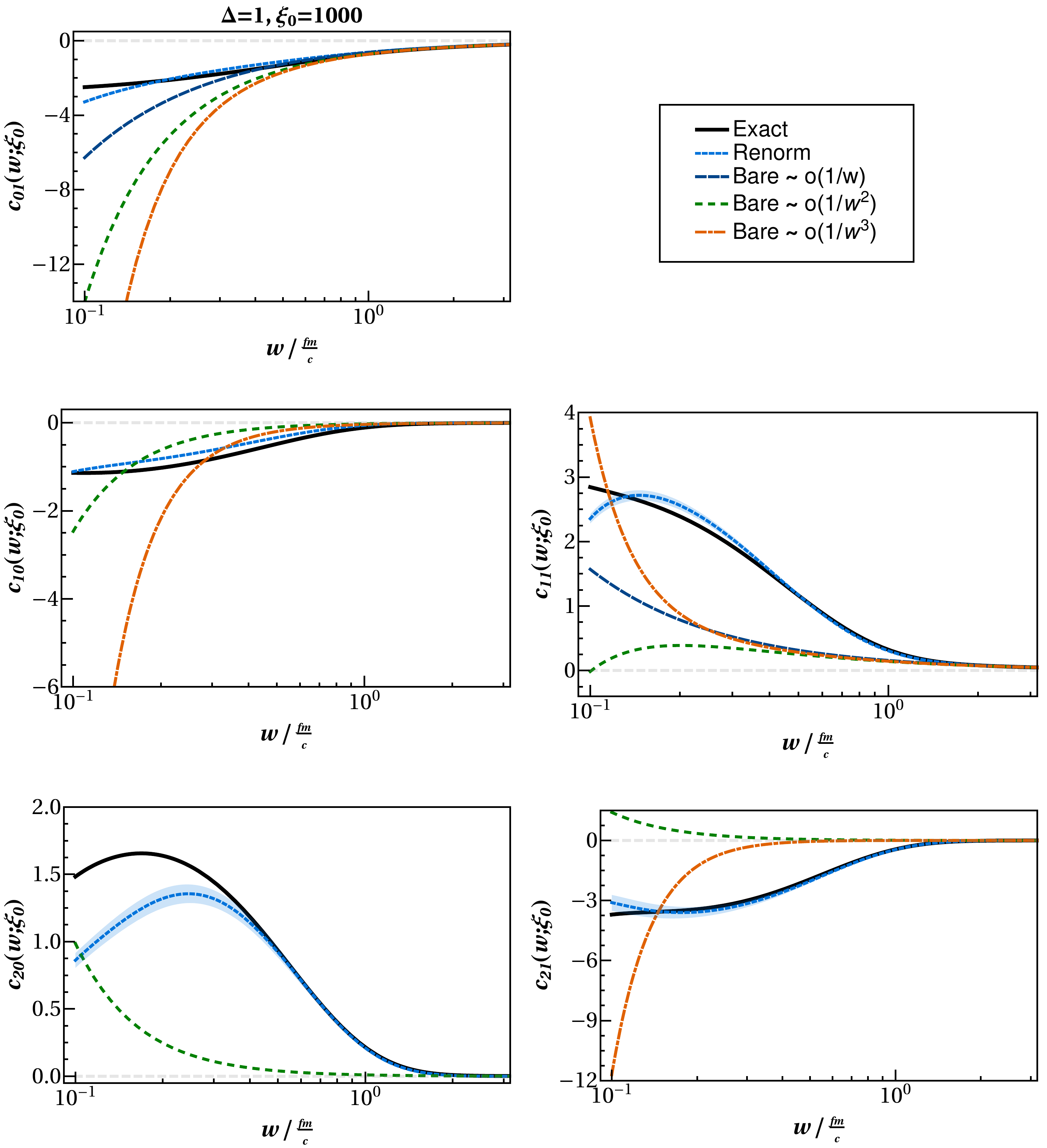}
\caption{Five lowest non-hydrodynamic moments for $\Delta=1$ system are computed using transseries, leading asymptotics and exact solution. The blue shaded area depicts the probable variation in the transseries due to standard deviation of integration constants $\sigma_i$. Each moment is calculated up to its leading asymptotic order and renormalized by transasymptotic matching including all the  transmonomials up to $15$th order.
}
\label{fig:transseries_Delta1}
\end{center}
\end{figure}

In this appendix, we extend the dynamical system in Eq.~\eqref{eq:ODEsBjor} to include general relaxation time, $\tau_r = \theta_0/T^{1-\Delta}$ with $0 \le \Delta \le 1$, which is very straightforward. The generalized system has the same number of asymptotic fixed points with their stability being identical to the case $\Delta=0$ as discussed in Sect.~\ref{subsec:fixedpoint}. So except for some small quantitative alterations in the flow lines around UV and IR, the qualitative picture of the phase portrait remains the same. In other words, there are no changes in global dynamics.

The generalized dynamical system is described by 
\be
\frac{d {\bf c}}{d \tau}
 &=& - \frac{2}{3 \tau} \left( {\frak B} {\bf c} + c_{01}{\frak D} {\bf c} + {\bf A}  \right) - \frac{T^{1-\Delta}}{\theta_0}{\bf c} ,  \label{eq:diff_tau1_gen} \\
 \frac{d T}{d \tau} &=& - \frac{T}{3\tau} \left( 1 + \frac{ c_{01}}{10} \right). \label{eq:diff_tau2_gen}
 \ee
As in the main bulk of the paper, we change to the coordinate $w = \tau T^{1-\Delta}$ and consider the dynamical system to be of the form
\be
\frac{d {\bf c}}{d w} &=& - \left({1+\Delta/2-\frac{1-\Delta}{20}c_{01}}\right)^{-1}
\left[  \hat{\Lambda} {\bf c} + \frac{1}{w} \left( {\frak B} {\bf c} +  c_{01} {\frak D} {\bf c} + {\bf A}  \right) \right],
\ee
where the matrices and vectors involved here have the same definitions as in \eqref{eq:dif_RH_cnl}.
Being again compatible with Costin's prepared form, we conclude that the transseries can be explained with the same form used to explore the solutions of the $\Delta=0$ model defined by \eqref{eq:ansatz_c}. By defining an invertible matrix $U$ for the diagonalization purposes again, one finds the recursive relation involving the transseries data as
\be
&& 20 \left[\left( (1+\Delta/2)\left( {\bf m} \cdot \tilde{\bf b} -k \right) + b_{i}    \right) \tilde{u}^{({\bf m})}_{i,k} + \left( \frac{3}{2 \theta_0} - (1+\Delta/2){\bf m} \cdot {\bf S} \right) \tilde{u}^{({\bf m})}_{i,k+1}  \right]  + 20\tilde{A}_i \,  \delta_{k,0} \delta_{{\bf m},{\bf 0}} \nl
&& - \sum_{|{\bf m}^\prime| \ge {\bf 0}}^{\bf m} \left[ (1-\Delta) \sum_{k^\prime=0}^{k} \left(
   {\bf m}^\prime \cdot \tilde{\bf b}  -k^\prime \right) \, u^{({\bf m}-{\bf m}^\prime)}_{1,k-k^\prime} \tilde{u}^{({\bf m}^\prime)}_{i,k^\prime} -20 \sum_{i^\prime=1}^I \sum_{k^\prime=0}^k u^{({\bf m}-{\bf m}^\prime)}_{1,k-k^\prime} \tilde{\frak D}_{ii^\prime}\tilde{u}^{({\bf m}^\prime)}_{i^\prime,k^\prime} 
  - (1-\Delta){\bf m}^{\prime} \cdot {\bf S}
  \sum_{k^\prime=0}^{k+1} u^{({\bf m}-{\bf m}^\prime)}_{1,k-k^\prime+1} \tilde{u}^{({\bf m}^\prime)}_{i,k^\prime} \right] = 0,  \label{eq:evo_nk_ngen_genDel} \nl
\ee
with the IR data $S_i$ and $\tilde{b}_i$ in the transseries being
\be
S_i = \frac{3}{(2+\Delta) \theta_0}, \qquad \tilde{b}_i = - \frac{1}{1+\Delta/2} \left( b_i - \frac{1-\Delta}{5(1+\Delta/2)} \right),
\ee
where $b_i$ are the eigenvalues of the matrix ${\frak B}$.

In addition, the transaymptotic mathing is obtained in the following, where $\hat{\zeta}_i = \partial/\partial \log {\zeta}_i$. Summing over ${\bf m}$ yields the transasymptotic matching for the general $\Delta$ as
\be
&& 20 \left[ \left(  (1+\Delta/2) (\tilde{\bf b} \cdot   \hat{\bm{\zeta}}   -k ) + b_{i}   \right) \tilde{C}_{i,k} - (1+\Delta/2) {\bf S} \cdot \hat{\bm{\zeta}}  \tilde{C}_{i,k+1}  +  \frac{3}{2 \theta_0} \tilde{C}_{i,k+1}  \right] + 20 \tilde{A}_i   \delta_{k,0} \nl
&& - (1-\Delta) \sum_{k^\prime=0}^{k} C_{1,k-k^\prime} \left( \tilde{\bf b} \cdot \hat{\bm{\zeta}}  -k^\prime \right) \tilde{C}_{i,k^\prime}
+ 20 \sum_{i^\prime=1}^I \sum_{k^\prime=0}^{k} C_{1,k-k^\prime} \tilde{\frak D}_{ii^\prime} \tilde{C}_{i^\prime,k^\prime}
+  (1-\Delta) \sum_{k^\prime=0}^{k+1} C_{1,k-k^\prime+1}     {\bf S} \cdot \hat{\bm{\zeta}} \tilde{C}_{i,k^\prime}= 0, \label{eq:transasymptotic_genDel}
\ee

The transseries and leading order asymptotics of five moments $c_{01},\dots, c_{21}$ for $\Delta=1$ system are depicted in Fig. \ref{fig:transseries_Delta1}. To avoid repeating ourselves here, we refer the interested reader to the explanations given in the text prior to  Fig.~\ref{fig:transseries_Delta0}.

\subsection{Transasymptotic matching for $c_{01}$ of the $N=0, L=1, \Delta=1$ dynamical system}
\label{sec:transasymptotic_Delta1}

As a result of Eq.~\eqref{eq:transasymptotic_genDel}, the mode-to-mode coupling term $c_{01}\frac{d c_{nl}}{d w}$ vanishes, and therefore the transasymptotic matching effectively becomes a finite expansion of the transmonomials packaged into $\zeta$ at each order. The first five transseries coefficients
are then found to be exactly given by
\bea
\tilde C_{01,0}(\zeta) &=& \zeta,\\
\tilde C_{01,1}(\zeta) &=& -\frac{2\theta_0}{45} (3 \zeta^2 - 16\zeta + 60), \\
\tilde C_{01,2}(\zeta) &=& \frac{2\theta_0^2}{4725} \left( 42 \zeta^3 + 17 \zeta^2 -3300 \right), \\
\tilde C_{01,3}(\zeta) &=& -\frac{8\theta_0^3}{1488375} \left( 441 \zeta^4 + 2709 \zeta^3 + 55283 \zeta^2+219600\right),\\
\tilde C_{01,4}(\zeta) &=& \frac{4 \theta^4}{1406514375} \left( 111132 \zeta^5 + 1320354 \zeta^4 + 28776951 \zeta^3 + 290390942 \zeta^2 -698220000 
\right).
\eea

\section{Review of the exact solution of the RTA-BE}
\label{app:RTABE}
In this section, we briefly touch down on some aspects of the exact RTA-BE that have been taken advantage of throughout this paper. For a more complete and detailed explanation of this solution, we refer the reader to Refs.~\cite{Baym:1984np,Florkowski:2013lya,Florkowski:2013lza}. 

The exact solution of the RTA-BE~\eqref{eq:RTABoltzmann} is
\be
\label{eq:sol}
f\left(\tau,p_T,p_\varsigma\right)\,=\,D\left(\tau,\tau_0\right)\,f_0\left(\tau_0,p_T,p_\varsigma\right)\,+\,\int_{\tau_0}^{\tau}\,\frac{d\tau'}{\tau_r(\tau')}\,D(\tau,\tau')\,f_{eq.}\left(\tau',p_T,p_\varsigma\right)\,,
\ee
where $D(\tau_2,\tau_1)=\exp\left[-\int_{\tau_1}^{\tau_2}d\tau'/\tau_r(\tau')\,\right]$ and $f_0\left(\tau_0,p_T,p_\varsigma\right)$ is the initial distribution function at $\tau=\tau_0$. Here, we consider the RS distribution function~\cite{Romatschke:2003ms} 
\be
\label{eq:RSansatz}
f_0\left(\tau_0,p_T,p_\varsigma\right)=\exp\left[{\Lambda_0}^{-1}{\sqrt{p_T^2\,+\,(1+\xi_0)\,(p_\varsigma/\tau_0)^2}}\right]\,,
\ee
where $\Lambda_0$ and $\xi_0$ are the initial temperature and initial momentum anisotropy along the $\varsigma$ direction, respectively. With \eqref{eq:RSansatz} at hand, the energy-momentum conservation law leads to the following integral equation~\cite{Baym:1984np,Florkowski:2013lza,Florkowski:2013lya}:
\be
\label{eq:tempeq}
\begin{split}
T^4(\tau)\,=\,T^4_0\,D(\tau,\tau_0)\,\frac{\mc R_{200}\left(\tau,\tau_0/\sqrt{1+\xi_0}\right)}{\mc R_{200}\left(\tau_0,\tau_0/\sqrt{1+\xi_0}\right)}\,+\,\int_{\tau_0}^{\tau}\,\frac{d\tau'}{\tau_r(\tau')}\,T^4(\tau')\,D(\tau,\tau')\,\frac{\mc R_{200}(\tau,\tau')}{\mc R_{200}(\tau,\tau)}\,.
\end{split}
\ee
The functions $\mc R_{nqr}$ appearing in the previous expression are given by~\cite{Molnar:2016gwq}
\be
\label{eq:Rnqr}
\mc R_{nqr}\left(\tau,\tau'\right)=\int_0^\pi\,d\theta\,\frac{\cos^r\theta\,\sin^{2q+1}\theta}{\left[\left(\frac{\tau}{\tau'}\right)^2\,\cos^2\theta+\sin^2\theta\right]^{\frac{n+2}{2}}}.
\ee
Eq.~\eqref{eq:tempeq} is solved with the help of the iterative methods discussed in Refs.~\cite{Banerjee:1989by,Florkowski:2013lza,Florkowski:2013lya,Denicol:2014tha}.
By plugging the exact solution~\eqref{eq:sol} in the definition of the moments $\cnl$ in~\eqref{eq:momnl}, one finds 
\bs
\label{eq:exact-cnl}
\beal
\cnl(\tau)=2\pi^2(4l+1)&\left[\,D(\tau,\tau_0)\,\mc B_{nl}\left(\tau,\tau_0;1+\xi_0,\frac{T_0}{\mc R^{1/4}(\xi_0)\,T(\tau)}\right)
+\,\int_{\tau_0}^\tau\,\frac{d\tau'}{\tau_r(\tau')}\,D(\tau,\tau')\,\mc B_{nl}\left(\tau,\tau';1,\frac{T(\tau')}{T(\tau)}\right)
\right]\,,
\end{align}
\es
where
\be
\label{eq:bint2}
\begin{split}
\mc B_{nl}\left(\tau,\tau';\lambda_\parallel,\frac{\Lambda(\tau')}{T(\tau)}\right)&=\frac{\Gamma(n+1)}{\Gamma(n+4)}\,\int_0^\pi\,\frac{d(\cos\theta)}{(2\pi)^2}\,\mc P_{2l}(\cos\theta)
\,\int_0^\infty\,dr\,r^3\,\mc L^{(3)}_n(r)\,\exp\left[-\frac{T(\tau)}{\Lambda(\tau')}\,r\,\sqrt{\sin^2\theta+\lambda_{\parallel}\left(\frac{\tau}{\tau'}\right)^2\cos^2\theta}\right]\,\\
&=\frac{1}{(2\pi)^2}\,\left(\frac{\Lambda(\tau')}{T(\tau)}\right)^4\,
\,\int_{-1}^1\,dx\,\mc P_{2l}(x)\,
\frac{\left[\left(1+\left(\lambda_\parallel\left(\frac{\tau}{\tau'}\right)^2-1\right)x^2\right)^{1/2}-\frac{\Lambda(\tau')}{T(\tau)}\right]^n}{\left[1+\left(\lambda_\parallel\left(\frac{\tau}{\tau'}\right)^2-1\right)x^2\right]^{(n+4)/2}}.
\end{split}
\ee
In Eq.~\eqref{eq:exact-cnl} the reader must bear in mind that the Landau matching condition for the energy density at $\tau=\tau_0$ when using the RS distribution function~\eqref{eq:RSansatz} implies $T_0=\mc R^{1/4}(\xi_0)\Lambda_0$, where the function $\mc R(\xi)$ is defined by~\cite{Martinez:2010sc}
\be
\label{eq:Rfun}
R(\xi)=\frac{1}{2}\left[\frac{1}{1+\xi}+\frac{\arctan(\sqrt{\xi})}{\sqrt{\xi}}\right]\,.
\ee
The moments $\bar{M}^{nl}$~\eqref{eq:mnm} can now be evaluated from the exact solution~\eqref{eq:sol} as follows
\be
\label{eq:mnmexact}
\bar M^{nm} = \frac{M^{nm}}{M^{nm}_{eq}} = \frac{ D(\tau,\tau_0) F^{nm}_{0}(\tau,\tau_0)+\int_{\tau_0}^\tau \frac{d\tau^\prime}{\tau_r(\tau^\prime)} D(\tau,\tau^\prime) F^{nm}_{eq}(\tau,\tau^\prime)}{F^{nm}_{eq}(\tau)},
\ee
where
\bs 
\label{eq:mnmexactfactor}
\beal
F^{nm}_0(\tau,\tau^\prime) &=\frac{\Gamma(n+2m+2)}{(2\pi)^2}R(\xi_0)^{-\frac{n+2m+2}{4}}T(\tau^\prime)^{n+2m+2}\int_{-1}^1 dx \frac{x^{2m}}{\left(\sqrt{1+\xi_0 x^2}\sqrt{1+\left(\left(\frac{\tau}{\tau^\prime}\right)^2-1\right)x^2}\right)^{n+2m+2}}, \\
F^{nm}_{eq}(\tau,\tau^\prime) &= \frac{\Gamma(n+2m+2)}{(2\pi)^2}T(\tau^\prime)^{n+2m+2}\int_{-1}^1 dx {x^{2m}}{\left(\sqrt{1+\left(\left(\frac{\tau}{\tau^\prime}\right)^2-1\right)x^2}\right)^{-(n+2m+2)}}.
\end{align}
\es
The exact solutions for both $\cnl$ and $\bar{M}_{nl}$ are determined numerically by having the expression for the temperature from Eq.~\eqref{eq:tempeq}. 
\section{Hydrodynamization of soft and hard modes for $\Delta=1$ system }
\label{app:thermalizationDelta1}
%
\begin{figure}[!htpb]
    \centering
    \includegraphics[width=0.9\linewidth]{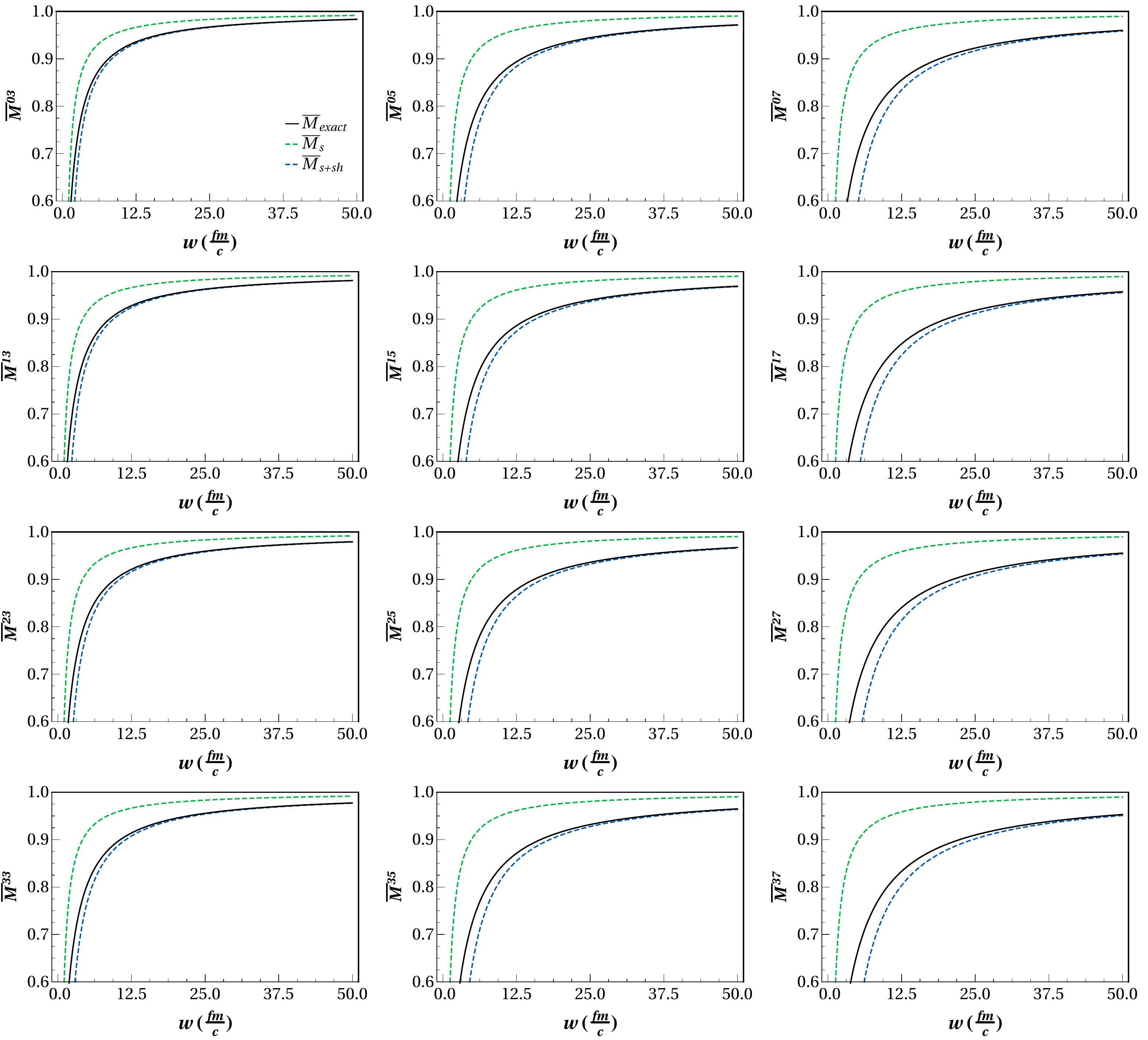}\\
\caption{Time evolution of several normalized moments $\bar M^{nm}$ ($n>0,l\geq 1$)in the case where the power law dependence of the relaxation time is a constant, i.e., $\tau_r=\theta_0$. In this plot we present the numerical results of the exact RTA solution~\eqref{eq:mnmexact} (black line), soft~\eqref{eq:mnm-NS} (green dot-dashed line) and soft+semi-hard~\eqref{eq:mnm-c11} (blue dashed line) limits. The initial conditions are the same as in Fig.~\ref{fig:thermalization}.
}
\label{fig:thermalizationDelta1}
\end{figure}
\begin{figure}[!htpb]
    \centering
    \includegraphics[width=0.9\linewidth]{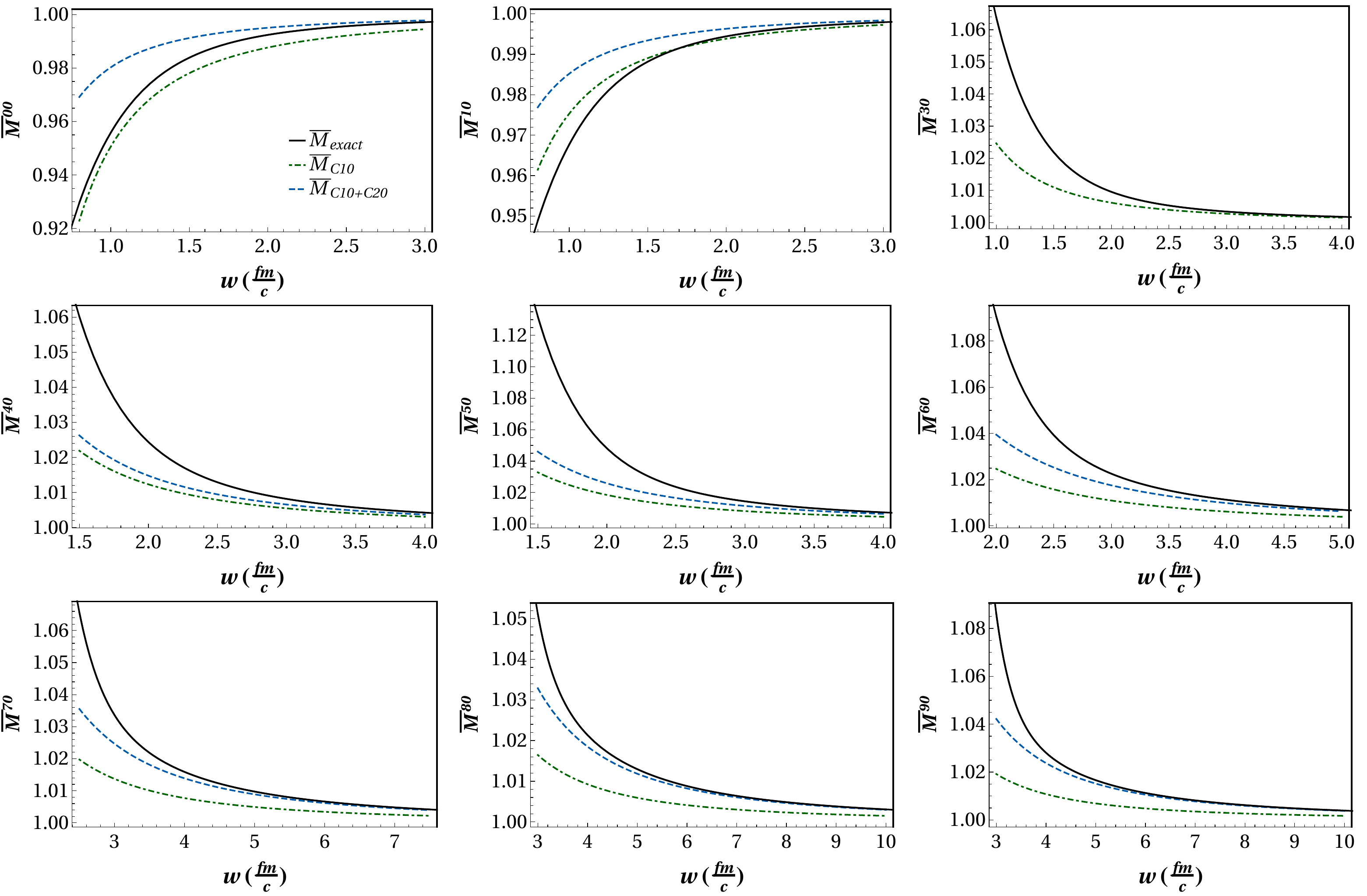}\\
\caption{Time evolution of different normalized moments $\bar M^{n0}$ in the case where the power law dependence of the relaxation time is a constant, i.e., $\tau_r=\theta_0$. In this plot we present the numerical results of the exact RTA solution~\eqref{eq:mnmexact} (black line)and the hard limits given by Eqs.~\eqref{eq:mn0-cna} (green dot-dashed line) and ~\eqref{eq:mn0-cnb} (blue dashed line). The saturation bounds are set to at the same level as before. The initial conditions are the same as in Fig.~\ref{fig:thermalization}.}
\label{fig:thermalizationDelta1_Mn0}
\end{figure}

\begin{figure}[!htpb]
\begin{center}
\includegraphics[width=0.9\linewidth]{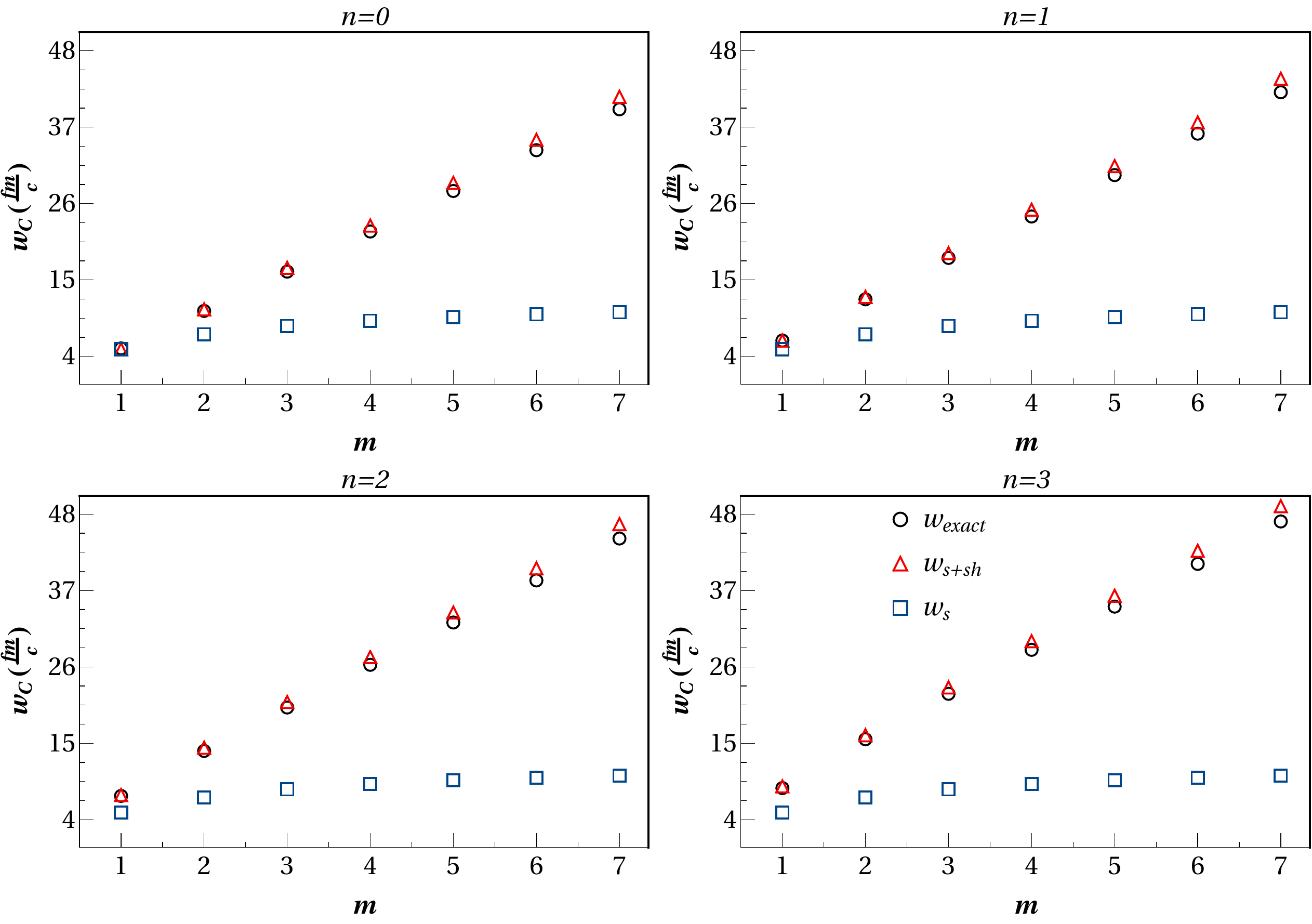}\\
\caption{Saturation value of $w_c$ vs. $m$ for different values of $n=\{0,1,2,3\}$ (top right, top left, bottom right and bottom left panels, respectively) for the exact RTA (black circle), soft (blue square) and soft+semi-hard (red triangle) regimes.}
\label{fig:thermtimeDelta1}
\end{center}
\end{figure}
\begin{figure}[!htpb]
\begin{center}
\includegraphics[width=0.5\linewidth]{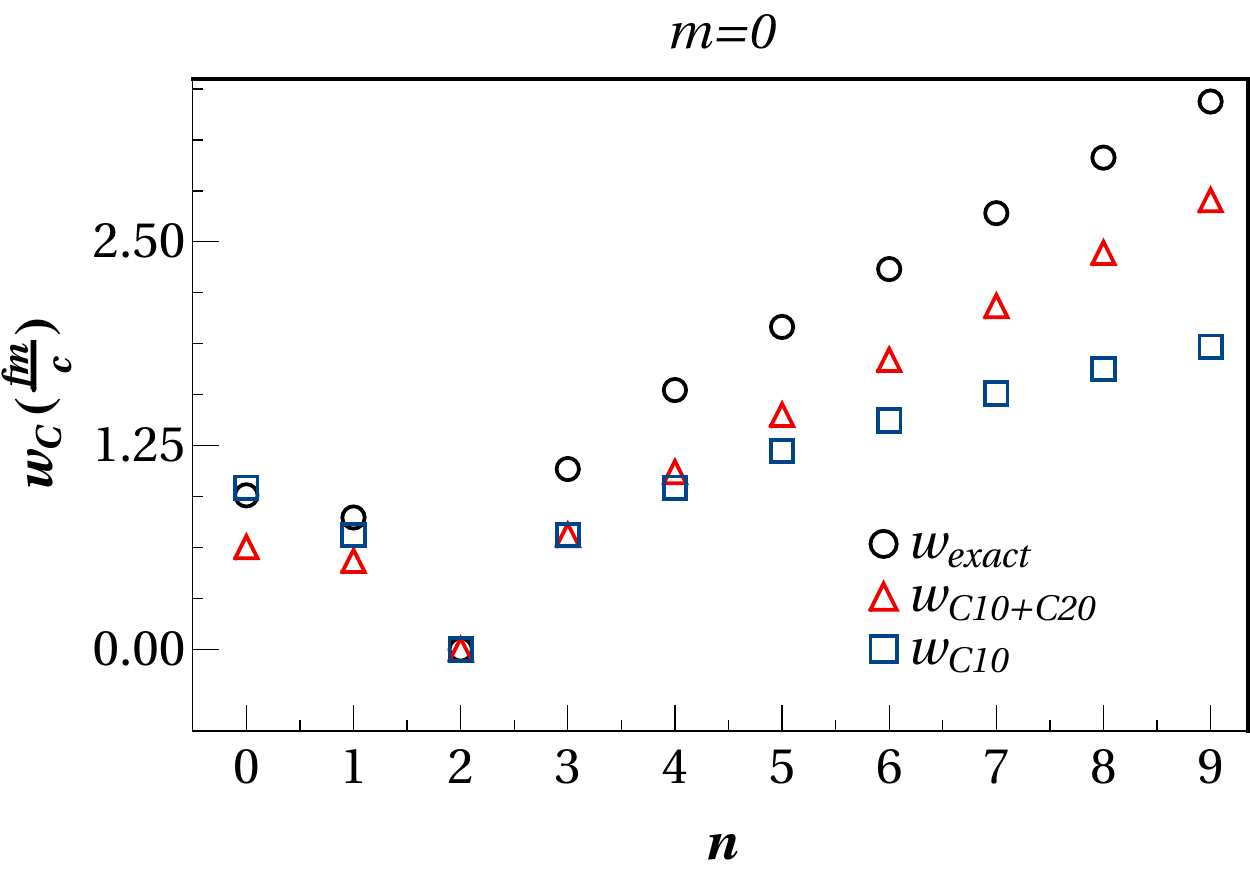}\\
\caption{Saturation value of $w_c$ vs. $n$ of the moments $\bar M^{n0}$ for the exact RTA (black circle) and hard limits,  Eqs.~\eqref{eq:mn0-cna} (blue square) and~\eqref{eq:mn0-cnb}(red triangle).}
\label{fig:thermtimeDelta1_Mn0}
\end{center}
\end{figure}
In Sect.~\ref{sec:hydrodyn} we analyzed the hydrodynamization process of the normalized moments $\bar M^{nm}$ for the conformal system, i.e., $\tau_r=\theta_0/T$. Here, we give the results of hydrodynamization when $\tau_r=\theta_0/T^{1-\Delta}$ by fixing $\Delta=1$. For this choice, the  variable $w=\tau (fm/c)$ becomes dimensionful. The soft and semi-hard limits of the normalized moments are given by the expressions in Eqs.~\eqref{eq:mnm-NS} and~\eqref{eq:mnm-c11}, respectively. Also, the exact moments $\bar M^{nm}$ are given by Eq.~\eqref{eq:mnmexact}. 

The results of the hydrodynamization of the normalized moments are presented in Figs.~\ref{fig:thermalizationDelta1}-\ref{fig:thermtimeDelta1_Mn0}. From these plots, we arrive at the conclusions already discussed in Sect~\ref{sec:hydrodyn}: the larger $n$ and $m$, the longer it takes for $\bar M^{nm}$ ($n>0,l\geq 1$) to thermalize while $\bar M^{n0}$ ($n\geq 0$). Furthermore, the IR behavior is uniquely determined by both non-hydrodynamic modes $c_{01}$ and $c_{11}$ for $\bar M^{nm}$ ($n>0,l\geq 1$) while the non-hydrodynamic modes $c_{10}$ and $c_{20}$ for $\bar M^{nm}$ ($n\geq 0$).  
\begin{figure}[!htpb]
\begin{center}
\includegraphics[width=0.8\linewidth]{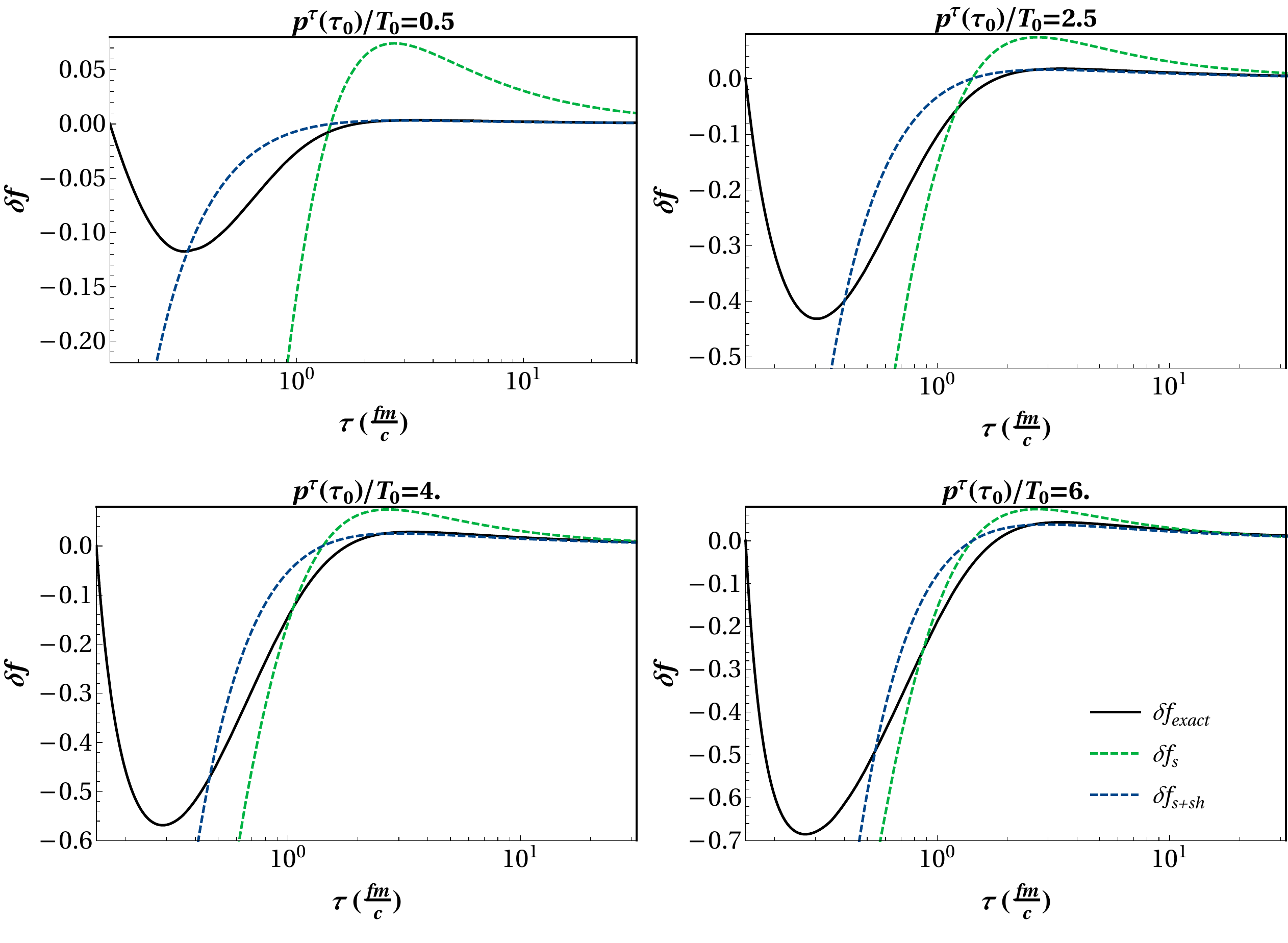}\\
\caption{Time evolution of the deviation from the distribution function~\eqref{eq:sol} for the exact RTA-BE solution when $\tau_r=\theta_0$. The black lines represent the exact result~\eqref{eq:sol} while the green dot-dashed and blue dashed lines correspond to the CNS~\eqref{eq:deltaf-cns} and ENS~\eqref{eq:deltaf-ens} limits. The initial conditions are the same as in Fig.~\ref{eq:deltaf-ens}.}
\label{fig:distributionf_Delta1}
\end{center}
\end{figure}
When calculating the deviation from equilibrium by following the same procedure discussed in Sect.~\ref{sec:hydrodyn} we reach the same conclusions presented in there: the soft regime does not capture the late-time asymptotics correctly while there is an extremely good match between the soft+semi-hard regime and exact results. This non-trivial double check gives a solid evidence of the importance of considering not only the effective shear mode $c_{01}$ but also the new non-hydrodynamic mode $c_{11}$.

\section{Exact solution vs. truncation of the dynamical system}
\label{app:impacttruncation}
In the numerical analysis presented in this work, we used various truncations in the number of moments in the original dynamical system \eqref{eq:cnlevol}.  Both the exact solution to RTA-BE and the solution to the truncated dynamical system share the same IR fixed point but as discussed in Sect.~\ref{sec:global}, the solution of the truncated dynamical system cannot fully agree with the exact solution at early times due to the missing corrections due to the truncation. For simplicity we take $\Delta =0$ in Eq.~\eqref{eq:rela-time}. 

\begin{figure}[!htpb]
    \centering
        \includegraphics[width=\linewidth]{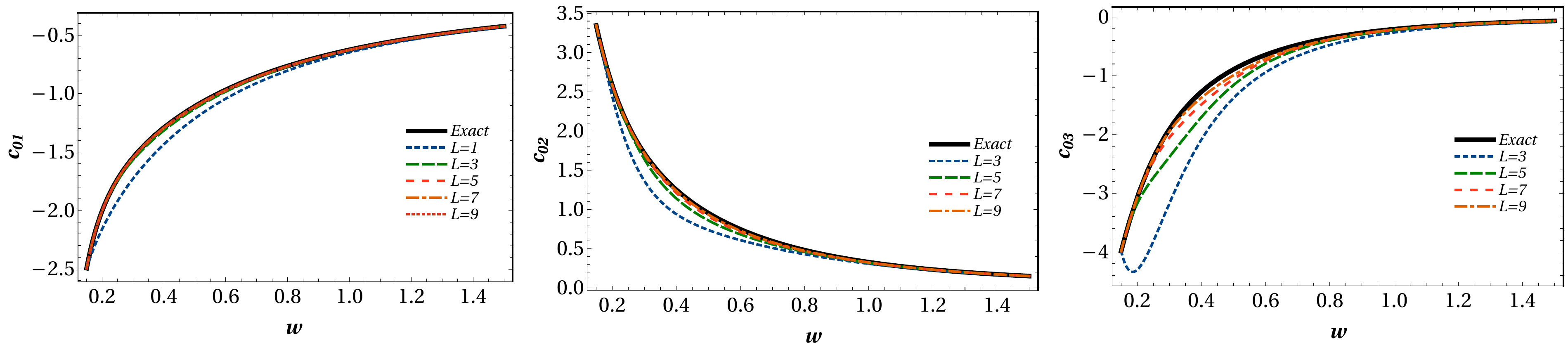}
        \vspace{1em}
        \includegraphics[width=\linewidth]{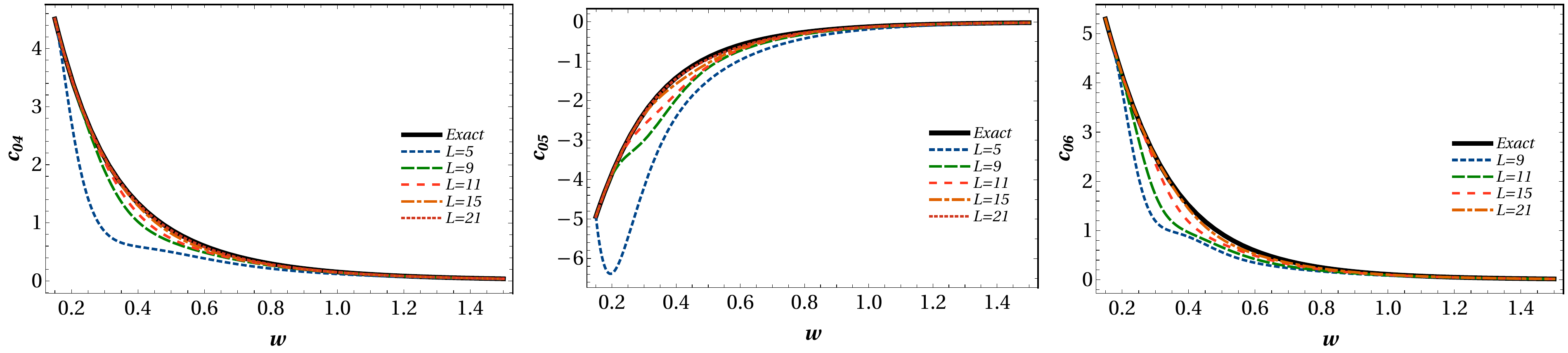}
\caption{Impact of truncation on $c_{01},\dots,c_{06}$. Exact solution to RTA-BE is shown with a black line. Dashed lines correspond to $c_{0l}$ in a truncated system of $L$-moments.}
\label{fig:truncation_c0l}
\end{figure}

\begin{figure}
\centering
\includegraphics[width=\linewidth]{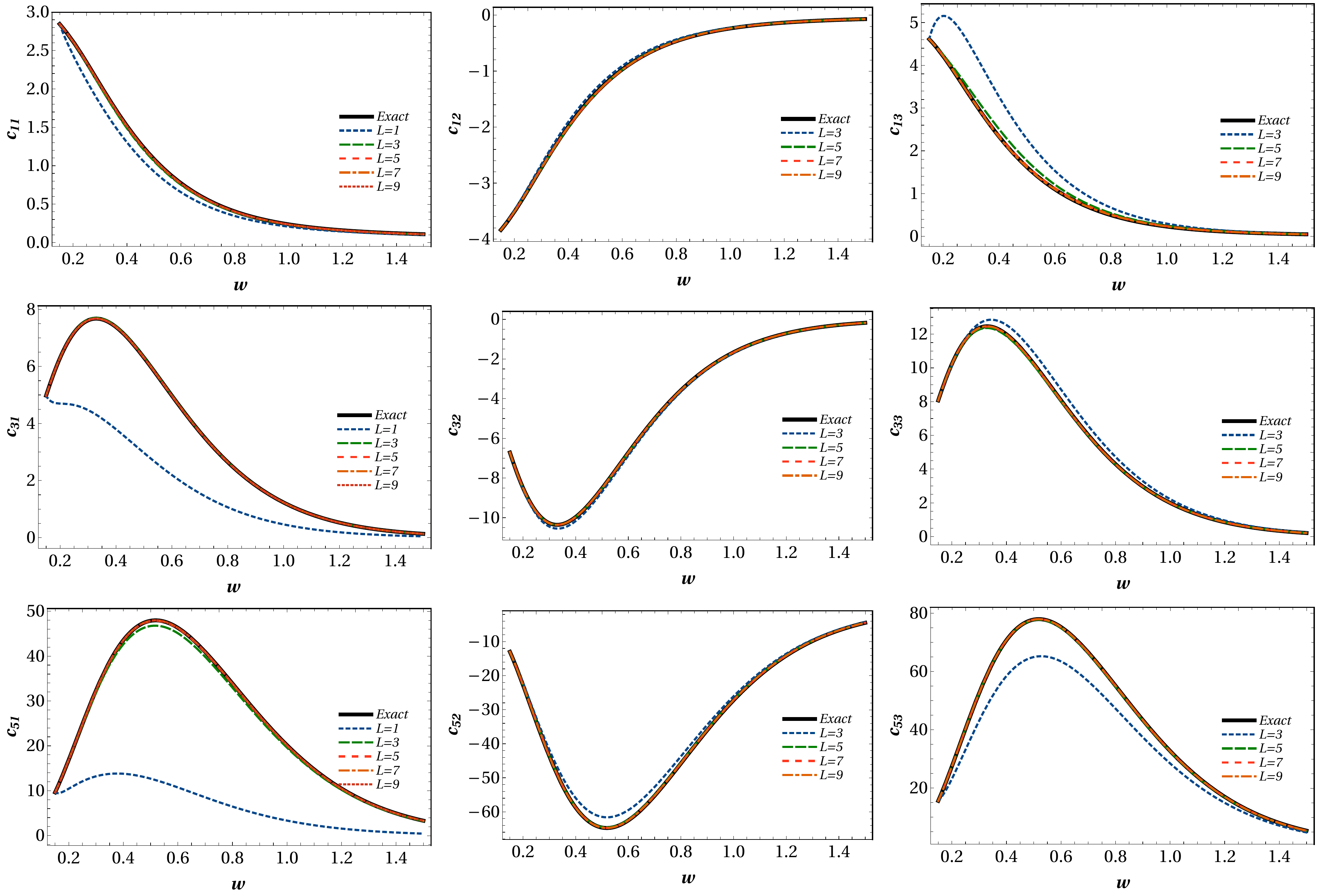}
\caption{Exact solutions for $c_{nl}$ of both RTA-BE and truncated system. Dashed lines represent $c_{nl}$ in a truncated system composed of $N=5,~L$-moments. The initial conditions for the exact $\cnl$~\eqref{eq:bint2} are $\tau_0=0.25~\mathrm{fm/c},T_0=0.6~\mathrm{GeV}$ and $\xi_0=1000$. }
\label{fig:truncation_N5_L3}
\end{figure}

In this section, we test how good the truncation schemes would match the exact RTA-BE result. We compare the moments $\cnl$ obtained from Eq.~\eqref{eq:mnmexact} by solving RTA-BE exactly against solutions to the truncated dynamical system in~\eqref{eq:momweq}, where the truncation is controlled by the number of involved moments. For this reason, we solve numerically the dynamical system in~\eqref{eq:cnlevol} by varying its dimension from $1$ to $L$ and compare against exact result~\eqref{eq:momnl} by using the same set of initial values of the moments $\cl$. The initial conditions for the exact $\cnl$~\eqref{eq:bint2} are chosen to be $\tau_0=0.25~\mathrm{fm/c},T_0=0.6~\mathrm{GeV}$ and $\xi_0=1000$. In Fig.~\ref{fig:truncation_c0l} we present the numerical results for our findings by considering the initial conditions $c_{0l}(w_0)=c_{0l}(w_0;\xi_0)$. When the dimension of the truncated dynamical system increases, they get closer to the exact solutions of the full RTA-BE. We notice that the larger the index $l$, the more moments are needed in order to match the exact solution of RTA-BE. 
In Fig. \ref{fig:truncation_N5_L3}, the exact moments $c_{nl}\quad (n=1,3,5)$ are compared to the solutions of the truncated system. This figure encodes the information of the non-linear structure of the original dynamical system: the evolution equation of $\cnl$ couples only to the moments of the same or lower order in $n$, and at the same time it couples to the next moment $c_{n,l+1}$. So $\cnl$ in Fig.~\ref{fig:truncation_N5_L3} become closer to the exact RTA result if $L$ is increased for a fixed value of $n$. We comment that the numerical results presented in Fig.~\ref{fig:truncation_c0l} and~\ref{fig:truncation_N5_L3} do not depend on the particular choice of initial conditions as long as we are bound to the basin of attraction of 
the invariant manifold of the truncated system.
%

\bibliography{lyapunov_transportcoeff}
\end{document}